\documentclass[11pt, a4paper]{article}
\pdfoutput=1
\usepackage{jheppub}
\usepackage{booktabs}
\usepackage{subcaption}
\usepackage[T1]{fontenc}
\usepackage{float}
\usepackage{bm}
\usepackage{latexsym, amsmath, amstext, amssymb, amsfonts, amscd, bm, array, multirow, amsbsy, mathrsfs}
\usepackage{amsthm}
\usepackage{t1enc}
\usepackage{indentfirst}
\usepackage{pb-diagram}
\usepackage{graphicx}
\usepackage{enumerate}
\usepackage{color}
\usepackage[all]{xy}
\usepackage{hyperref}
\hypersetup{colorlinks=false,pdfborderstyle={/S/U/W 0}}

\newcommand{\sslash}{\mathbin{/\mkern-6mu/}}

\usepackage{url}

\DeclareMathOperator{\arccot}{arccot}

\newcommand{\bl}{\begin{Lemma}}
	\newcommand{\el}{\end{Lemma}}
\newcommand{\bt}{\begin{Theorem}}
	\newcommand{\et}{\end{Theorem}}
\newcommand{\bd}{\begin{definition}}
	\newcommand{\ed}{\end{definition}}

\newcommand{\Aut}{\mathrm{Aut}}

\newcommand{\eqdef}{\stackrel{{\rm def.}}{=}}

\DeclareFontFamily{U}{rsf}{}
\DeclareFontShape{U}{rsf}{m}{n}{<5> <6> rsfs5 <7> <8> <9> rsfs7 <10-> rsfs10}{}
\DeclareMathAlphabet\Scr{U}{rsf}{m}{n}

\def\cU{\mathcal{U}}

\def\Z{\mathbb{Z}}
\def\C{\mathbb{C}}
\def\R{\mathbb{R}}

\def\H{\mathbb{H}}

\def\SL{\mathrm{SL}}
\def\PSL{\mathrm{PSL}}

\def\dd{\mathrm{d}}

\def\vol{\mathrm{vol}}

\def\bN{\boldsymbol{\mathcal{N}}}

\def\balpha{{\boldsymbol{\alpha}}}

\def\bt{\mathbf{t}}

\newcommand{\be}{\begin{equation*}}
\newcommand{\ee}{\end{equation*}}
\newcommand{\ben}{\begin{equation}}
\newcommand{\een}{\end{equation}}
\newcommand{\beqa}{\begin{eqnarray*}}
	\newcommand{\eeqa}{\end{eqnarray*}}
\newcommand{\beqan}{\begin{eqnarray}}
\newcommand{\eeqan}{\end{eqnarray}}

\newcommand{\nn}{\nonumber}

\newcommand{\Tr}{\mathrm{Tr}}

\def\Hess{\mathrm{Hess}}

\def\cR{{\mathcal R}}

\def\cC{{\mathcal C}}

\def\tphi{\tilde{\varphi}}

\newcommand{\cM}{\mathcal{M}}

\def\SO{\mathrm{SO}}

\def\cD{\mathcal{D}}

\def\hSigma{\widehat{\Sigma}}

\def\cN{\mathcal{N}}
\def\cG{{\mathcal{G}}}

\def\cC{\mathcal{C}}

\def\G_2{\mathrm{G_2}}

\def\cL{\mathcal{L}}

\def\fD{\mathfrak{D}}
\def\fM{\mathfrak{M}}

\def\P{\mathbb{P}}

\def\mD{\mathbf{D}}

\def\G{\mathrm{G}}

\def\Im{\mathrm{Im}}

\def\hPhi{\hat{\Phi}}

\def\mD{\mathbb{D}}
\def\fD{\mathfrak{D}}

\def\grad{\mathrm{grad}}

\def\bd{\boldsymbol{\dd}}

\def\rS{\mathrm{S}}

\def\area{\mathrm{area}}

\def\Iso{\mathrm{Iso}}
\def\Re{\mathrm{Re}}
\def\Im{\mathrm{Im}}

\def\i{\mathbf{i}}
\def\sc{\mathrm{sc}}

\def\fB{\mathfrak{B}}

\def\tPhi{{\tilde \Phi}}
\def\tvarphi{{\tilde \varphi}}

\def\cK{\mathcal{K}}

\def\CP{\mathbb{C}\mathbb{P}}

\def\rC{\mathrm{C}}

\def\tv{\tilde{v}}
\def\balpha{\boldsymbol{\alpha}}
\def\bPhi{\boldsymbol{\Phi}}
\def\btPhi{\boldsymbol{\tPhi}}
\def\bH{\mathbf{H}}
\def\tcR{\tilde{\cR}}
\def\fB{\mathfrak{B}}

\def\fM{\mathfrak{M}}

\title{Generalized two-field $\alpha$-attractor models from the hyperbolic triply-punctured sphere}

\author{Elena Mirela Babalic$^{1,2}$, Calin Iuliu Lazaroiu$^{1,2}$}

\affiliation{$^1$ Center for Geometry and Physics, Institute for Basic
  Science, Pohang 37673, Republic of Korea\\
$^2$ Horia Hulubei National Institute for Physics and Nuclear Engineering (IFIN-HH), 
Reactorului 30, POB-MG6, Bucharest-Magurele 077125, Romania}

\emailAdd{mbabalic@theory.nipne.ro, calin@ibs.re.kr}

\abstract{We study generalized two-field $\alpha$-attractor models
whose rescaled scalar manifold is the triply-punctured sphere endowed
with its complete hyperbolic metric, whose underlying complex manifold
is the modular curve $Y(2)$. Using an explicit embedding into the end
compactification, we compute solutions of the cosmological evolution
equations for a few globally well-behaved scalar potentials,
displaying particular trajectories with inflationary behavior as well
as more general cosmological trajectories of surprising complexity. In
such models, the orientation-preserving isometry group of the scalar
manifold is isomorphic with the permutation group on three elements,
acting on $Y(2)$ as the group of anharmonic transformations. When the
scalar potential is preserved by this action, $\alpha$-attractor
models of this type provide a geometric description of two-field
``modular invariant $j$-models'' in terms of gravity coupled to a
non-linear sigma model with topologically non-trivial target and with
a finite (as opposed to discrete but infinite) group of
symmetries. The precise relation between the two perspectives is
provided by the elliptic modular function $\lambda$, which can be
viewed as a field redefinition that eliminates almost all of the
countably infinite unphysical ambiguity present in the Poincar\'e
half-plane description of such models.}

\begin{document}

\maketitle

\pagebreak

% \vskip .6in

\section*{Introduction}
\label{Intro}

Reference \cite{genalpha} introduced a very large class of two-field
cosmological models called {\em generalized two-field
$\alpha$-attractors}. These form a tractable sub-family of the
extremely wide class of two-field cosmological models, which -- along
with general multi-field models -- are a subject of active theoretical
interest (see, for example, \cite{PT1,PT2,
m1,m2,m3,m4,m5,m6,Sch01,Sch02,Sch1,Sch2,Gong,Dias1,Dias2,Mulryne,c1,BBJMV}).
Indeed, multi-field models are easier to produce than one field models
in fundamental theories of gravity and future observations at higher
precision that current data \cite{Planck} (which at present are well
accounted for by one-field models) may allow one to detect multi-field
effects in the future.

Generalized two-field $\alpha$-attractors are derived from
four-dimensional gravity coupled to a non-linear sigma model whose
scalar manifold $\Sigma$ is a borderless, connected, oriented and
non-compact smooth surface, endowed with a complete metric $\cG$ of
constant negative curvature, with a scalar potential given by a smooth
real-valued function $\Phi$ defined on $\Sigma$. The cosmological
models derived from such a theory provide a wide generalization of the
two-field version of ordinary $\alpha$-attractors \cite{alpha5,Escher}
(for which $(\Sigma,\cG)$ is -- up to a constant rescaling of the
metric -- given by the hyperbolic disk). In a naive one-field
truncation, such generalized models have the same universal behavior
as one-field $\alpha$-attractors \cite{alpha3,alpha4} for those
special trajectories for which the fields undergo slow motion along
geodesics flowing from an end of $\Sigma$ where the scalar potential
is ``well-behaved'' and ``locally maximal'' toward the compact core of
$\Sigma$ (see \cite{genalpha} for details). This property justifies
the name `generalized two-field $\alpha$-attractors' which was given
in \cite{genalpha} to such models. The role of the parameter $\alpha$
is played by the quantity $\frac{1}{3|K|}$, where $K<0$ is the
Gaussian curvature of $\cG$. Equivalently, we have $\cG=3\alpha G$,
where $G$ is a complete hyperbolic metric on $\Sigma$, i.e. a complete
metric of constant Gaussian curvature equal to $-1$. As shown in
\cite{genalpha}, such models can be studied using uniformization
theory (see \cite{unif,Katok,Beardon,FN} for an introduction). They
allow for multipath inflation when the scalar potential or the
topology of $\Sigma$ are sufficiently nontrivial and they have
intricate cosmological dynamics already when $(\Sigma,G)$ is an
elementary hyperbolic surface \cite{elem}. We shall freely use certain
results and notions from the geometry of hyperbolic surfaces, which we
summarized in \cite[Appendices B,C,D]{genalpha} for the convenience of
cosmologists. We refer the reader to loc. cit. for relevant
information.  As explained in the present paper, those special
two-field generalized $\alpha$-attractors for which the underlying
complex manifold of $(\Sigma,G)$ is a modular curve can be related
through uniformization to the modular-invariant cosmological models
considered in \cite{Sch01,Sch1,Sch2}. The reader should notice that
the underlying complex curve of a hyperbolic surface $(\Sigma,G)$ of a
generalized two-field $\alpha$-attractor model need {\em not} be a
modular curve. Indeed, such a hyperbolic surface may have infinite
area and may have ends whose hyperbolic type is not that of a cusp
end. The simplest such examples are provided by the elementary
hyperbolic surfaces (namely the hyperbolic disk, hyperbolic punctured
disk and hyperbolic annuli), which were considered in detail in
reference \cite{elem}. Even for the simple case of ``modular
$j$-inflation'' (which was considered in \cite{Sch1,Sch2}), the
relation to two-field generalized $\alpha$-attractors is somewhat
subtle, as we explain in the present paper.

In this paper, we consider the case when $\Sigma$ is the
triply-punctured sphere (equivalently, the doubly-punctured complex
plane) endowed with its unique complete hyperbolic metric $G$. With
respect to this metric, the three punctures of the sphere are located
at infinite distance from the compact core, determining three
hyperbolic cusp regions \cite{Borthwick} of infinite length but
finite area. The hyperbolic area of the triply-punctured sphere is
finite and equals $2\pi$. Such a surface can be viewed as a
degeneration of a hyperbolic pair of pants, in the limit when each of
the cuff lengths of the latter tends to zero. When viewed as a smooth
non-compact complex (or affine algebraic) curve, the hyperbolic
triply-punctured sphere coincides with one of the classical
non-compact modular curves \cite{DS,Shimura}, usually denoted by
$Y(2)$. It is uniformized by a non-Abelian surface group
$\Gamma(2)\subset \PSL(2,\Z)$ which is freely generated by two
parabolic elements, namely the principal congruence subgroup of
$\PSL(2,\Z)$ at level two\footnote{In some references, the symbol
$\Gamma(2)$ is used instead to denote the principal congruence
subgroup of $\SL(2,\Z)$ at level two. In this paper, we only use it to
denote the corresponding subgroup of $\PSL(2,\Z)$.}. The
uniformization map from the Poincar\'e half-plane is given
\cite{Ahlfors} by the classical elliptic modular function $\lambda$.

The end (a.k.a. Ker\'ekj\'art\'o-Stoilow) compactification of $Y(2)$ 
is the unit sphere $\rS^2$. The
conformal compactification (which in this case is usually denoted
by $X(2)$) is the unit sphere endowed with its unique
orientation-compatible complex structure, i.e. the Riemann sphere
$\C\P^1$. An explicit embedding of $Y(2)$ into the Riemann sphere is
obtained by restricting the inverse of the stereographic projection. A
smooth scalar potential defined on $Y(2)$ is globally well-behaved in
the sense of \cite{genalpha} iff it is induced through this projection
by a smooth real-valued map $\hPhi$ defined on the sphere. Using these
observations, one can describe globally well-behaved scalar potentials
defined on $Y(2)$ by expanding $\hPhi$ into spherical harmonics.

As explained in \cite{genalpha}, cosmological trajectories of the
generalized $\alpha$-attractor model defined by $Y(2)$ can be obtained
by first computing trajectories of an appropriately lifted model
defined on the Poincar\'e half-plane and then projecting the latter to
$Y(2)$ through the uniformization map. Using this method, we compute
cosmological trajectories for a few globally well-behaved scalar
potentials. In particular, we display trajectories which have
inflationary behavior for small enough times. We find that the model
exhibits rich cosmological dynamics already in the absence of a scalar
potential, which becomes even more complex when the scalar potential
is included. We also find inflationary trajectories which produce
between 50 and 60 efolds, thus showing that such models are
potentially relevant to observational cosmology.

The lift \cite{genalpha} of the two-field generalized
$\alpha$-attractor models based on $Y(2)$ to the Poincar\'e half-plane
can be viewed as modular-invariant cosmological models in the sense of
\cite{Sch1,Sch2} for the modular group $\Gamma(2)$. The description
provided by the lifted model, while convenient for some purposes,
contains a countably infinite ambiguity which signals the fact that
the physically-relevant scalar field is not valued in the Poincar\'e
half-plane but in the triply-punctured sphere $Y(2)$.  The
uniformization map $\lambda$ can be viewed as a (countable to one)
field redefinition which eliminates this unphysical ambiguity, thus
affording easier understanding of the physics near the cusp ends of $Y(2)$.

The orientation-preserving isometry group of the hyperbolic
triply-punctured sphere is isomorphic with the permutation group on
three elements, acting on $Y(2)$ as the anharmonic subgroup $\fB$ of
$\PSL(2,\C)$, which permutes the three punctures. The triply-punctured
sphere can be written as a six-fold branched cover $q:Y(2)\rightarrow
\C$ of the complex plane in such a way that the composition $q\circ
\lambda$ coincides with the modular $j$ function. The anharmonic
action permutes the branches of this cover, hence the topological
quotient $Y(2)/\fB$ can be identified with the complex plane. Since
$q$ is ramified above two points of the complex plane, the hyperbolic
metric of $Y(2)$ descends to a {\em singular} metric on the
topological quotient $\C=Y(2)/\fB$.  When the scalar potential $\Phi$
is preserved by the anharmonic action, the generalized
$\alpha$-attractor model with scalar manifold $Y(2)$ has an $S_3$
symmetry and the corresponding lifted model is a ``$j$-model'' in the
sense of \cite{Sch1,Sch2}, being invariant under the entire
classical modular group $\PSL(2,\Z)$.  In that case, the field
redefinition provided by the elliptic modular function $\lambda$
eliminates an infinite $\Gamma(2)$-ambiguity, replacing it with a
finite $\fB\simeq S_3$ symmetry. The latter cannot be eliminated
directly within the framework of generalized two-field
$\alpha$-attractor models, since taking the quotient through the
anharmonic action of $S_3$ would lead to a singular metric on the
complex plane, which is disallowed in the usual construction of
non-linear sigma models by the principle of conservation of energy.

The paper is organized as follows. Section \ref{sec:alpha} briefly
recalls the definition of generalized $\alpha$-attractor models and
the lift of their cosmological equations of motion to the Poincar\'e
half-plane. It also explains when the lift to the Poincar\'e
half-plane can be interpreted as a modular-invariant cosmological
model in the sense of \cite{Sch01,Sch1,Sch2} and discusses symmetries
of such models. In section \ref{sec:geom}, we describe the end and
conformal compactifications as well as the hyperbolic geometry of
$Y(2)$, its symmetries and its uniformization. We also explain the
difference between $Y(2)$ (which coincides with the coarse moduli
space of elliptic curves with level $2$ structure) and the usual
coarse moduli space of elliptic curves (which equals the complex
plane), of which $Y(2)$ is a branched six-fold cover. This accounts
for the difference between the elliptic modular functions $\lambda$
and $j$, which differ by composition with a rational function. This is
standard mathematical material which we chose to explain in some
detail since it may be unfamiliar to cosmologists. Finally, we show
that modular-invariant cosmological models for the classical modular
group $\PSL(2,\Z)$ (see \cite{Sch1,Sch2}) are related by a countable
to one field redefinition to those generalized two-field
$\alpha$-attractor models with scalar manifold $Y(2)$ for which the
scalar potential is invariant under the action of the anharmonic
group. Section \ref{sec:traj} discusses globally well-behaved scalar
potentials on the triply-punctured sphere and presents examples of
numerically-computed cosmological trajectories. In particular, we
present examples of inflationary trajectories which produce between 50
and 60 efolds. In Section \ref{sec:gradflow}, we discuss the gradient
flow approximation near cusp ends and a class of scalar potentials for
which one can obtain any desired number of efolds for certain
`universal' cosmological trajectories nearby such ends. Section
\ref{sec:conclusions} concludes and suggests some directions for
further research. The appendices contain certain technical material on
the anharmonic action and on the uniformization of $Y(2)$.

\paragraph{Notations and conventions.} 
All manifolds considered are smooth, connected, oriented and
paracompact (hence also second-countable). All homeomorphisms and
diffeomorphisms considered are orientation-preserving. By definition,
a Lorentzian four-manifold has ``mostly plus'' signature.  The
Poincar\'e half-plane is the upper half-plane with complex coordinate
$\tau$:
\ben
\label{PoincareHP}
\H=\{\tau\in \C\,|\,\Im\tau>0\}~~,
\een
endowed with its unique complete metric of Gaussian curvature $-1$,
which is given by:
\be
\dd s^2_\H=\frac{1}{(\Im \tau)^2}|\dd \tau|^2~~.
\ee
The real coordinates on $\H$ are denoted by $x\eqdef \Re \tau$ and
$y\eqdef \Im \tau$. The complex coordinate on the hyperbolic {\em disk} is
denoted by $u$, while that on the twice-punctured complex plane is
denoted by $\zeta$. The symbol $\i$ denotes the imaginary unit. The
{\em rescaled Planck mass} is defined through:
\ben
\label{M0}
M_0\eqdef M\sqrt{\frac{2}{3}}~~,
\een
where $M$ is the reduced Planck mass. 

\section{Generalized $\alpha$-attractor models}
\label{sec:alpha}

In this section, we briefly recall the definition of generalized
two-field $\alpha$-attractor models given in \cite{genalpha} and their
lifts to the Poincar\'e half-plane. We also explain when this lift can
be interpreted as a modular-invariant cosmological model in the sense
of \cite{Sch01,Sch1,Sch2}. Finally, we discuss symmetries of such models.

\subsection{Definition of the models}

\noindent 
Let $(\Sigma, G)$ be a non-compact oriented, connected and complete
two-dimensional Riemannian manifold without boundary (the {\em scalar
  manifold}) and $\Phi:\Sigma\rightarrow \R$ be a smooth function (the
{\em scalar potential}). We assume that $(\Sigma,G)$ is {\em
  hyperbolic}, i.e. that $G$ has constant Gaussian curvature equal to
$-1$. We also assume that the fundamental group of $\Sigma$ is
finitely-generated. Let $\alpha$ be a positive constant. The
rescaled metric $\cG\eqdef 3\alpha G$ has constant Gaussian curvature
$K(\cG)=-\frac{1}{3\alpha}$.

Given an oriented four-manifold $X$ which supports Lorentzian metrics, the
Einstein-Scalar theory defined by $(\Sigma,\cG,\Phi)$ on
$X$ describes a Lorentzian metric $g$ on $X$ and a smooth map
$\varphi:X\rightarrow \Sigma$ through the action (see \cite{genalpha}
for the notations):
\ben
\label{S}
S[g,\varphi]=\int_X \left[\frac{M^2}{2} \mathrm{R}(g)-\frac{1}{2}\Tr_g \varphi^\ast(\cG)-\Phi\circ \varphi\right] \vol_g~~,
\een
where $\vol_g$ is the volume form of $(X,g)$, $\mathrm{R}(g)$ is the
scalar curvature of $g$ and $M$ is the reduced Planck mass. When $X$
is diffeomorphic with $\R^4$ and $g$ is a FLRW metric with flat
spatial section, solutions of the equations of motion of \eqref{S} for
which $\varphi$ depends only on the cosmological time $t$ define {\em
  generalized two-field $\alpha$-attractor models} \cite{genalpha}. We shall assume 
that $\Phi$ is non-negative, as appropriate for cosmological applications. 

Let $J$ be the unique orientation-compatible complex structure on
$\Sigma$ which has the property that $G$ is Hermitian (and hence
K\"ahler) with respect to $J$. Endowing $\Sigma$ with this complex
structure, let $\zeta$ be a local holomorphic coordinate on $\Sigma$,
defined on an open subset $\cU\subset \Sigma$. Since $G$ is Hermitian
with respect to $J$, we have $\dd s_G^2|_\cU=
\rho(\zeta,\bar{\zeta})^2|\dd \zeta|^2$ for some positive
function $\rho(\zeta,\bar{\zeta})>0$, known as the {\em hyperbolic
  density} of $G$ with respect to $\zeta$. Choosing a local chart
$(U,(x^\mu))$ of $X$ such that $\varphi(U)\subset \cU$ and setting
$\zeta(x)\eqdef \zeta(\varphi(x))$, the map $\varphi$ is described
locally by the complex-valued scalar field $\zeta(x)$ and the Lagrangian
density of \eqref{S} has the local form:
\ben
\label{cLcomplex}
\cL[g|_U,\varphi|_U]=\frac{M^2}{2} \mathrm{R}(g)-\frac{3\alpha}{2}\rho^2(\zeta,\bar{\zeta})
g^{\mu\nu}\partial_\mu \zeta\partial_\nu \bar{\zeta}- \Phi(\zeta,\bar{\zeta})~~.
\een

\subsection{Lift to the Poincar\'e half-plane}
\label{subsec:lift}

\noindent
The cosmological equations of motion of the generalized
two-field $\alpha$-attractor model defined by \eqref{S} can be lifted
\cite{genalpha} from $\Sigma$ to the Poincar\'e half-plane through the
$J$-holomorphic covering map $\pi_\H:\H\rightarrow \Sigma$ which
uniformizes \cite{unif} $(\Sigma,G)$ to $\H$. This presents
$(\Sigma,G)$ as the Riemannian quotient $\H/\Gamma$ (where
$\Gamma\subset \PSL(2,\R)$ is the uniformizing surface group
\cite{Katok,Beardon, FN}) and allows one to determine the cosmological
trajectories $\varphi(t)$ by projecting solutions $\tvarphi(t)\in \H$
of the following equations \cite[eq. (7.4)]{genalpha}:
\beqan
\label{elplane}
&& \ddot{x}-\frac{2}{y}\dot{x}\dot{y} +\frac{1}{M} \sqrt{\frac{3}{2}} \left[3\alpha \frac{\dot{x}^2+\dot{y}^2}{y^2}+2\tPhi(x,y)\right]^{1/2}\dot{x}+\frac{1}{3\alpha} y^2 \partial_x\tPhi(x,y)=0~~, \\
&& \ddot{y}+\frac{1}{y}(\dot{x}^2-\dot{y}^2)+\frac{1}{M} \sqrt{\frac{3}{2}} \left[3\alpha \frac{\dot{x}^2+\dot{y}^2}{y^2}+2\tPhi(x,y)\right]^{1/2}\dot{y}+\frac{1}{3\alpha} y^2\partial_y\tPhi(x,y)=0~~\nn
\eeqan
through the map $\pi_\H$. Here $\dot{~}\eqdef \frac{\dd}{\dd t}$, where $t$
is the cosmological time, and
$x=\Re\tau$, $y=\Im \tau$ are the Cartesian coordinates on $\H$ (and
we wrote $\tvarphi(t)=x(t)+\i y(t)$), while $\tPhi\eqdef \Phi\circ
\pi_\H:\H\rightarrow \R$ is the {\em lifted scalar potential}. As
shown in \cite{genalpha}, any cosmological trajectory $\varphi(t)$ of
the generalized two-field $\alpha$-attractor model defined by \eqref{S} can be
written as $\varphi(t)=\pi_\H(\tvarphi(t))$ for some appropriate
solution $\tvarphi$ of \eqref{elplane} (which is determined by
$\varphi$ up to the action of $\Gamma$). In order to simplify
computations, it is convenient to eliminate the Planck mass by writing
\eqref{elplane} in the equivalent form:
\beqan
\label{elplane0}
&& \ddot{x}-\frac{2}{y}\dot{x}\dot{y} +\left[3\balpha \frac{\dot{x}^2+\dot{y}^2}{y^2}+2\btPhi(x,y)\right]^{1/2}\dot{x}+\frac{1}{3\balpha} y^2 \partial_x\btPhi(x,y)=0~~, \nn\\
&& \ddot{y}+\frac{1}{y}(\dot{x}^2-\dot{y}^2)+\left[3\balpha \frac{\dot{x}^2+\dot{y}^2}{y^2}+2\btPhi(x,y)\right]^{1/2}\dot{y}+\frac{1}{3\balpha} y^2\partial_y\btPhi(x,y)=0~~,
\eeqan
where $\btPhi\eqdef \frac{\tPhi}{M_0}$, $\balpha\eqdef
\frac{\alpha}{M_0}$ and $M_0=M\sqrt{\frac{2}{3}}$ is the rescaled Planck
mass \eqref{M0}. Accordingly, the Hubble parameter can be written as
$H(t)=\sqrt{M_0} \, \bH(t)$, where (cf. \cite[eq. (2.8)]{genalpha}):
\ben
\label{bHubble}
\bH(t)=\frac{1}{3}\sqrt{3\balpha ||\dot{\tvarphi}(t)||^2_\H+2\btPhi(\tvarphi(t))}=\frac{1}{3}\sqrt{3\balpha \frac{\dot{x}(t)^2+\dot{y}(t)^2}{y(t)^2}+2\btPhi(x(t),y(t))}~~,
\een
and $||.||_\H$ denotes the norm of vectors tangent to $\H$, computed with respect 
to the Poincar\'e plane metric:
\be
\dd s_\H^2=\frac{\dd x^2+\dd y^2}{y^2}~~.
\ee 

\subsection{The inflation region of the tangent bundle}
\label{subsec:infregion}
\noindent The inflationary time periods of a trajectory $\varphi(t)$
(defined as the time intervals for which the FLRW scale factor $a(t)$
is a convex and increasing function of $t$) are given by the condition
(cf. \cite[eq. (2.12)]{genalpha}):
\ben
\label{infcond0}
H(t)<H_c(\varphi(t))~~, 
\een
where: 
\ben
\label{critHubble}
H_c(p)\eqdef \frac{1}{M}\sqrt{\frac{\Phi(p)}{2}}~~ 
\een
is the {\em critical Hubble parameter} at a point $p\in \Sigma$. We
have: $H_c(p)=\sqrt{M_0}\bH_c(p)$, with $\bH_c(p)\eqdef
\sqrt{\frac{\bPhi(p)}{3}}$, where defined $\bPhi\eqdef
\frac{1}{M_0}\Phi$. Condition \eqref{infcond0} is equivalent
\cite{genalpha} with:
\ben
\label{infcond}
||\dot{\varphi}(t)||^2_\cG<\Phi(\varphi(t))~~.
\een
This inequality states that a trajectory $\varphi(t)$ of the
$\alpha$-attractor model is inflationary for those times $t$ for which
the tangent vector $\dot{\varphi}(t)\in T_{\varphi(t)}\Sigma$ belongs
to the {\em inflation region} defined by $\Phi$ at parameter $\alpha$:
\be
\cR:=\cR_\alpha(\Phi)\eqdef \{v\in T\Sigma \,|\, ||v||_G < v_c(p)\}~~,
\ee
where the {\em critical speed} $v_c(p)\in T_p\Sigma$ at a point $p$ of
$\Sigma$ is defined through:
\be
v_c(p)\eqdef \sqrt{\frac{\Phi(p)}{3\alpha}}=\sqrt{\frac{\bPhi(p)}{3\balpha}}~~.
\ee
Notice that $\cR$ is an open disk bundle over $\Sigma$. The quantity
$v_c(p)$ gives the radius of the disk fiber of this bundle at a
point $p\in \Sigma$, computed with respect to the metric $G_p$ on
$T_p\Sigma$.

Condition \eqref{infcond} can be expressed as follows in terms of any
trajectory $\tvarphi$ of \eqref{elplane} which lifts $\varphi$ to the
Poincar\'e half-plane:
\ben
\label{infcondlifted}
||\dot{\tvarphi}(t)||^2_\H<\frac{\btPhi(\tvarphi(t))}{3\balpha}~,~\mathrm{i.e.}~~\dot{x}(t)^2+\dot{y}(t)^2 <\frac{y^2 \btPhi(x(t),y(t))}{3\balpha}~~.
\een
For any $\tau\in \H$, define the {\em lifted critical speed} at $\tau$ on the
Poincar\'e half-plane through:
\be
\tv_c(\tau)\eqdef \sqrt{\frac{\tPhi(\tau)}{3\alpha}}=\sqrt{\frac{\btPhi(\tau)}{3\balpha}}~~. 
\ee
We have $\tv_c=v_c\circ \pi_\H$, so $\tv_c$ is a $\Gamma$-invariant
function defined on $\H$.  The tangent bundle $T\H$ of $\H$ is trivial
and can be identified with $\H\times \R^2$. The {\em lifted inflation
  region} defined by $\tPhi$ at parameter $\alpha$ is the open subset
(an open disk bundle over $\H$) of the total space of $T\H$
defined through:
\be
\tcR:=\tcR_\alpha(\tPhi)\eqdef \{(\tau,\tv)\in T\H \,|\, ||\tv||_{\H,\tau}
 < \tv_c(\tau)\} \equiv \{ (x,y,v)\in \H\times \R^2\,|\, \tv_x^2+\tv_y^2
 <\frac{y^2 \tPhi(x,y)}{3\alpha} \}~~,
\ee
where we wrote $\tau=x+\i y$ and $\tv=\tv_x+\i \tv_y$ with real
$\tv_x,\tv_y$. Notice that $\tcR$ is invariant under the action of $\Gamma$ on $T\H$ 
and that $\cR$ coincides with the image of $\tcR$ through
the differential $\dd \pi_\H:T\H\rightarrow T\Sigma$ of the
uniformization map. A trajectory $\tvarphi$ of the lifted system
\eqref{elplane} projects to the inflationary portion of a trajectory
$\varphi=\pi_\H\circ \tvarphi$ of the $\alpha$-attractor model defined
by $(\Sigma,\cG)$ for those times $t$ for which the tangent vector
$\dot{\tvarphi}(t)$ belongs to $\tcR$. Notice that $v_c$ has the same 
level sets as $\Phi$ and that $\tv_c$ has the same level sets as $\tPhi$, 
though the values on the same level set generally differ. 

\subsection{Relation to modular-invariant cosmological models}

\noindent Equations \eqref{elplane} can be viewed as the cosmological
equations of motion of a ``lifted'' generalized two-field
$\alpha$-attractor model, for which $\Sigma$ is replaced by the
Poincar\'e half-plane $\H$, the metric on $\Sigma$ is replaced by
$3\alpha G_\H$ (where $G_\H$ is the Poincar\'e metric of $\H$) and the
scalar potential is replaced by the lifted potential $\tPhi$. Notice
that $\tPhi\eqdef \Phi\circ \pi_\H$ is $\Gamma$-invariant by
construction and that the action of $\Gamma$ on $\H$ is isometric
(recall that $\Gamma$ is a subgroup of $\PSL(2,\R)$, which is the
group of orientation-preserving isometries of the Poincar\'e
half-plane). This implies that the lifted model is $\Gamma$-invariant,
being similar to (but more general than) the type of model considered\footnote{See
\cite{Sch01,Sch02} for the general framework of ``automorphic
inflation''.} in \cite{Sch1,Sch2}, up to a rescaling of the Poincar\'e
metric by $3\alpha$. Note, however, that our uniformizing group
$\Gamma$ need not be a subgroup of $\PSL(2,\Z)$, since we do not limit
ourselves to arithmetic groups; in particular, $\Sigma$ need not be a
modular curve. Unlike \cite{Sch1,Sch2}, we do {\em not} view this
lifted model as being physical, but only as a technical tool for
studying the cosmological dynamics of the original generalized
two-field $\alpha$-attractor model defined by cosmological solutions
of \eqref{S}. In our approach, two distinct values of $\tvarphi$ which
are related through the action of $\Gamma$ are identified and they are
viewed as physically equivalent; it is only the projection
$\pi_\H\circ \tvarphi=\varphi$ which has a direct physical meaning.
Notice that this interpretation is consistent with putative embeddings
of our models into string theory. In such an embedding, $\Sigma$ would
be interpreted as a moduli space of internal string backgrounds, while
the Poincar\'e half-plane would arise as a Teichm\"uller space of the
same; the uniformizing group $\Gamma$ would then arise as a group of
discrete physical symmetries which identify distinct points of the
Teichm\"uller space and hence must be quotiented out in order to
correctly describe moduli field dynamics through an effective
nonlinear sigma model. In such putative string theory embeddings, it
would be appropriate to treat the {\em moduli field} $\varphi$ as the
physical field, rather than the ``Teichm\"uller field'' $\tphi$, which
is only an auxiliary object without direct physical significance in
the effective theory. As already shown in \cite{elem} and also
illustrated later in the present paper, the effect of the projection
$\pi_\H$ is quite dramatic. The uniformization map $\pi_\H$ can be
viewed as an ``$\infty$ to $1$'' field reparameterization which
eliminates a (discretely) infinite unphysical ambiguity affecting the
effective description of low energy\footnote{Low compared to the
string scale.}  physics which is present in the lifted model.

Suppose for definiteness that $(\Sigma,G)$ has finite hyperbolic area
(equivalently, that $\Gamma$ is a Fuchsian group of the first kind
\cite{Katok,Beardon}), as is the case for the scalar manifold $Y(2)$
studied latter in this paper. Then $(\Sigma,G)$ has only cusp ends and the
limit set $\Lambda$ of $\Gamma$ equals the entire conformal boundary
$\partial_\infty \H\simeq \rS^1\simeq \R\cup\{\infty\}$ of the
Poincar\'e half-plane. This implies \cite{Katok,Beardon} that each of
the orbits of the action of $\Gamma$ on the Poincar\'e half-plane has
$\Lambda=\partial_\infty \H$ as its set of accumulation points. When
$\Phi$ is not constant, it follows that the lifted potential $\tPhi$
has extremely complicated behavior near the conformal boundary. In
particular, any extremum of $\tPhi$ inside $\H$ is repeated a
countable number of times (at all points of its $\Gamma$-orbit) and
the $\Gamma$-images of any given extremum point accumulate near any
point of $\partial_\infty\H$. The preimages through the uniformization
map $\pi_\H$ of each of the cusp ideal points of $\Sigma$ form a
countable subset of $\partial_\infty \H$. As a consequence, the
cosmological trajectories of the lifted model have extremely
complicated behavior near the conformal boundary. Luckily, it is not
these lifted trajectories (or the lifted model itself) which are of
direct physical interest, but rather their projections to $\Sigma$
through the uniformization map.

A particular sub-class of hyperbolic surfaces of finite area arises
when $\Gamma$ is a finite index subgroup of $\PSL(2,\Z)$ (in which case
the complex manifold corresponding to $(\Sigma,G)$ is a modular
curve). One can construct special examples of two-field
$\alpha$-attractors of this type by taking $\Phi$ to be a functional
combination of modular functions. In this rather special situation,
the corresponding {\em lifted} model is of the type considered in
\cite{Sch01,Sch1,Sch2}. Notice that such two-field generalized
$\alpha$-attractor models are quite sparse within the class of all
two-field generalized $\alpha$-attractors, since arithmetic Fuchsian
groups are very special among all Fuchsian groups (just like modular
curves are very special among non-compact hyperbolic surfaces). For
example, any geometrically-finite hyperbolic surface of infinite area
necessarily has ends which are not of cusp type, hence such a surfaces
cannot be a modular curve. Simple examples of such hyperbolic surfaces
are the hyperbolic disk, hyperbolic punctured disk and hyperbolic
annulus (all of which have infinite hyperbolic area). The two-field
generalized $\alpha$-attractor models associated to the hyperbolic
punctured disk and hyperbolic annulus were studied in \cite{elem}.

\subsection{Symmetries}

\noindent Let $(\Sigma,G)$ be a geometrically-finite hyperbolic surface
uniformized by the surface group $\Gamma\subset \PSL(2,\R)$. The group
of orientation-preserving isometries of $(\Sigma,G)$ is given by:
\be
\Iso^+(\Sigma,G)=N(\Gamma)/\Gamma~~,
\ee
where: 
\be
N(\Gamma):=N_{\PSL(2,\R)}(\Gamma)\eqdef \{A\in \PSL(2,\R)|A\Gamma A^{-1}=\Gamma\}
\ee
denotes the normalizer of $\Gamma$ inside $\PSL(2,\R)$. Since $\Gamma$
is a normal subgroup of $N(\Gamma)$, we have:
\be
\Sigma/\Iso^+(\Sigma,G)=(\H/\Gamma)/(N(\Gamma)/\Gamma)\simeq \H/N(\Gamma)~~.
\ee
In the second quotient, the group $N(\Gamma)/\Gamma$ acts on $\Sigma$ through: 
\ben
\label{adef}
a\zeta\eqdef [A\tau]_\Gamma ~~\mathrm{for}~~a=[A]_\Gamma\in N(\Gamma)/\Gamma~~\mathrm{and}~~\zeta=[\tau]_\Gamma\in \Sigma~~,
\een
where $A\in N(\Gamma)$, $\tau\in \H$ and $[A]_\Gamma$, $[\tau]_\Gamma$
denote the equivalence classes under the left actions of $\Gamma$ on
$N(\Gamma)$ and on $\H$. It is easy to check that the action \eqref{adef}
is well-defined.

The `obvious' orientation-preserving symmetry group\footnote{More precisely, 
this is the group of `vertical' Noether symmetries.} of the $\alpha$-attractor
model defined by $(\Sigma,\cG,\Phi)$ is the subgroup of
$\Iso^+(\Sigma,G)$ given by \cite{genalpha}: 
\be
\Aut^+(\Sigma,G,\Phi)=\{ a \in \Iso^+(\Sigma,G)\, | \, \Phi(a \zeta)=\Phi(\zeta)~~\forall \zeta\in \Sigma\}~~.
\ee
This can be written as: 
\be
\Aut^+(\Sigma,G,\Phi)=[N(\Gamma)\cap \Aut^+(\H,G_\H,\tPhi)]/\Gamma~~,
\ee
where: 
\be
\Aut^+(\H,G_\H,\tPhi)=\{A\in \PSL(2,\R) \, | \, \tPhi(A \tau)=\tPhi(\tau)~~\forall \tau \in \H\}
\ee
is the orientation-preserving symmetry group of the lifted model,
which is a subgroup of $\PSL(2,\R)$.  In particular, $\Phi$ is
invariant under $\Iso^+(\Sigma,G)$ iff $\tPhi$ is invariant under the
action of $N(\Gamma)$ on $\H$ by fractional linear transformations.
In this case, we have $\Aut^+(\Sigma,G,\Phi)=\Iso^+(\Sigma,G)$ and $
N(\Gamma)\subset \Aut^+(\H,G_\H,\tPhi)$.

\paragraph{Remark.} 
Suppose that $N(\Gamma)$ contains elliptic elements. Then the action
of $N(\Gamma)$ on $\H$ and the action of $\Iso^+(\Sigma,G)$ on
$\Sigma$ have orbits whose points have non-trivial stabilizer. Due to
this fact, the Poincar\'e metric of $\H$ and the hyperbolic metric $G$
of $\Sigma$ do not descend to a non-singular metric on the topological
quotient $S\eqdef \H/N(\Gamma)\simeq \Sigma/\Iso^+(\Sigma,G)$. Rather,
they descend to an orbifold metric on the {\em orbifold} quotient
$\H\sslash N(\Gamma)\simeq \Sigma\sslash \Iso^+(\Sigma,G)$, which has
underlying space $S$. If $s_1,\ldots, s_k\in S$ denote the singular
points of this orbifold quotient, then the orbifold metric can be
viewed as a metric defined on $S\setminus \{s_1,\ldots,s_k\}$ which
has conical singularities at each of the points $s_1,\ldots, s_k$. We
will see an example of this in the next section.

\section{The hyperbolic triply-punctured sphere}
\label{sec:geom}

In this section, we summarize the geometry of the hyperbolic
triply-punctured sphere $Y(2)$, its end and conformal
compactifications and its uniformization to the Poincar\'e half-plane
and to the hyperbolic disk. We also discuss the orientation-preserving
isometry group of $Y(2)$ and the topological and orbifold quotients of
$Y(2)$ by this group. Finally, we show that modular-invariant
cosmological models for the classical modular group $\PSL(2,\Z)$ can
be related by an infinite to one field redefinition to those
generalized two-field $\alpha$-attractor models with scalar manifold
$Y(2)$ for which the scalar potential is invariant under the action of
the anharmonic group. The mathematical results summarized in this
section on the geometry of $Y(2)$ are well-known in the literature on
modular curves and uniformization theory, but we give a rather
detailed account for the benefit of cosmologists and in order to
clarify the precise relation to the ``modular invariant $j$-models''
of \cite{Sch1,Sch2}. Some technical details are relegated to the
appendices.

\subsection{The modular curve $Y(2)$}

\noindent 
The unit sphere $\rS^2=\{(x_1,x_2,x_3)\in \R^3 \, | \,
x_1^2+x_2^2+x_3^2=1\}$ admits a unique\footnote{Up to
  orientation-preserving diffeomorphism.} orientation-compatible
complex structure $I$, which makes the complex manifold $(\rS^2,I)$
biholomorphic with the Riemann sphere $\C\P^1$. Let $p_1,p_2,p_3$ be
any three distinct points of $\rS^2\simeq \CP^1$. By definition, a
{\em triply-punctured Riemann sphere} is the surface:
\be
Y(2)\eqdef \C\P^1\setminus \{p_1,p_2,p_3\}~~,
\ee
endowed with the complex structure $J=I|_{Y(2)}$ induced from $\C\P^1$. This
complex manifold of complex dimension one
is a classical example of a non-compact modular curve.  It is uniquely
determined up to biholomorphism, since three distinct points of
$\CP^1$ can be moved together in arbitrary position by acting with an
element of the biholomorphism group of $\C\P^1$, which is isomorphic
with the M\"obius group $\PSL(2,\C)$. Up to such a transformation, one
can therefore take:
\be
p_1=\nu\eqdef (0,0,1)~~,~~p_2=\sigma\eqdef(0,0,-1)~~,~~p_3=p\eqdef (1,0,0)~~,
\ee
where $\nu$ and $\sigma$ are the north and south poles of $\rS^2$. 

Let $\psi$ and $\theta$ be spherical coordinates on $\rS^2$, thus:
\ben
\label{sphcoord}
x_1=\sin\psi\cos\theta~~,~~x_2=\sin\psi\sin\theta~~,~~x_3=\cos\psi~~,
\een
where $\psi\in [0,\pi]$ and $\theta\in [0,2\pi)$. Then the points
  $\nu$ and $\sigma$ correspond respectively to $\psi=0$ and $\psi=\pi$,
  while $p$ corresponds to $\psi=\frac{\pi}{2}$ and $\theta=0$. The
  stereographic projection from the north pole $\nu$:
\ben
\label{sterproj}
\zeta=\cot\left(\frac{\psi}{2}\right)e^{\i\theta} \in \C
\een
is a biholomorphism from $ \C\P^1\setminus \{\nu\}$ to the
complex plane $\C$ with complex coordinate $\zeta$. This
maps $\nu$ to the point at infinity (which is not part
of $\C$) and $\sigma$ to the origin $\zeta=0$ of the complex plane. It
maps $p$ to the point $\zeta=1$. As a consequence, one can identify: 
\be
Y(2)=\C\setminus \{0,1\}~~.
\ee

\subsection{The end compactification}

\noindent 
Since $Y(2)$ is a planar surface, its end (a.k.a. Ker\'ekj\'art\'o-Stoilow
\cite{Richards,Stoilow}) compactification $\widehat{Y(2)}$ coincides
with the unit sphere $\rS^2$. An explicit embedding
$Y(2)\hookrightarrow \rS^2$ is given by:
\ben
\label{ee} 
\psi=2\arccot(|\zeta|)~~,~~\theta=\arg(\zeta)~~.  
\een 
In particular, the punctures $p_1=\nu\equiv \infty$, $p_2=\sigma \equiv
0$ and $p_3=p\equiv 1$ coincide with the ideal points of
$\widehat{Y(2)}$ and hence can be identified with the ends of $Y(2)$.

\subsection{The conformal compactification}

\noindent 
Since $\widehat{Y(2)}\simeq \rS^2$ admits a unique
orientation-compatible complex structure $I$, the prolongation
\cite{Haas,Maskit} (see also \cite[Appendix C]{genalpha}) complex structure $J$ of $Y(2)$
coincides with $I$.  Hence the conformal compactification
\cite{genalpha} of $Y(2)$ coincides with $(\rS^2,I)\simeq \C\P^1$.
The latter also coincides with the compact modular curve $X(2)$.

\subsection{The hyperbolic metric}

\begin{figure}[H]
\centering \includegraphics[width=90mm]{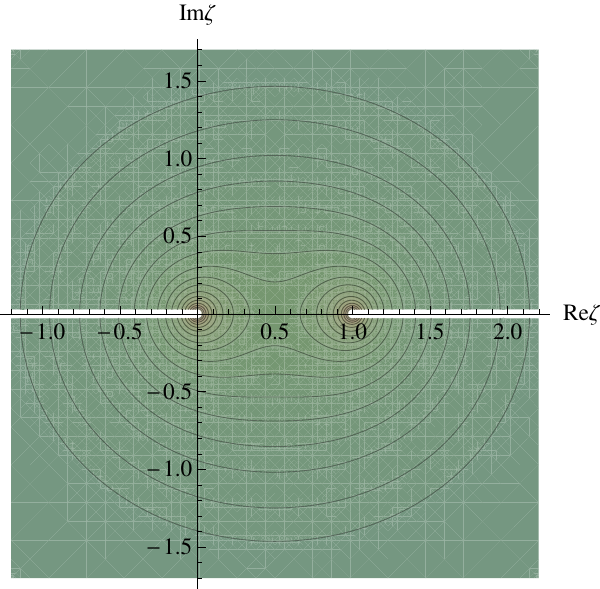}
\caption{Plot of the hyperbolic density $\rho$ on the twice-punctured
  plane $Y(2)$. The function $\rho(\zeta,\bar{\zeta})$ grows to
  infinity at each of the punctures $\zeta=0,1$ and tends to zero for
  $\zeta\rightarrow \infty$. The white slots are due to the branch
  cuts of $\cK$, being an artifact produced by the
  limited numerical precision of the graphics software used.}
\label{fig:rho}
\end{figure}

\noindent 
The surface $Y(2)$ admits a uniquely-determined complete hyperbolic
metric $G$ which is K\"ahler with respect to its complex structure.
This metric was determined explicitly in \cite{Agard} and studied
further in \cite{SolV, SV}. It is given by (see Figure \ref{fig:rho}): 
\ben
\label{metric}
\dd s_G^2=\rho(\zeta,\bar{\zeta})^2|\dd \zeta|^2~~,~\mathrm{with}~~\rho(\zeta,{\bar \zeta})=\frac{\pi}{8|\zeta(1-\zeta)|} \, \frac{1}{\Re [\cK(\zeta)\cK(1-\bar{\zeta})]}~~,
\een
where:
\be
\cK(\zeta)=\int_{0}^1 \frac{\dd t}{\sqrt{(1-t^2)(1-\zeta t^2)}}=\frac{\pi}{2} ~_2F_1\left(\frac{1}{2},\frac{1}{2};1 \Big|\zeta \right)
\ee
is the complete elliptic integral of the first kind, continued
analytically to the complex plane with cut along the interval
$[1,+\infty)$ of the real axis. The metric \eqref{metric} differs from
  the restriction to $Y(2)$ of the Euclidean metric of the plane by
  the conformal factor $\rho^2$, which has the effect of pushing each of
  the points $\zeta=0,1,\infty$ to infinite distance. As a result, the
  hyperbolic surface $(Y(2), G)$ looks like a sphere with three
  infinitely-long cusps. Notice that each cusp has finite hyperbolic area, even
  though it is infinitely long. In particular, the surface $Y(2)$ has
  finite hyperbolic area (equal to $2\pi$).

In principle, the cosmological model can be studied directly on
$\C\setminus \{0,1\}$ using the explicit metric
\eqref{metric}. However, it is more convenient to lift to $\H$ or to
$\mD$ as explained in \cite{genalpha} and recalled in Subsection
\ref{subsec:lift}. Among other advantages, this will allow us to
illustrate the effect of projecting through the uniformization map and
hence the difference between the behavior of the generalized
two-field $\alpha$-attractor model defined by $Y(2)$ and that of the lifted
model defined on the Poincar\'e half-plane.

\subsection{Uniformization to the Poincar\'e half-plane}

\paragraph{The group $\Gamma(2)$.}

\noindent 
The hyperbolic surface $(Y(2),G)$ is uniformized to the Poincar\'e
half-plane $\H$ with complex coordinate $\tau$ (see
\eqref{PoincareHP}) by the principal congruence
subgroup\footnote{Notice that we consider $\Gamma(2)$ as a subgroup of
  $\PSL(2,\Z)$ rather than of $\SL(2,\Z)$.} $\Gamma(2)\subset
\PSL(2,\Z)$, which is defined as the kernel of the surjective group
morphism $\mu_2:\PSL(2,\Z)\rightarrow \PSL(2,\Z_2)$ given by:
\be
\mu_2\left(\left[\begin{array}{cc} a & b \\ c & d \end{array}\right]\right)\eqdef \left[\begin{array}{cc} a\!\!\mod \!2 &~b\!\!\mod \!2\\ c\!\!\mod \!2&~d\!\!\mod \! 2\end{array}\right]~~.
\ee
In particular, $\Gamma(2)$ is a normal subgroup of $\PSL(2,\Z)$. It
consists of all matrices $A=\left[\begin{array}{cc} a & b \\ c &
    d \end{array}\right]\in \PSL(2,\Z)$ for which $a$ and $d$ are odd
and $b$ and $c$ are even. This non-Abelian group is freely generated
by the two parabolic elements:
\ben
\label{gens}
P_\infty=\left[\begin{array}{cc} 1 &2\\0 &1\end{array}\right]~~,~~ P_0=\left[\begin{array}{cc} 1 &0\\ -2&1\end{array}\right]~~,
\een
which act as: 
\be
P_\infty(\tau)= \tau+2~~,~~P_0(\tau)=\frac{\tau}{1-2\tau}~~.
\ee
We have $\pi_1(Y(2))\simeq \Gamma(2)$. Notice that
$P_0^{-1}=\left[\begin{array}{cc} 1 &0\\ 2&1\end{array}\right] $
(which acts as $P_0^{-1}(\tau)=\frac{\tau}{2\tau+1}$) and that
$\Gamma$ is also generated by $P_0^{-1}$ and $P_\infty$.  It is
known\footnote{More generally, the normalizer inside $\PSL(2,\R)$ of
  any principal congruence subgroup $\Gamma(N)$ of $\PSL(2,\Z)$ equals
  $\PSL(2,\Z)$. See \cite[Corollary 3.6 (vi)]{Zemel}. This statement also 
follows from \cite[Theorem 6]{Newman}.}  that the
normalizer of $\Gamma(2)$ inside $\PSL(2,\R)$ equals $\PSL(2,\Z)$:
\be
N(\Gamma(2))\eqdef N_{\PSL(2,\R)}(\Gamma(2))=\PSL(2,\Z)~~.
\ee
We also have $N_{\PSL(2,\Z)}(\Gamma(2))=\PSL(2,\Z)$, since $\Gamma(2)$
is a normal subgroup of $\PSL(2,\Z)$.

\paragraph{The anharmonic action and the anharmonic group.}

\noindent The group $\PSL(2,\Z_2)$ is isomorphic with the permutation group
$S_3$ on three elements. This group has an isometric action on $Y(2)$
known as the {\em anharmonic action}, which permutes the three
punctures of $Y(2)$ (i.e. the three cusp ends of the hyperbolic 
triply-punctured sphere).  Moreover, $\PSL(2,\Z_2)$ can be identified with
the quotient $\PSL(2,\Z)/\Gamma(2)$ and there exists a canonical lift
of $\PSL(2,\Z_2)$ to a subgroup $\fB$ of $\PSL(2,\Z)$ known as the
{\em anharmonic group}.  Viewing the anharmonic group as a subgroup of
the M\"obius group $\PSL(2,\C)$, this allows one to describe the
anharmonic action of $\PSL(2,\Z_2)\simeq S_3$ on $Y(2)$ as the
restriction to $Y(2)\subset \C\P^1$ of the action of six M\"obius
transformations of $\C\P^1$. We refer the reader to Appendix
\ref{app:anharmonic} for an account of this construction and for
details regarding the anharmonic action on $Y(2)$.

\paragraph{The uniformizing map.}

\noindent 
For $\Sigma=Y(2)$, the holomorphic covering map $\pi_\H:\H\rightarrow \Sigma$ 
is given \cite[Chapter 7]{Ahlfors} by the elliptic modular lambda function
$\lambda: \H\rightarrow Y(2)$, which is defined through:
\be
\lambda(\tau)=\frac{\wp_\tau(\frac{1+\tau}{2})-\wp_\tau(\frac{\tau}{2})}{\wp_\tau(\frac{1}{2})-\wp_\tau(\frac{\tau}{2})}~~,
\ee
where:
\be
\wp_\tau(z)\eqdef \frac{1}{z^2}+\sum_{(m,n)\in \Z^2\setminus \{ (0,0)\}} \left(\frac{1}{(z-m-n\tau)^2}-\frac{1}{(m+n\tau)^2}\right)
\ee
is the Weierstrass elliptic function of modulus $\tau$. The function
$\lambda(\tau)$ is invariant under the action of $\Gamma(2)$ on $\H$:
\be
\lambda(\frac{a\tau+b}{c\tau+d})=\lambda(\tau)~~,~~\forall A=\left[\begin{array}{cc} a & b \\ c & d \end{array}\right]\in \Gamma(2)
\ee
and satisfies:
\be
\lambda(\tau+1)=\frac{\lambda(\tau)}{\lambda(\tau)-1}~~,~~\lambda(-\frac{1}{\tau})=1-\lambda(\tau)~~.
\ee
The fundamental polygons for uniformization of $Y(2)$ to the
Poincar\'e half-plane and to the Poincar\'e disk can be found in
Appendix \ref{app:fpolygons}, to which we refer the interested reader
for details. The same appendix explains the construction of canonical
coordinates near each cusp end of $Y(2)$ (see \cite{genalpha}).
 
\subsection{A multivalued inverse of $\lambda$}

\noindent A multivalued inverse $\mu$ of the holomorphic covering map
$\lambda$ is given as follows in terms of hypergeometric functions:
\ben
\label{mu}
\tau=\mu(\zeta)=\i \frac{~_2F_1(\frac{1}{2},\frac{1}{2};1|1-\zeta)}
{~_2F_1(\frac{1}{2},\frac{1}{2};1|\zeta)}~~.
\een
This multivalued analytic function has monodromy around the punctures
$\zeta=0,1,\infty$, generated respectively by the transformations
$P_0$, $P_1\eqdef P_\infty P_0$ and $P_\infty$.
In particular, a preimage of the point $\zeta=-1$ is given by: 
\be
\tau_0=\i \frac{~_2F_1(\frac{1}{2},\frac{1}{2},1|2)}{~_2F_1(\frac{1}{2},
\frac{1}{2};1|-1)}=1+\i~~.
\ee
This multivalued inverse of $\lambda$ is useful for identifying
preimages in $\H$ of points on $Y(2)$ and was used in various
computations presented in Section \ref{sec:traj}.

\subsection{Presentation of $Y(2)$ as a branched cover of the complex plane}

\noindent 
Let $q:\C\rightarrow \overline{\C}=\C\sqcup\{\infty\}\simeq \C\P^1$ be the
rational function defined through:
\be
z=q(\zeta)\eqdef 256 \frac{(1-\zeta+\zeta^2)^3}{\zeta^2(1-\zeta)^2}=256 \frac{(\zeta-e^{\frac{\i\pi}{3}})^3(\zeta-e^{-\frac{\i\pi}{3}})^3}{\zeta^2(1-\zeta)^2}~~, 
\ee
which has poles at $\zeta=0$ and $\zeta=1$. This extends to a
map $\bar{q}:\C\P^1\rightarrow \C\P^1$ upon setting
$\bar{q}(\infty)=\infty$.  The map $\bar{q}$ is invariant under the
anharmonic action \eqref{anharmonic} of $\PSL(2,\Z_2)$. Its
ramification points are $z=0$, $z=2^6 3^3=1728$ and $z=\infty$ and we have:
\begin{itemize}
\itemsep 0.0em
\item $\bar{q}^{-1}(\{\infty\})=\{0,1,\infty\}$, each preimage point
  having ramification index two
\item $\bar{q}^{-1}(\{0\})=\{e^{-\frac{\i\pi}{3}}, e^{+
  \frac{\i\pi}{3}}\}$, each preimage point having ramification index
  three
\item $\bar{q}^{-1}(\{1728\})=\{-1,\frac{1}{2},2\}$, each preimage
  point having ramification index two.
\end{itemize}
These special level sets of $\bar{q}$ coincide with the short orbits
of the action of $\fB$ on $X(2)$ shown in Tables \ref{table:A},
\ref{table:Afix2} and \ref{table:Afix3} of Appendix
\ref{app:anharmonic}.  In particular, $\bar{q}$ presents $Y(2)$ as a
degree six branched cover of the complex plane with complex coordinate
$z$, ramified at the points $z=0$ and $z=1728$; it is the projection
map of the topological quotient $Y(2)\rightarrow Y(2)/\fB=\C$.

\subsection{Quotients of $Y(2)$ by the anharmonic group $\fB$}

\noindent Klein's $j$-function $j:\H\rightarrow \C$ (which is invariant under
the whole classical modular group $\PSL(2,\Z)$) is related to the
$\lambda$ function through:
\be
j(\tau)=(q\circ\lambda)(\tau)=256\frac{(1-\lambda(\tau)+\lambda(\tau)^2)^3}{\lambda(\tau)^2(1-\lambda(\tau))^2}~~.
\ee
This function presents $\H$ as an $\infty:1$ {\em branched} cover of
the complex plane. The preimage $j^{-1}(\{z\})$ of any point $z\in \C$
is a full orbit of the action of $\PSL(2,\Z)$ on $\H$.  The
topological quotient $\H/\PSL(2,\Z)=Y(2)/\fB$ can be identified with
the complex plane $\C$ with coordinate $z$ using the $j$-function.

One can construct a good orbifold whose underlying space is the
complex plane by taking the {\em orbifold} quotient $\fM \eqdef
\H\sslash \PSL(2,\Z)=Y(2)\sslash \fB$ instead of the topological
quotient. This quotient has orbifold points located at $z=0$ and
$z=1728$, with isotropy groups $\Z_3$ and $\Z_2$ respectively. It has
a natural compactification given by the good orbifold
$\overline{\fM}\eqdef \C\P^1\sslash\fB$, which has a further singular
point at $z=\infty$ with isotropy group $\Z_2$.

\paragraph{Remark.}
The complex plane $\C$ coincides with the coarse moduli space of
elliptic curves.  The (uncompactified) moduli stack of elliptic curves
with one marked point is $\cM=\H\sslash\SL(2,\Z)\simeq Y(2)\sslash
(\rC\times \fB)$, where $\rC=\{-1,1\}\simeq \Z_2$ is the center of
$\SL(2,\Z)$. Thus $\fM$ is an orbifold two-fold cover of $\cM$. On the
other hand, the non-compact modular curve $Y(2)$ coincides with the
coarse moduli space of elliptic curves with level two
structure\footnote{Given a natural number $N>1$, a level $N$ structure
  on an elliptic curve $E$ defined over the complex numbers is a basis
  $(e_1,e_2)$ of $H_1(E,\Z_N)$ such that the intersection number
  $e_1\cdot e_2\in \Z_N$ equals ${\hat 1}\in \Z_N$. See \cite{Hain}
  for an introduction.}. The (uncompactified) moduli stack of elliptic
curves with level two structure is the quotient $\cM[2]=Y(2)\sslash
\rC$, which is a six-fold cover of $\cM$.

\subsection{The orbifold hyperbolic metric induced on $\fM$}
\label{subsec:singmetric}

\noindent 
Since $\PSL(2,\Z)$ contains elliptic elements
(which have fixed points on $\H$), the Poincar\'e metric $G_\H$ does
{\em not} descend through $j=q\circ \lambda$ to an ordinary Riemannian
metric on the topological quotient $\C=\H/\PSL(2,\Z)=Y(2)/\fB$ (in
fact, $\C$ admits no complete hyperbolic metric which is compatible
with its usual complex structure). Equivalently, the hyperbolic metric
$G$ of $Y(2)$ does not descend to an ordinary metric on the quotient, 
because the action of $\fB$ on $Y(2)$ has orbits with
non-trivial isotropy.

However, $G_\H$ (or $G$) does descend to a hyperbolic {\em orbifold}
metric on $\fM$. This can be viewed as a metric defined on
$\C\setminus \{0,1728\}$, with conical singularities at $z=0$ and
$z=1728$. It has the form:
\be
\dd s^2_0=\rho_0(z,\bar{z})^2|\dd z|^2
\ee
where: 
\be
\rho_0(q(\zeta),\overline{q(\zeta)})=\frac{\rho(\zeta,\bar{\zeta})}{|q'(\zeta)|}~~\mathrm{with}~~q'(\zeta)=\frac{(\zeta+1)(\zeta-2)(2\zeta-1)(1-\zeta+\zeta^2)^2}{\zeta^3(\zeta-1)^3}~~.
\ee 
Notice that $\rho_0$ tends to $0$ for $z\rightarrow \infty$ (i.e. for
$\zeta=0,1$) and that it tends to $+\infty$ at the orbifold points $z=0$ and
$z=1728$ (i.e. when $\zeta=e^{\pm \frac{\i\pi}{3}}$ and when $\zeta\in \{-1,1/2,2\}$).

\subsection{Description of modular-invariant $j$-models as lifts of 
generalized two-field $\alpha$-attractor models}

\noindent Since the orbifold metric $\rho_0$ has conical singularities
at finite distance, it cannot be used to define an $\alpha$-attractor
model in the sense of \cite{genalpha}. In particular, the
$\PSL(2,\Z)$-invariant models (the ``modular invariant $j$-models'')
defined in \cite{Sch1,Sch2} on the Poincar\'e half-plane do {\em
not} descend to generalized two-field $\alpha$-attractor models (in the sense
of \cite{genalpha}) defined on the topological quotient
$\H/\PSL(2,\Z)=Y(2)/\fB=\C$. However, those models do descend (upon
taking the quotient by $\Gamma(2)$ rather than by $\PSL(2,\Z)$) to
generalized two-field $\alpha$-attractor models defined on $Y(2)$, whose scalar
potential $\Phi$ is invariant under the action of the anharmonic group
$\fB$. Notice that a generalized two-field $\alpha$-attractor model defined on
$Y(2)$ typically lifts through $\lambda$ to a model which is invariant
only under the action of the subgroup $\Gamma(2)$ rather than under
the action of the full classical modular group $\PSL(2,\Z)$, since the
scalar potential $\Phi$ of an $\alpha$-attractor model with scalar
manifold $Y(2)$ need not be invariant under the action of $\fB$. As a
consequence, the models of \cite{Sch1,Sch2} do not provide the most
general lift of our $Y(2)$ models to the Poincar\'e half-plane. To
describe the most general lift, one must require invariance of the
half-plane model only under $\Gamma(2)$ rather than under
$\PSL(2,\Z)$.

\section{Cosmological trajectories}
\label{sec:traj}

In this section, we present numerical examples of cosmological
trajectories in generalized two-field $\alpha$-attractor models defined by
$Y(2)$ and in the corresponding lifted models, for certain simple but
natural choices of globally well-behaved scalar potentials, with
$\alpha=\frac{M_0}{3}$.

\subsection{Globally well-behaved scalar potentials on $Y(2)$}

\noindent 
Let $\rho=|\zeta|$ and $\theta=\arg(\zeta)$. The embedding \eqref{ee}
into the end compactification shows that a smooth potential
$\Phi:Y(2)=\C\setminus \{0,1\}\rightarrow \R$ is globally well-behaved
iff there exists a smooth map $\hPhi:X(2)=\C\P^1\simeq
\rS^2\rightarrow \R$ such that:
\be
\Phi(\rho,\theta)=\hPhi(2\arccot(\rho),\theta)~~.
\ee
Expanding $\hPhi$ into real (a.k.a. tesseral) spherical harmonics $Y_{lm}$: 
\be
\hPhi(\psi,\theta)=\sum_{l=0}^\infty\sum_{m=-l}^l D_{lm}Y_{lm}(\psi,\theta)
\ee
(where $D_{lm}$ are real constants) gives the uniformly-convergent expansion: 
\be
\Phi(\rho,\theta)=\sum_{l=0}^\infty\sum_{m=-l}^l D_{lm}Y_{lm}(2\arccot(\rho),\theta)~~.
\ee
Some simple choices for $\hPhi$ are as follows: 
\begin{itemize}
\itemsep 0.0em
\item The following linear combinations of the $s$ ($Y_{00}$) and
  $p_z$ ($Y_{10}$) orbitals:
\ben
\label{hPhipm}
\hPhi_+(\psi)=M_0\cos^2\left(\frac{\psi}{2}\right)~~,~~\hPhi_-(\psi)=M_0\sin^2\left(\frac{\psi}{2}\right)~~,
\een
\item The following linear combination of the $s$ ($Y_{00}$) and $p_x$
  ($Y_{11}$) orbitals:
\ben
\label{hPhi0}
\hPhi_0(\psi,\theta)=M_0(1+\sin\psi\cos\theta)=M_0(1+x_1)~~.
\een
\item The following linear combination of the $s$ ($Y_{00}$), $d_{z^2}$
  ($Y_{20}$) and $d_{x^2-y^2}$ ($Y_{22}$) orbitals:
\be
\hPhi_1(\psi,\theta)=M_0[1-\sin^2(\psi)\cos^2(\theta)+\cos^2(\psi)]=M_0\left[1-x_1^2+x_3^2\right]~~.
\ee
\end{itemize}
These choices give:
\beqan
\label{YPot}
&& \Phi_+=M_0\frac{\rho^2}{1+\rho^2}~~,~~\Phi_-=M_0\frac{1}{1+\rho^2}~~,~~\Phi_0=M_0\left[1+\frac{2\rho\cos\theta}{1+\rho^2}\right]~~,\nn\\
&& \Phi_1=M_0\left[1-\frac{4\rho^2}{(1+\rho^2)^2}\cos^2\theta+\frac{(1-\rho^2)^2}{(1+\rho^2)^2}\right]~~.
\eeqan
Notice that all of these potentials are compactly Morse in the sense
of \cite{genalpha}.  The extrema of the extended potentials are shown
in Table \ref{table:Extrema}. 

\begin{table}[H]
\centering
\begin{tabular}{|c|c|c|c|c|}
\toprule extremum in coords. $(\psi,\theta)$ & $\psi=0$ & $\psi=\pi$ &
$(\psi,\theta)=(\frac{\pi}{2},0)$ &
$(\psi,\theta)=(\frac{\pi}{2},\pi)$ \\ \midrule extremum in coord. $\zeta$ &
$\zeta=\infty$ & $\zeta=0$ & $\zeta=1$ & $\zeta=-1$\\ 
\midrule\midrule 
$\hPhi_+/M_0$ & max (1) & min (0) & none (1/2) & none (1/2) \\ 
\hline 
$\hPhi_-/M_0$ & min (0) & max (1) & none (1/2) & none (1/2) \\ 
\hline 
$\hPhi_0/M_0$ & none (1) & none (1) & max (2) & min (0) \\ 
\hline 
$\hPhi_1/M_0$ & max (2) & max (2) & min (0) & min (0) \\ 
\bottomrule
\end{tabular}
\caption{Extrema of extended potentials. In parantheses, we indicate
  the value of $\hPhi/M_0$ at the extremum point. Notice that only
  $\Phi_0$ and $\Phi_1$ have extrema inside $Y(2)$. Also notice that
  each of the cusp points is an extremum point of $\hPhi_1$.}
\label{table:Extrema}
\end{table}

Since $\tau_0=1+\i\in \H$ is an inverse image of the point $\zeta=-1$,
the lifted potentials $\tPhi_0$ and $\tPhi_1$ have local minima (equal
to zero) at the point $\tau_0$ and at each of its images through the
action of $\Gamma(2)$. These local minima accumulate toward any
point on the conformal boundary $\partial_\infty \H$. The level sets
of the potentials $\Phi_\pm$, $\Phi_0$ and $\Phi_1$ and of their lifts
to $\H$ are shown in Figures \ref{fig:PhiPlus}, \ref{fig:PhiMinus},
\ref{fig:Phi0} and \ref{fig:Phi1} of the next subsection. Notice that 
none of these potentials is invariant under the anharmonic group $\fB$.

\subsection{Some examples of cosmological trajectories}

\noindent In this subsection, we present four numerically computed
trajectories (drawn in black, red, magenta and yellow) for each of the
potentials $\Phi_\pm, \Phi_0$ and $\Phi_1$. These trajectories were
chosen such that their lifts to $\H$ have the initial conditions at
cosmological time $t=0$ shown in Table \ref{table:InCond},
irrespective of the choice of the scalar potential.

\begin{table}[H]
\centering
\begin{tabular}{|c|c|c|}
\toprule
trajectory & $\tau_0$ & $\tv_0$ \\
\midrule\midrule
black  & $0.4+0.5\i$  & $0.3+\i$ \\
\hline
red  & $1.4+0.5\i$ & $0.1+0.2\i$  \\
\hline
magenta & $0.2+0.7\i$ & $0.7+0.5\i$  \\
\hline
yellow & $0.3+0.5\i$  & $0$\\
\bottomrule
\end{tabular}
\caption{Initial conditions for four trajectories on the Poincar\'e half-plane.}
\label{table:InCond}
\end{table}

\noindent For each of the four scalar potentials, Table \ref{table:InCondInf}
shows which of these initial conditions belong to the inflation region
of the corresponding lifted potential. Trajectories which satisfy 
this condition produce a cosmological scale factor $a(t)$ which is 
a convex and increasing function of $t$ for some 
interval starting at $t=0$. Notice that this condition cannot be satisfied 
when $\Phi=0$. 

\vspace{3mm}

\begin{table}[H]
\centering
\begin{tabular}{|c|c|c|c|c|}
\toprule trajectory & $\Phi_+$ & $\Phi_-$ & $\Phi_0$ & $\Phi_1$
\\ \midrule
\midrule 
black & no & no & no & no \\ 
\hline 
red & yes & no & yes & yes \\ 
\hline 
magenta & no & no & yes & no \\ 
\hline 
yellow & yes & yes & yes & yes \\ 
\bottomrule
\end{tabular}
\caption{Inflationary character of the initial conditions of
  Table \ref{table:InCond} with respect to the scalar potentials
  $\Phi_\pm$, $\Phi_0$ and $\Phi_1$. The table indicates ``yes''
  when $(\tau_0,\tv_0)$ belongs to the inflation region of $\tPhi$ and
  ``no'' otherwise.}
\label{table:InCondInf}
\end{table}

\

\paragraph{Trajectories for $\Phi=0$.}

\begin{figure}[H]
\centering
\begin{minipage}{.49\textwidth}
\centering \includegraphics[width=.98\linewidth]{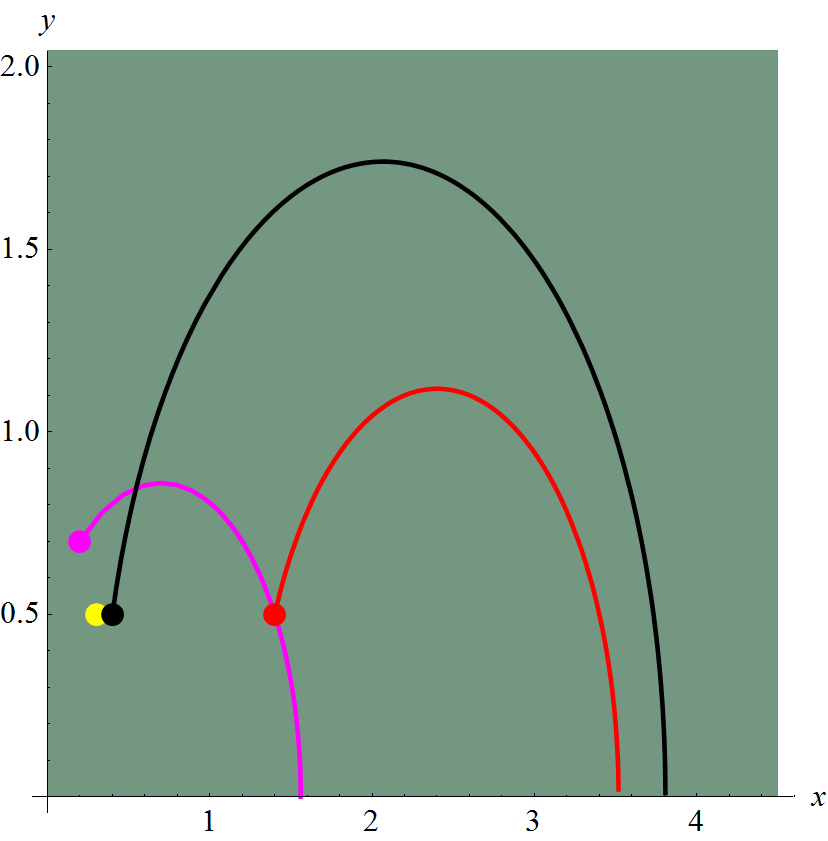}
\subcaption{Trajectories for $\tPhi=0$ on the Poincar\'e
  half-plane. The solution shown in yellow is stationary.}
\label{fig:TrajNoPot}
\end{minipage}\hfill
\begin{minipage}{.47\textwidth}
\centering \includegraphics[width=.99\linewidth]{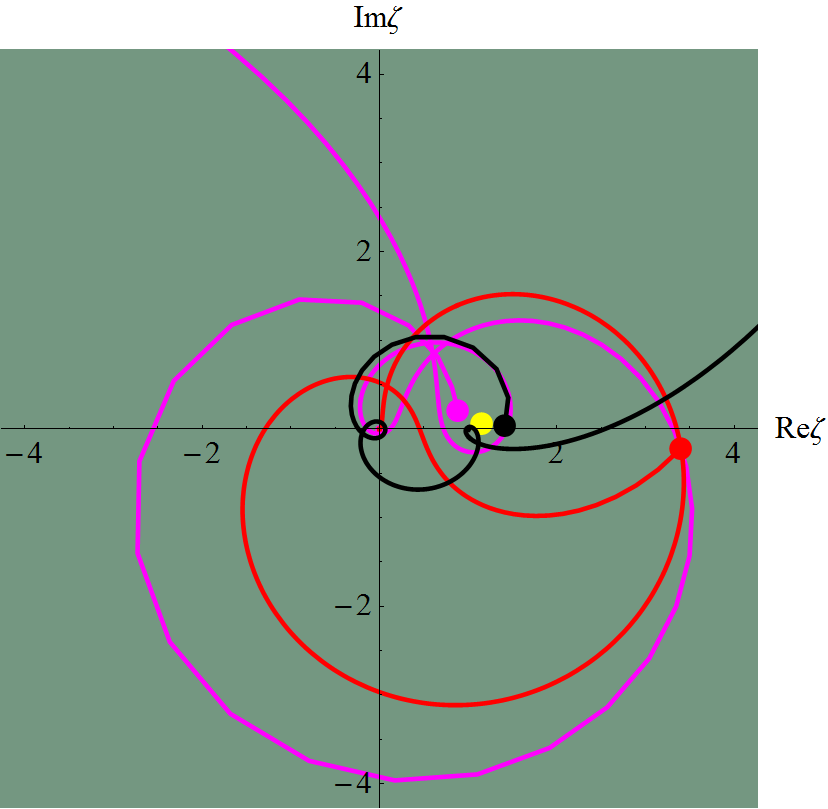}
\vskip 5mm
\subcaption{Projection to $Y(2)$ of the trajectories shown at the left.}
\label{fig:TrajNoPotProj}
\end{minipage}
\caption{Numerical solutions for $\Phi=0$ and $\alpha=\frac{M_0}{3}$.}
\label{fig:TrajNoPotAll}
\end{figure}

\noindent 
To understand the effect of the hyperbolic geometry, it is instructive
to consider first the case when the scalar potential vanishes
identically. In this situation, the yellow trajectory (which has
vanishing initial speed) remains stationary for all values of the
cosmological time, as can be seen by inspecting the system
\eqref{elplane} and using the Picard-Lindel\"of theorem. Any other
trajectory on $Y(2)$ ultimately evolves toward one of the cusp points,
while its lift to the Poincar\'e half-plane evolves toward the
conformal boundary $\partial_\infty\H$. Hence each cusp point exerts
an attractive ``effective force''. As shown in Figure
\ref{fig:TrajNoPotAll}, each projected trajectory spirals in a
complicated manner around the cusp points of $Y(2)$, until eventually
``falling'' deep inside one of the cusp neighborhoods, where it
evolves when $t\rightarrow +\infty$ toward the corresponding ideal
point. In the figures, the initial point of each trajectory is shown
as a thick dot.

\

\paragraph{Trajectories for $\Phi_+$.}

\begin{figure}[H]
\centering
\vskip -0.5em
\begin{minipage}{.48\textwidth}
\vskip 0.5em
\centering \includegraphics[width=.98\linewidth]{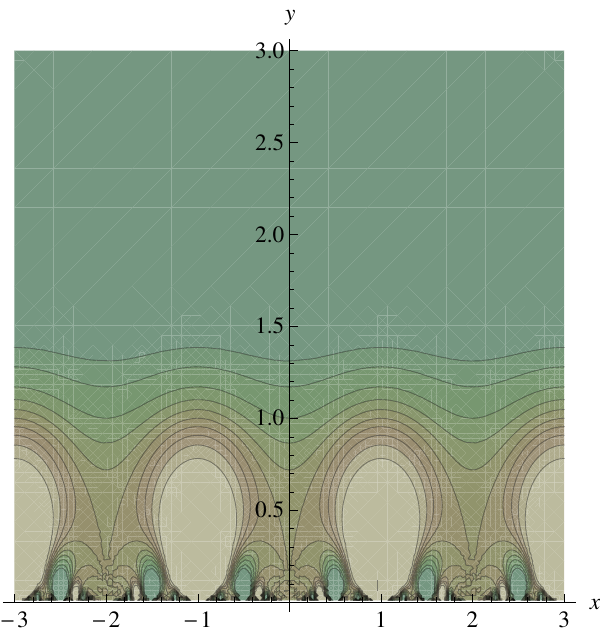}
\vskip 0.3em
\subcaption{Level plot of $\tPhi_+/M_0$ on the Poincar\'e half-plane.}
\label{fig:PhiPlusLifted}
\end{minipage}\hfill
\begin{minipage}{.48\textwidth}
\vskip 0.3em
\centering \includegraphics[width=1.03\linewidth]{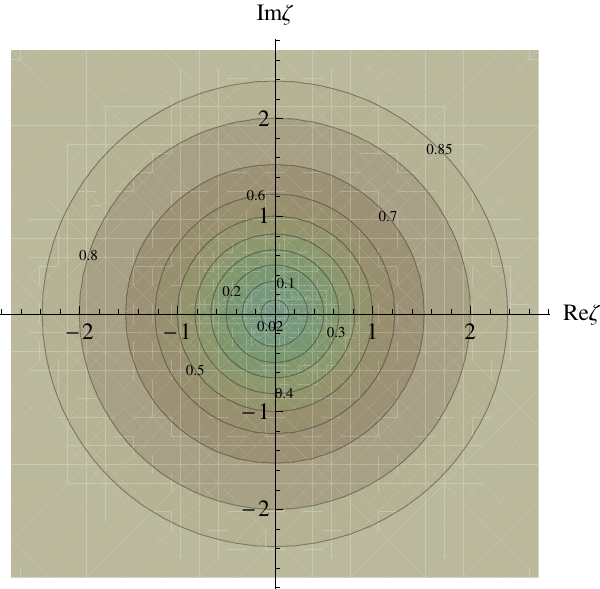}
\vskip 1em
\subcaption{Level plot of $\Phi_+/M_0$ on the twice punctured plane.}
\label{fig:PhiPlusProj}
\end{minipage}
\vskip 0.2em
\caption{Level plots of $\tPhi_+/M_0$ and $\Phi_+/M_0$. Darker
  tones indicate lower values.}
\label{fig:PhiPlus}
\end{figure}

\noindent The level sets of the potential $\Phi_+$ and of its lift
$\tPhi_+$ to the Poincar\'e half-plane are shown in Figure
\ref{fig:PhiPlus}. This potential has a vanishing infimum at the cusp
$\zeta=0$ and tends to its supremum (which equals $M_0$) for
$\xi\rightarrow \infty$. 

\begin{figure}[H]
\centering
\begin{minipage}{.47\textwidth}
\centering \includegraphics[width=1.0\linewidth]{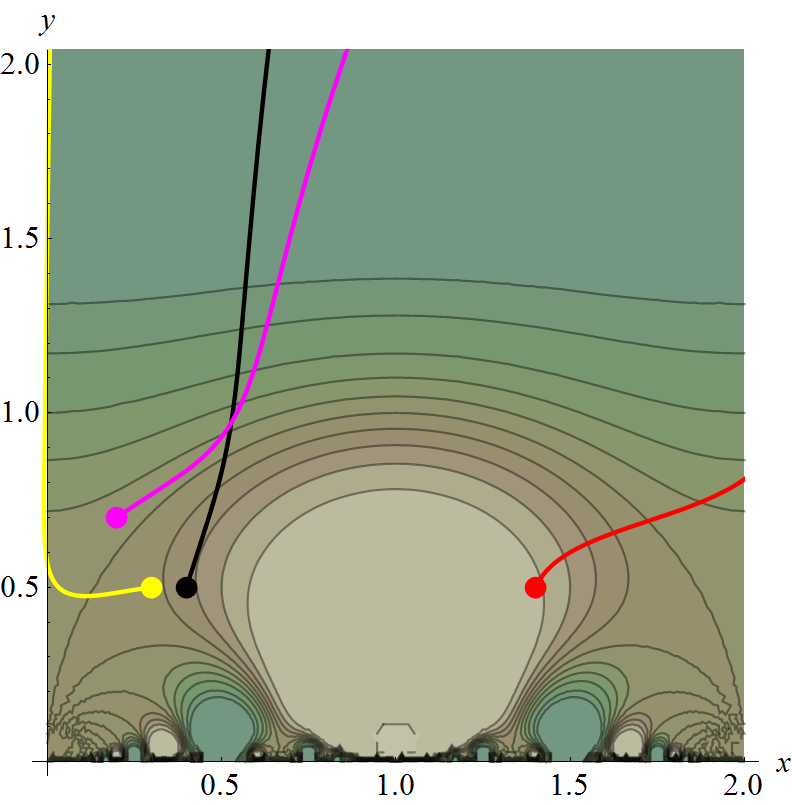}
\vskip -0.1em
\subcaption{Trajectories for $\tPhi=\tPhi_+$ on the Poincar\'e
  half-plane.}
\label{fig:TrajPhiPlus}
\end{minipage}\hfill
\begin{minipage}{.47\textwidth}
\centering \includegraphics[width=.98\linewidth]{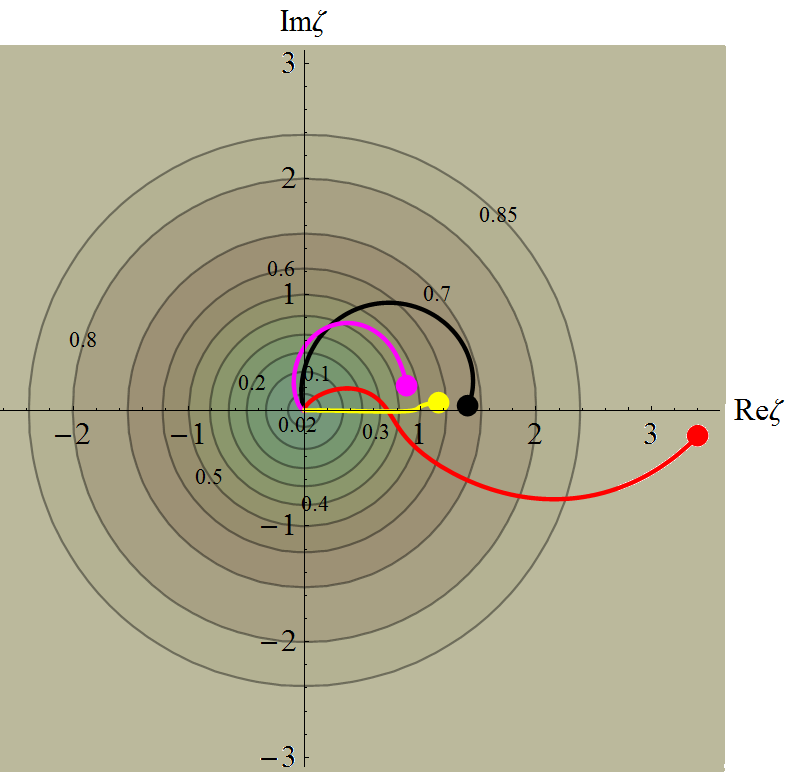}
\vskip 1em
\subcaption{Projection to $Y(2)$ of the trajectories shown at the left.}
\label{fig:TrajPhiPlusProj}
\end{minipage}
\caption{Numerical solutions for $\Phi=\Phi_+$ and $\alpha=\frac{M_0}{3}$.}
\label{fig:TrajPhiPlusAll}
\end{figure}

\noindent The four trajectories whose lifts have the initial
conditions given in Table \ref{table:InCond} are shown in Figure
\ref{fig:TrajPhiPlusAll}. Due to the effect of the potential, the
projected yellow trajectory (which starts with initial velocity zero)
evolves toward the cusp point at $\zeta=0$, as do the other three
trajectories. For clarity, Figure \ref{fig:TrajPhiPlus} shows only a
small portion of the trajectories on the Poincar\'e half-plane.

\begin{figure}[H]
\centering
\begin{minipage}{.47\textwidth}
\centering \includegraphics[width=.98\linewidth]{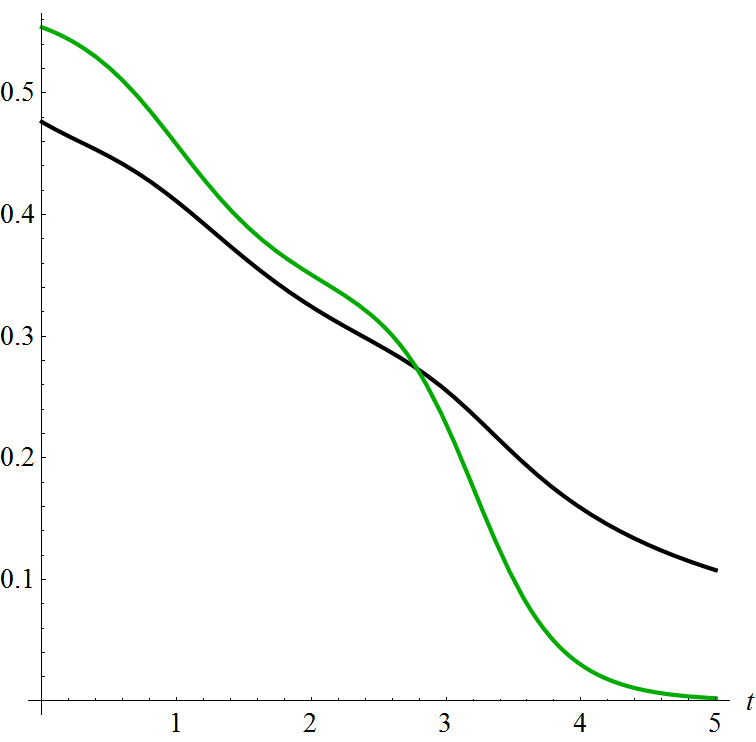}
\subcaption{Plot of $H(t)/\sqrt{M_0}$ and $H_c(t)/\sqrt{M_0}$ for the red trajectory.}
\label{fig:HPhiPlus1}
\end{minipage}\hfill
\begin{minipage}{.47\textwidth}
\centering \includegraphics[width=.98\linewidth]{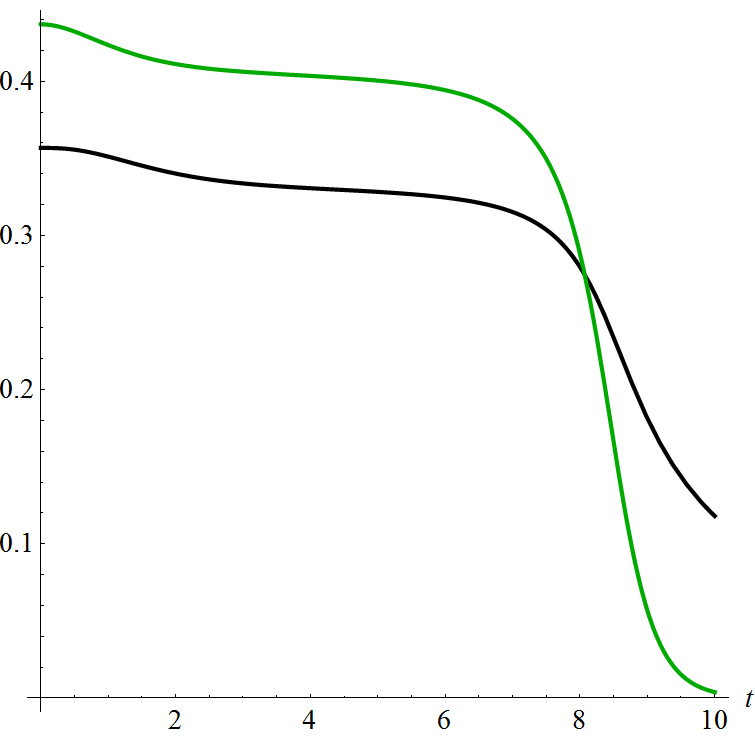}
\subcaption{Plot of $H(t)/\sqrt{M_0}$ and $H_c(t)/\sqrt{M_0}$ for the yellow trajectory.}
\label{fig:HPhiPlus2}
\end{minipage}
\caption{Plot of $H(t)/\sqrt{M_0}$ (black) and $H_c(t)/\sqrt{M_0}$ (green) for the red and
  yellow trajectories with $\Phi=\Phi_+$ and $\alpha=\frac{M_0}{3}$.}
\label{fig:HPhiPlus}
\end{figure}

\noindent
Figure \ref{fig:HPhiPlus} shows the evolution of the Hubble parameter
$H(t)$ for the red and yellow trajectories, comparing it with the
critical Hubble parameter $H_c(t)\eqdef H_c(\varphi(t))$ along that
trajectory.
For the clarity of the figure we represent here the evolution of 
$H(t)$ and $H_c(t)$ only up to $t<5s$.
As clear from the figure, the tangent vectors of these
trajectories lie within the inflation region of $\Phi_+$ for some time
interval starting at $t=0$, thus displaying the behavior expected in
inflation.

\paragraph{Trajectories for $\Phi_-$.}

\begin{figure}[H]
\centering
\vskip -0.5em
\begin{minipage}{.47\textwidth}
\vskip 0.5em
\centering \includegraphics[width=1.03\linewidth]{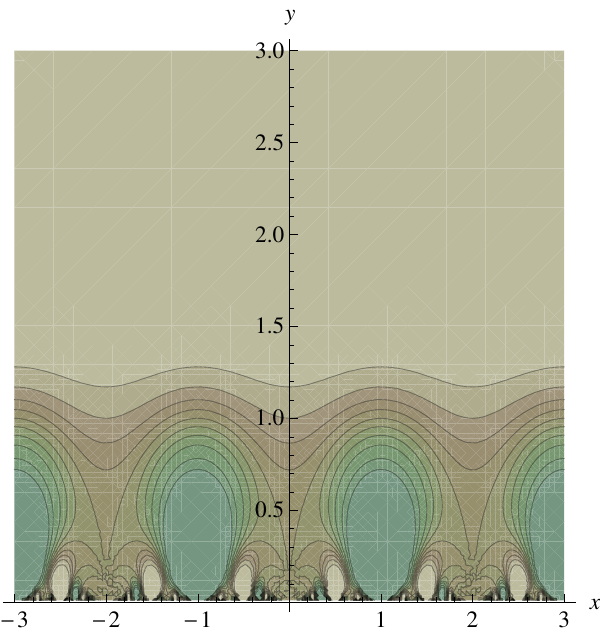}
\vskip 0.5em
\subcaption{Level plot of $\tPhi_-/M_0$ on the Poincar\'e half-plane.}
\label{fig:PhiMinusLifted}
\end{minipage}\hfill
\begin{minipage}{.47\textwidth}
\vskip 0.2em
\centering \includegraphics[width=1.08\linewidth]{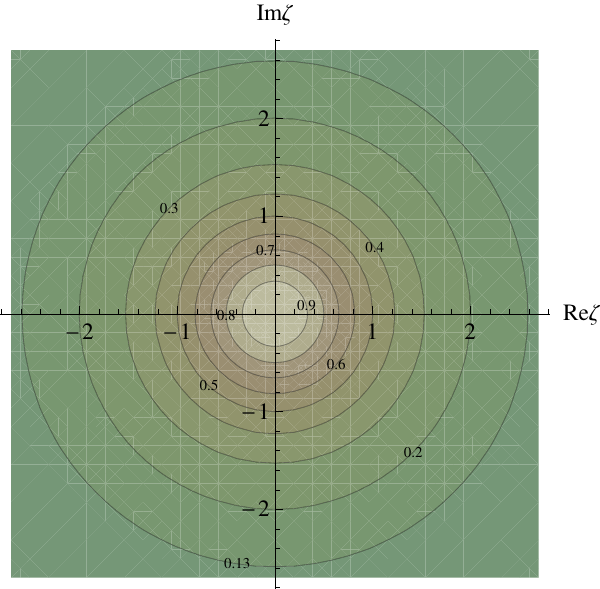}
\vskip 1em
\subcaption{Level plot of $\Phi_-/M_-$ on the twice punctured plane.}
\label{fig:PhiMinusProj}
\end{minipage}
\vskip 0.5em
\caption{Level plots of $\tPhi_-/M_0$ and $\Phi_-/M_0$. Darker
  tones indicate lower values.}
\label{fig:PhiMinus}
\end{figure}

\noindent
The level sets of the potential $\Phi_-$ and of its lift $\tPhi_-$ to
the Poincar\'e half-plane are shown in Figure \ref{fig:PhiMinus}. This
potential has a supremum (which equals $M_0$) at the puncture $\zeta=0$
and tends to its vanishing infimum for $\zeta\rightarrow +\infty$.

\begin{figure}[H]
\centering
\begin{minipage}{.47\textwidth}
\vskip -0.5em
\centering \includegraphics[width=1.03\linewidth]{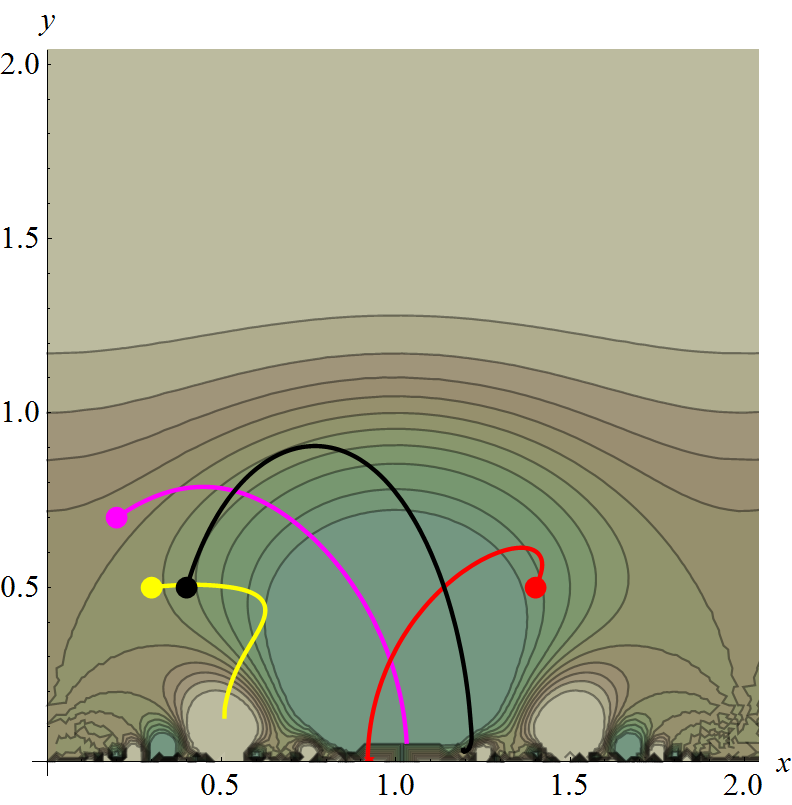}
\vskip -0.1em
\subcaption{Trajectories for $\tPhi=\tPhi_-$ on the Poincar\'e
  half-plane.}
\label{fig:TrajPhiMinus}
\end{minipage}\hfill
\begin{minipage}{.47\textwidth}
\vskip 0.5em
\centering \includegraphics[width=1.06\linewidth]{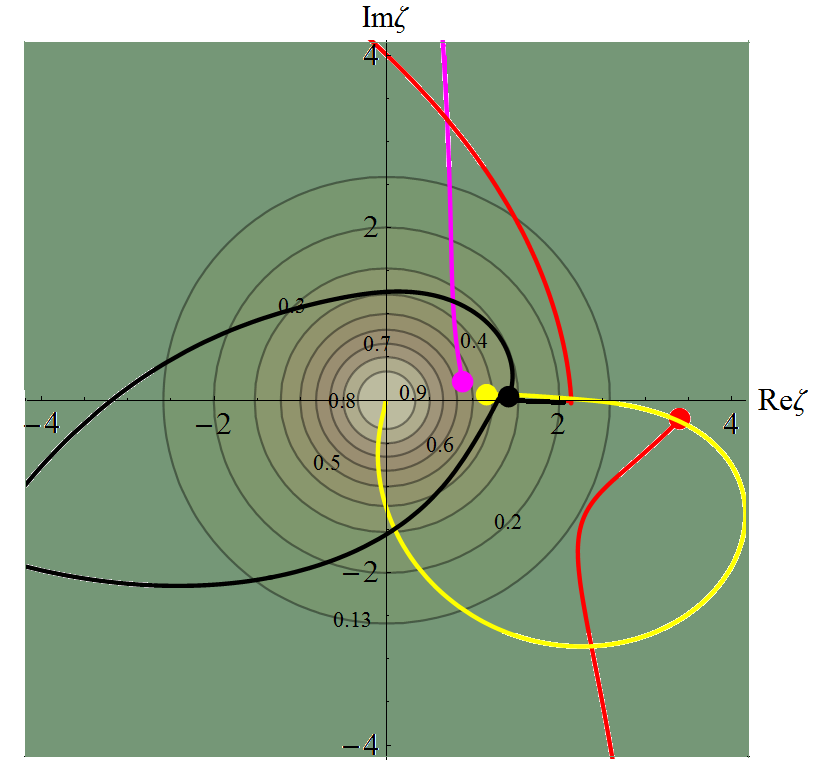}
\vskip 0.5em
\subcaption{Projection to $Y(2)$ of the trajectories shown at the left. For clarity, we 
show only the region around the puncture $\zeta=0$.}
\label{fig:TrajPhiMinusProj}
\end{minipage}
\caption{Numerical solutions for $\Phi=\Phi_-$ and $\alpha=\frac{M_0}{3}$.}
\label{fig:TrajPhiMinusAll}
\end{figure}

\begin{figure}[H]
\centering
\begin{minipage}{.47\textwidth}
\centering \includegraphics[width=1.02\linewidth]{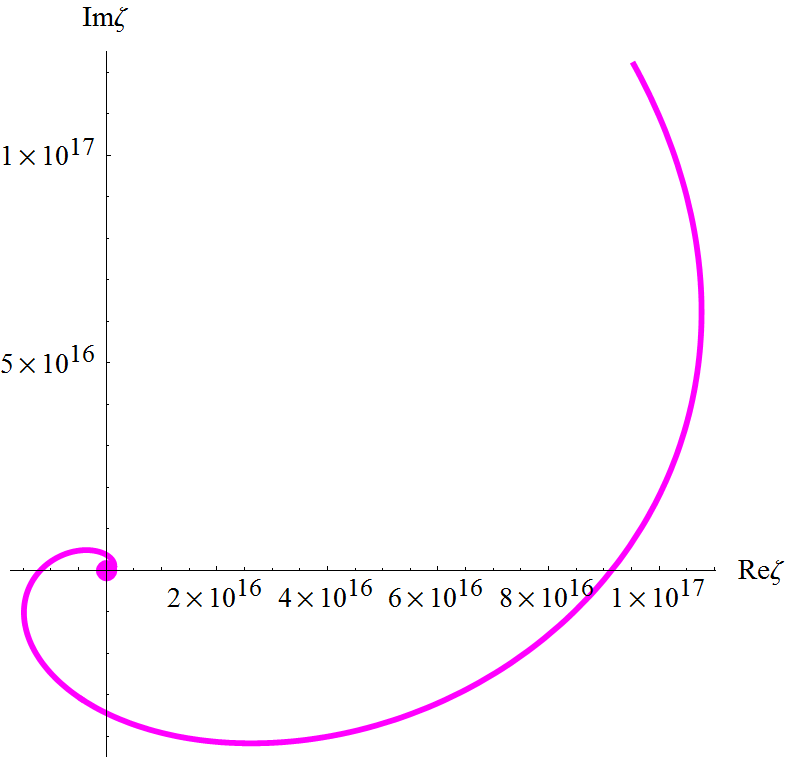}
\subcaption{Projected magenta trajectory for the potential $\Phi_-$.}
\label{fig:TrajDetailMagenta}
\end{minipage}\hfill
\begin{minipage}{.47\textwidth}
\centering \includegraphics[width=1.05\linewidth]{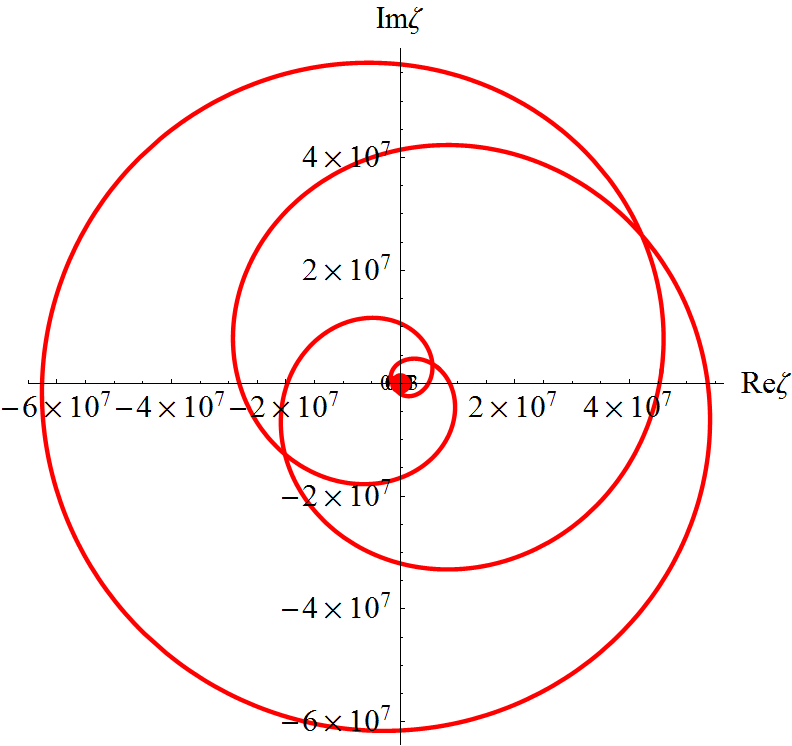}
\vskip 1.1em
\subcaption{Projected red trajectory  for the potential  $\Phi_-$.}
\label{fig:TrajDetailRed}
\end{minipage}
\caption{Large scale view of the red and magenta trajectories for the potential $\Phi_-$ with $\alpha=\frac{M_0}{3}$.}
\label{fig:TrajDetailRedMagenta}
\end{figure}

\begin{figure}[H]
\centering
\begin{minipage}{.47\textwidth}
\centering
\includegraphics[width=1.03\linewidth]{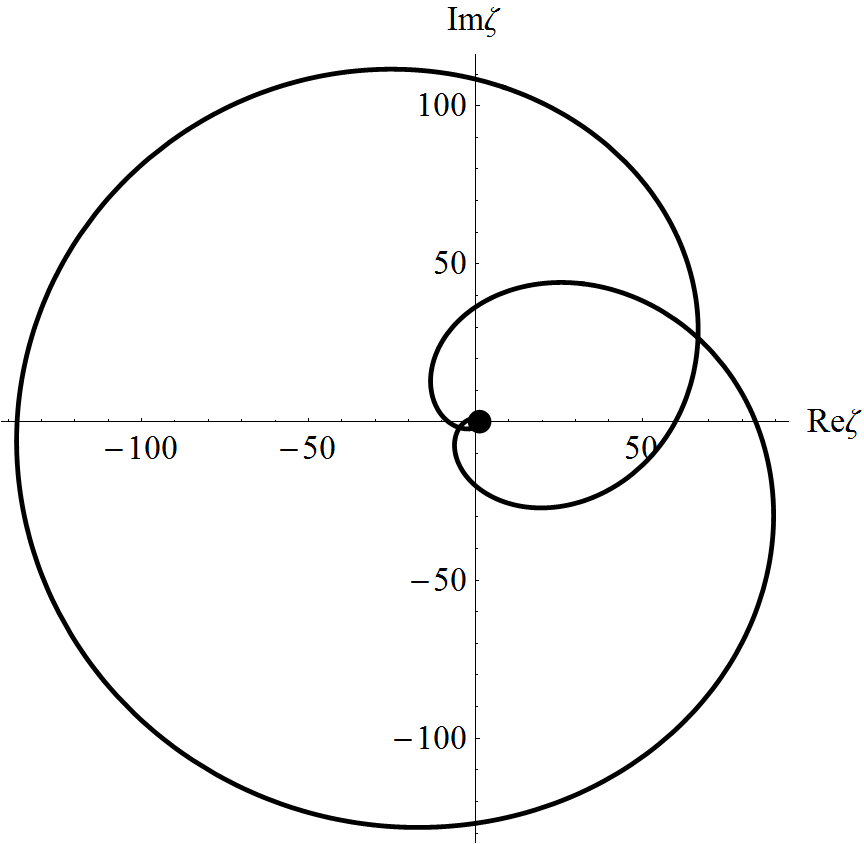}
\vskip 1.0mm \subcaption{Large scale view of the projected black trajectory for the potential
  $\Phi_-$}
\label{fig:TrajDetailBlack1}
\end{minipage}\hfill
\begin{minipage}{.47\textwidth}
\centering
\includegraphics[width=1.03\linewidth]{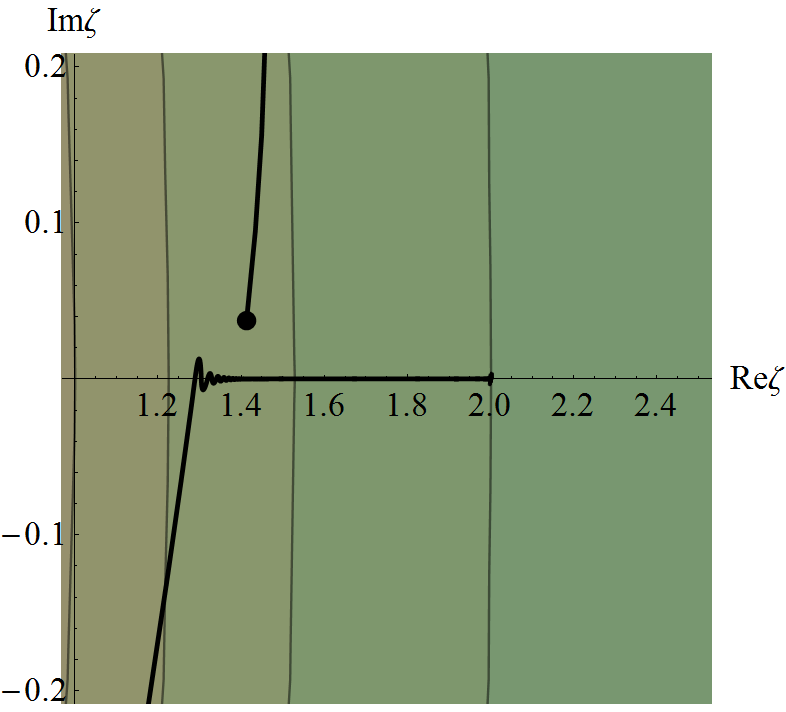}
\vskip 5.0mm \subcaption{Detail of the beginning and end of the
  projected black trajectory for the potential $\Phi_-$}
\label{fig:TrajDetailBlack2}
\end{minipage}
\caption{The projected black trajectory for the potential $\Phi_-$ with $\alpha=\frac{M_0}{3}$.}
\label{fig:TrajDetailBlack}
\end{figure}

\noindent Figure \ref{fig:TrajPhiMinusAll} shows the lifted
trajectories with initial conditions given in Table \ref{table:InCond}
and their projections through the uniformization map. Despite the
repulsive force produced by the potential (which is counterbalanced by
the attractive effective force produced by the hyperbolic geometry of
the cusp), the yellow trajectory evolves toward the cusp point $\zeta=0$. 
As shown in Figure \ref{fig:TrajDetailMagenta}, the projected magenta 
trajectory evolves toward the cusp at infinity, while the projected red and
 black trajectories (which, for clarity, are not fully shown in Figure 
\ref{fig:TrajPhiMinusProj}) evolve slowly toward a point located beyond 
the cusp, at $\zeta=2$, where they appear to stop after winding 
around it a few times in a complicated manner. For more details of
 these two trajectories see Figure \ref{fig:TrajDetailRed}  and 
Figure \ref{fig:TrajDetailBlack}. In this example, the repulsive force
exerted by the potential is weaker that the effective attraction
produced by the cusp at the origin due to the hyperbolic metric.

\

\begin{figure}[H]
\centering
\begin{minipage}{.47\textwidth}
\centering \includegraphics[width=1.0\linewidth]{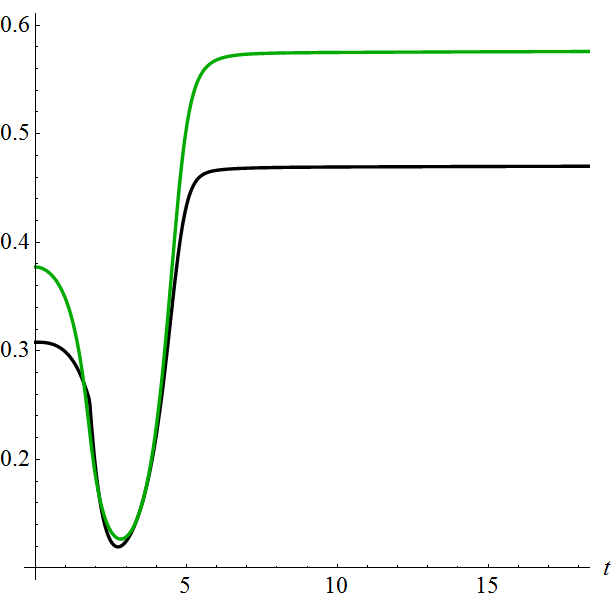}
\subcaption{Plot of $H(t)/\sqrt{M_0}$ and $H_c(t)/\sqrt{M_0}$ for the
  yellow trajectory. This trajectory produces inflation for small $t$,
  then exits the inflation regime and re-enters it at some later
  time.}
\label{fig:HPhiMinus1}
\end{minipage}\hfill
\begin{minipage}{.47\textwidth}
\centering \includegraphics[width=1.02\linewidth]{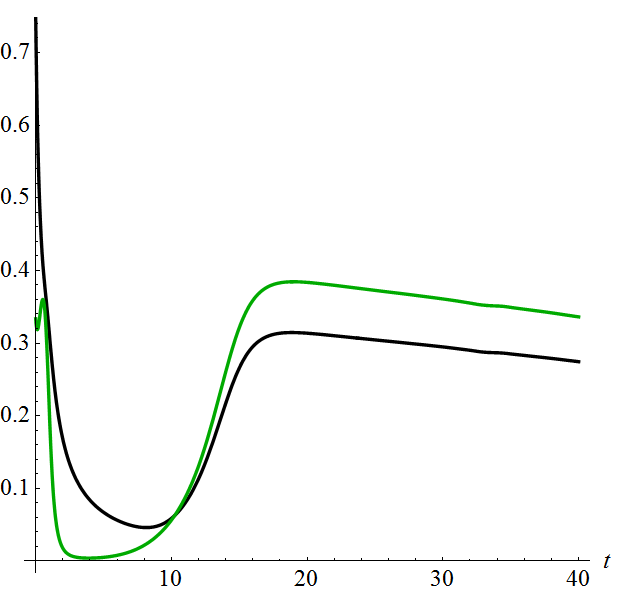}
\subcaption{Plot of $H(t)/\sqrt{M_0}$ and $H_c(t)/\sqrt{M_0}$ for the
  black trajectory. This trajectory does not produce inflation for
  small $t$, but it enters the inflation regime at later times.}
\label{fig:HPhiMinus2}
\end{minipage}
\vskip 0.5em
\caption{Plot of $H(t)/\sqrt{M_0}$ (black) and $H_c(t)/\sqrt{M_0}$
  (green) for the yellow and black trajectories with $\Phi=\Phi_-$ and
  $\alpha=\frac{M_0}{3}$.}
\label{fig:HPhiMinus}
\end{figure}

\noindent Figure \ref{fig:HPhiMinus} shows the evolution of the Hubble
parameter $H(t)$ along the yellow and black trajectories.

\paragraph{Trajectories for $\Phi_0$.}

\begin{figure}[H]
\centering
\vskip -0.5em
\begin{minipage}{.47\textwidth}
\vskip 0.5em
\centering \includegraphics[width=1\linewidth]{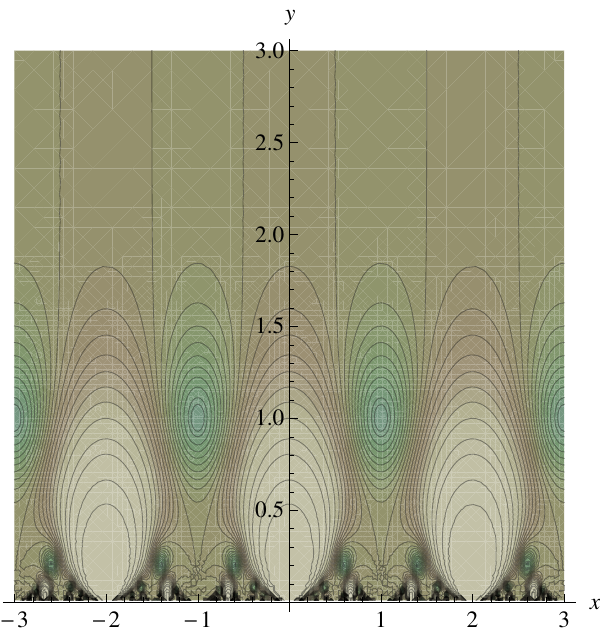}
\vskip 0.5em
\subcaption{Level plot of $\tPhi_0/M_0$ on the Poincar\'e half-plane.}
\label{fig:Phi0Lifted}
\end{minipage}\hfill
\begin{minipage}{.47\textwidth}
\vskip 0.3em
\centering \includegraphics[width=1.04\linewidth]{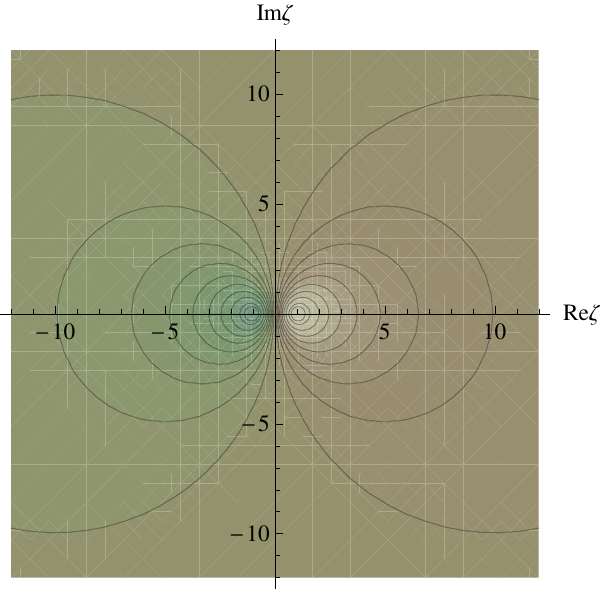}
\vskip 1em
\subcaption{Level plot of $\Phi_0/M_0$ on the twice punctured plane.}
\label{fig:Phi0Proj}
\end{minipage}
\vskip 0.5em
\caption{Level plots of $\tPhi_0/M_0$ and $\Phi_0/M_0$. Darker
  tones indicate lower values.}
\label{fig:Phi0}
\end{figure}

\noindent The level sets of the potential $\Phi_0$ and of its lift
$\tPhi_0$ to the Poincar\'e half-plane are shown in Figure
\ref{fig:Phi0}. This potential has a supremum (which equals $2M_0$) at
the cusp $\zeta=1$ and a vanishing minimum at the point $\zeta=-1$,
which lies inside $Y(2)$.

\begin{figure}[H]
\centering
\begin{minipage}{.47\textwidth}
\centering \includegraphics[width=1.03\linewidth]{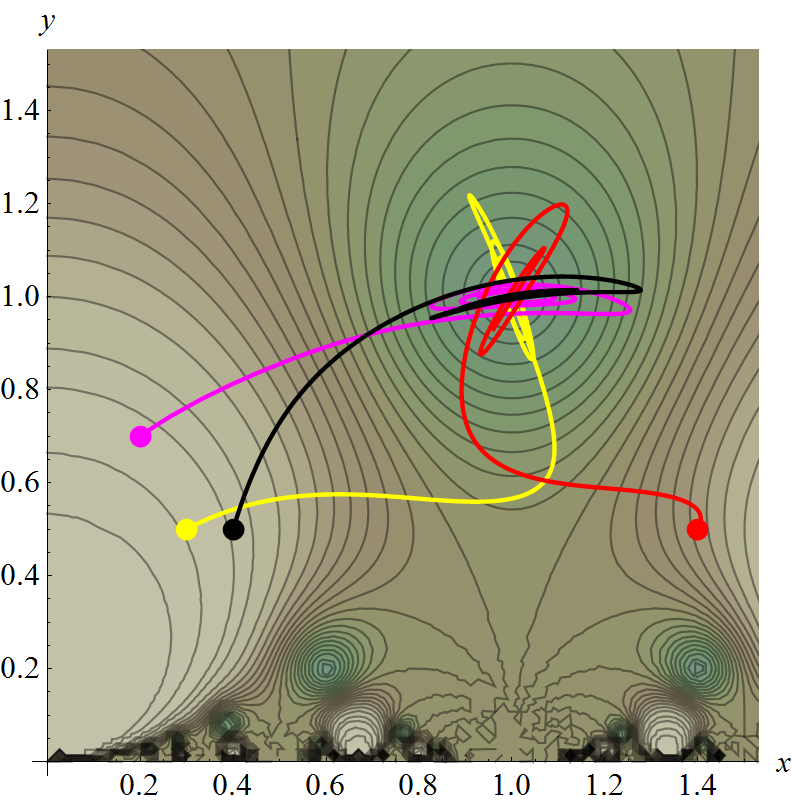}
\vskip 0.2em
\subcaption{Trajectories for $\tPhi=\tPhi_0$ on the Poincar\'e
  half-plane.}
\label{fig:TrajPhi0}
\end{minipage}\hfill
\begin{minipage}{.47\textwidth}
\centering \includegraphics[width=1.06\linewidth]{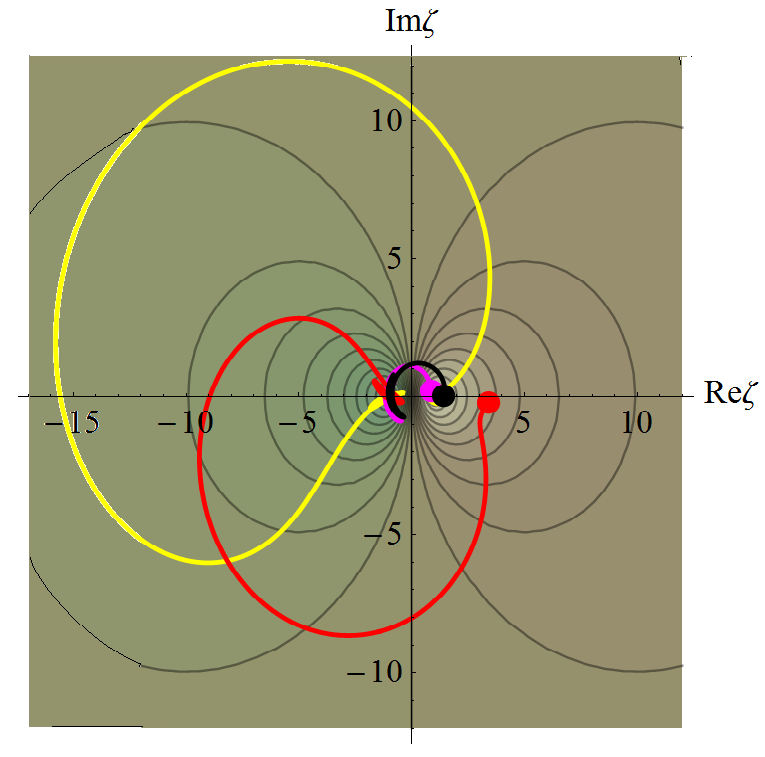}
\vskip 0.5em
\subcaption{Projection to $Y(2)$ of the trajectories shown at the left.}
\vskip 0.5em
\label{fig:TrajPhi0Proj}
\end{minipage}
\caption{Numerical solutions for $\Phi=\Phi_0$ and $\alpha=\frac{M_0}{3}$.}
\label{fig:TrajPhi0All}
\end{figure}

\begin{figure}[H]
\centering \includegraphics[width=.47\linewidth]{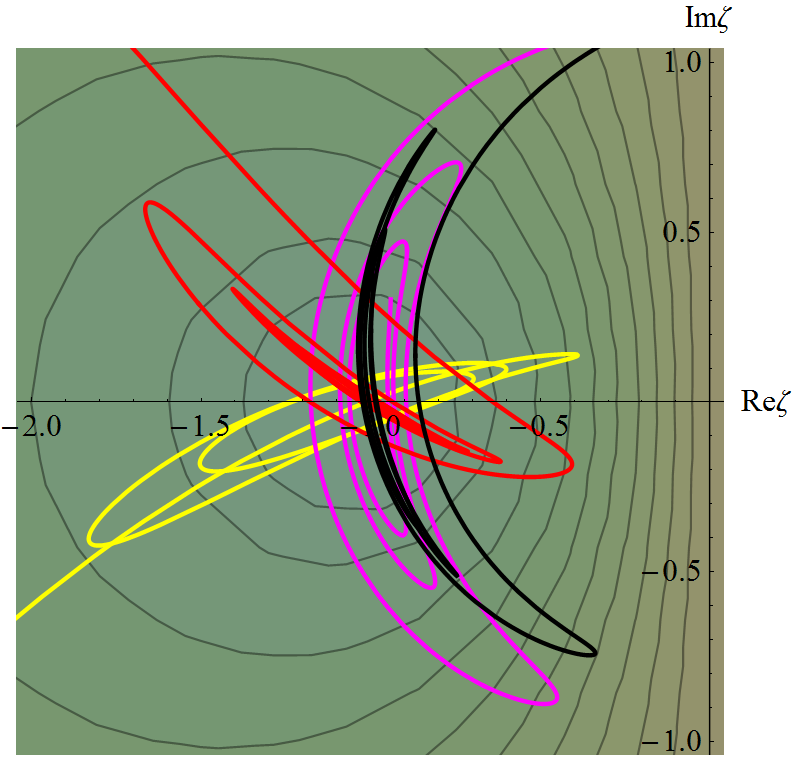}
\caption{Detail of the projected trajectories close to the minimum of $\Phi_0$, for $\alpha=\frac{M_0}{3}$.}
\label{fig:TrajPhi0ProjLarge}
\end{figure}

\begin{figure}[H]
\centering
\begin{minipage}{.47\textwidth}
\centering \includegraphics[width=1.0\linewidth]{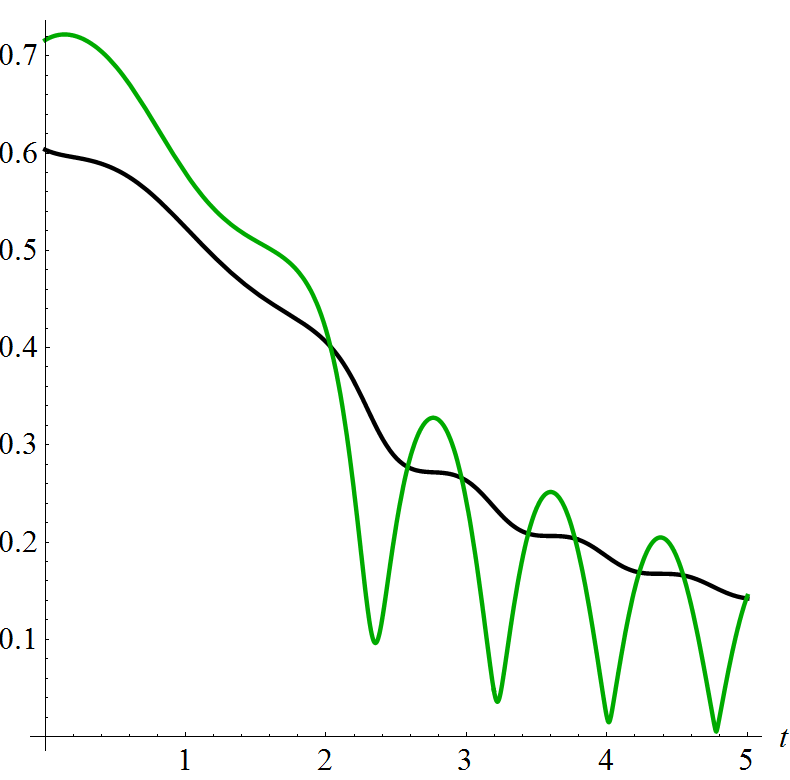}
\subcaption{Plot of $H(t)/\sqrt{M_0}$ and $H_c(t)/\sqrt{M_0}$ for the red trajectory.}
\label{fig:HPhi01}
\end{minipage}\hfill
\begin{minipage}{.47\textwidth}
\centering \includegraphics[width=1.0\linewidth]{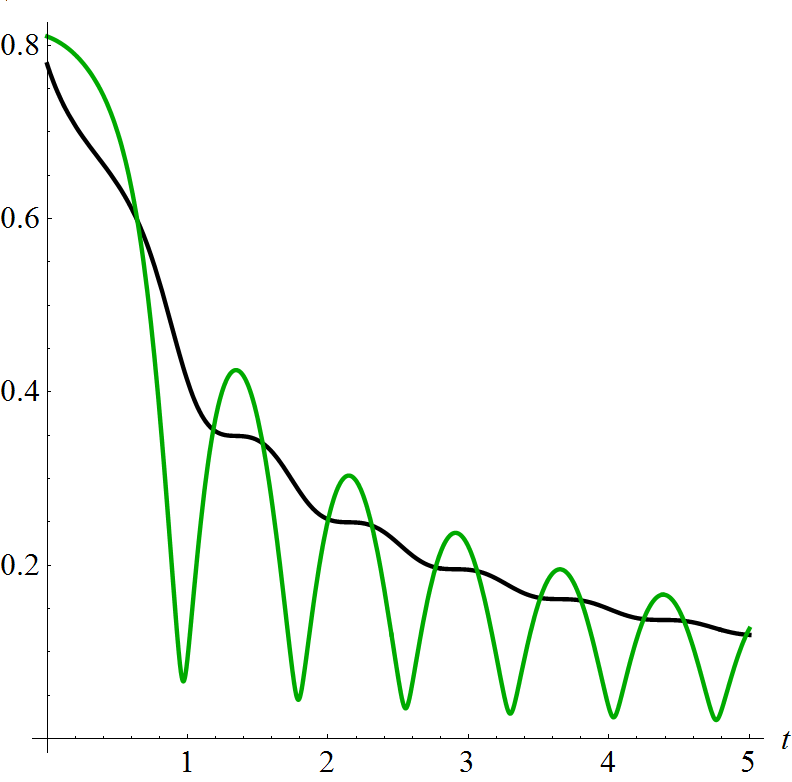}
\subcaption{Plot of $H(t)/\sqrt{M_0}$ and $H_c(t)/\sqrt{M_0}$ for the magenta trajectory.}
\label{fig:HPhi02}
\end{minipage}
\caption{Plot of $H(t)/\sqrt{M_0}$ (black) and $H_c(t)/\sqrt{M_0}$ (green) for the red and
  magenta trajectories with $\Phi=\Phi_0$ and $\alpha=\frac{M_0}{3}$.}
\label{fig:HPhi0}
\end{figure}

\noindent Figure \ref{fig:TrajPhi0All} shows the four lifted and
projected trajectories with initial conditions given in Table
\ref{table:InCond}. All four trajectories eventually evolve toward the
minimum point $\zeta=-1$ of $\Phi_0$, after spiraling around it in a
complicated manner (see the detail of these trajectories shown in
Figure \ref{fig:TrajPhi0ProjLarge}). Figure \ref{fig:HPhi0} shows 
the time evolution of the Hubble parameter along the red and magenta 
trajectories, both of which lead to inflation for some time interval 
starting at $t=0$. 

\paragraph{Trajectories for $\Phi_1$.}

\begin{figure}[H]
\centering
\vskip -0.5em
\begin{minipage}{.47\textwidth}
\vskip 0.5em
\centering ~~\includegraphics[width=.98\linewidth]{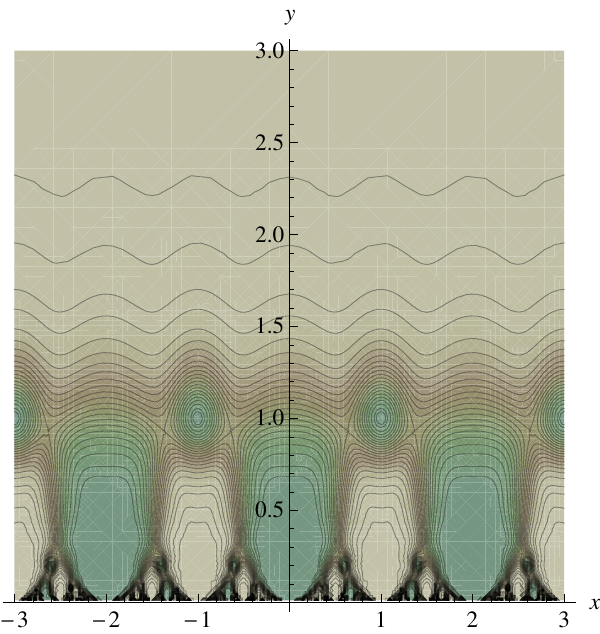}
\vskip 0.5em
\subcaption{Level plot of $\tPhi_1/M_0$ on the Poincar\'e half-plane.}
\label{fig:Phi1Lifted}
\end{minipage}\hfill
\begin{minipage}{.47\textwidth}
\centering ~~~~\includegraphics[width=1.02\linewidth]{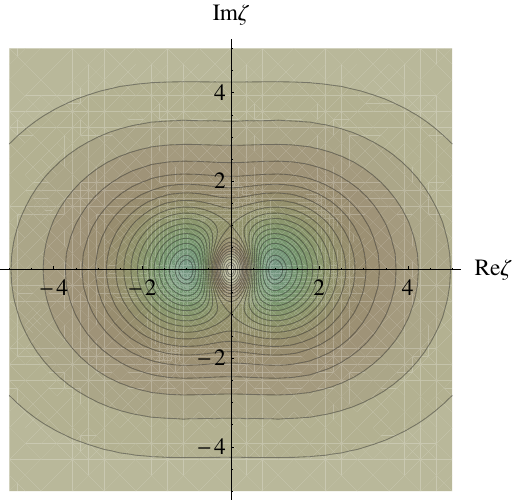}
\vskip 1em
\subcaption{Level plot of $\Phi_1/M_0$ on the twice punctured plane.}
\label{fig:Phi1Proj}
\end{minipage}
\vskip 0.3em
\caption{Level plots of $\tPhi_1/M_0$ and $\Phi_1/M_0$. Darker
  tones indicate lower values.}
\label{fig:Phi1}
\end{figure}

\noindent
The level sets of the potential $\Phi_1$ and of its lift $\tPhi_1$ to
the Poincar\'e half-plane are shown in Figure \ref{fig:Phi1}. 

\begin{figure}[H]
\centering
\begin{minipage}{.47\textwidth}
\centering \includegraphics[width=1.0\linewidth]{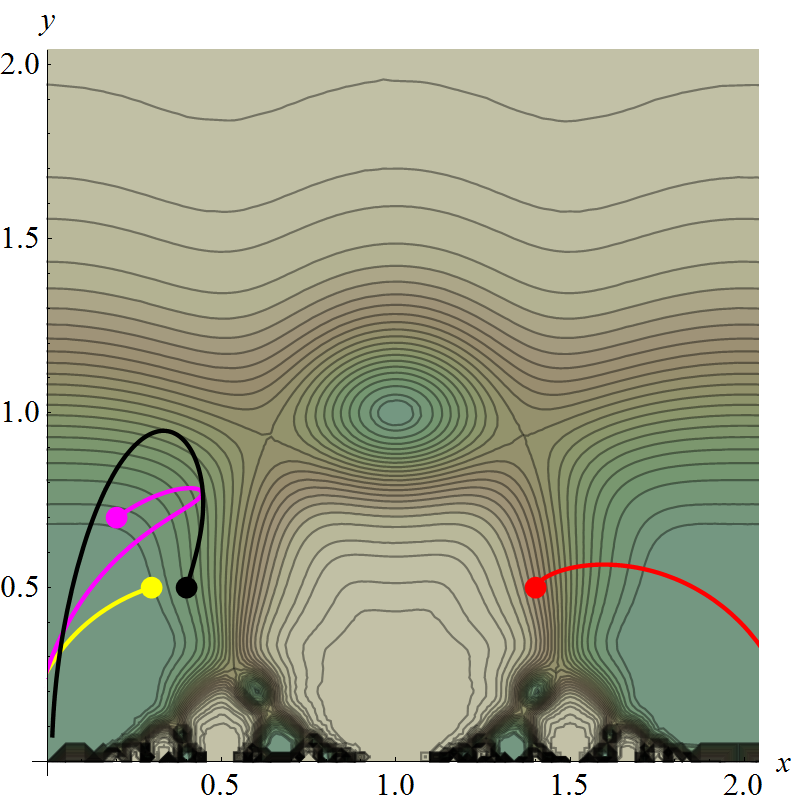}
\subcaption{Trajectories for $\tPhi=\tPhi_1$ on the Poincar\'e
  half-plane.}
\label{fig:TrajPhi1}
\end{minipage}\hfill
\begin{minipage}{.47\textwidth}
\centering \!\!\!\!\includegraphics[width=1.07\linewidth]{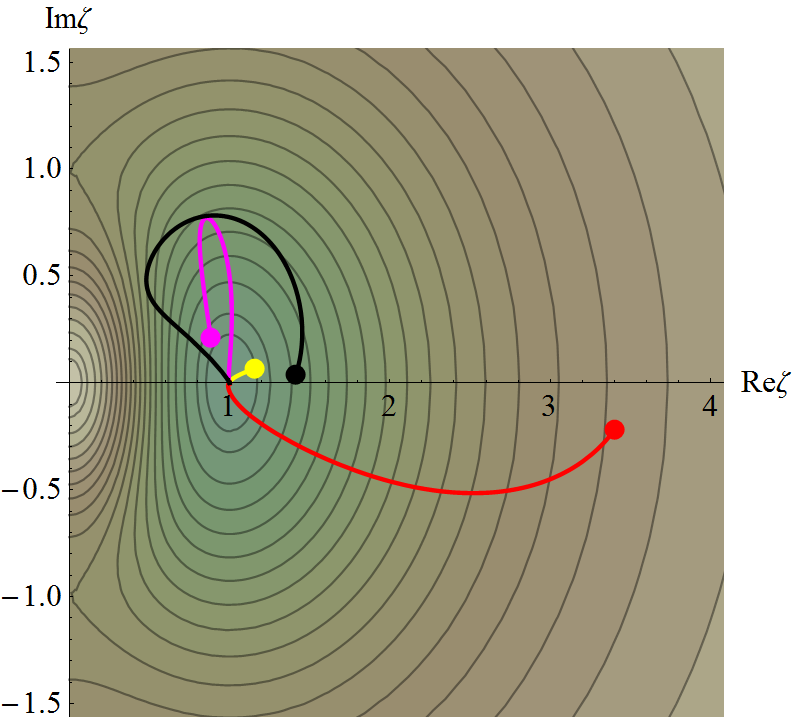}~~
\vskip 0.7em
\subcaption{Projection to $Y(2)$ of the trajectories shown at the left.}
\label{fig:TrajPhi1Proj}
\end{minipage}
\caption{Numerical solutions for $\Phi=\Phi_1$ and $\alpha=\frac{M_0}{3}$.}
\label{fig:TrajPhi1All}
\end{figure}

\noindent At the
cusp points, this potential tends either to its vanishing infimum (for
$\zeta=1$) or to its supremum, which equals $2M_0$ (for
$\zeta=0,\infty$). Furthermore, it has a minimum (equal to zero) at
the point $\zeta=-1$, which lies inside $Y(2)$. 

The four lifted
trajectories with initial conditions given in Table \ref{table:InCond}
and their projections to $Y(2)$ are shown in Figure
\ref{fig:TrajPhi1All}.  All four projected trajectories evolve toward
the cusp point $\zeta=1$, due to the combined effect of the potential
(which has an infimum there) and of the effective attractive force
produced by the cusp. 

\begin{figure}[H]
\centering
\begin{minipage}{.47\textwidth}
\centering \includegraphics[width=1.0\linewidth]{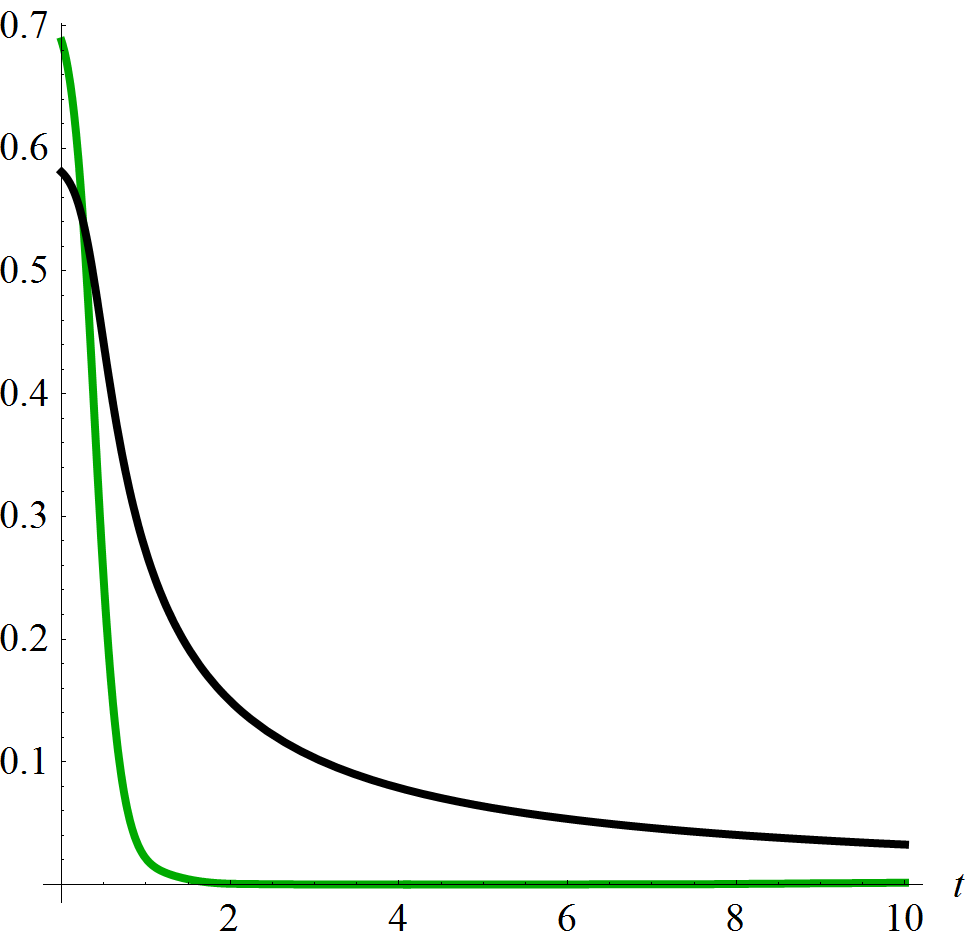}
\subcaption{Plot of $H(t)/\sqrt{M_0}$ and $H_c(t)/\sqrt{M_0}$ for the red trajectory.}
\label{fig:HPhi11}
\end{minipage}\hfill
\begin{minipage}{.47\textwidth}
\vskip 0.3em
\centering \includegraphics[width=1.02\linewidth]{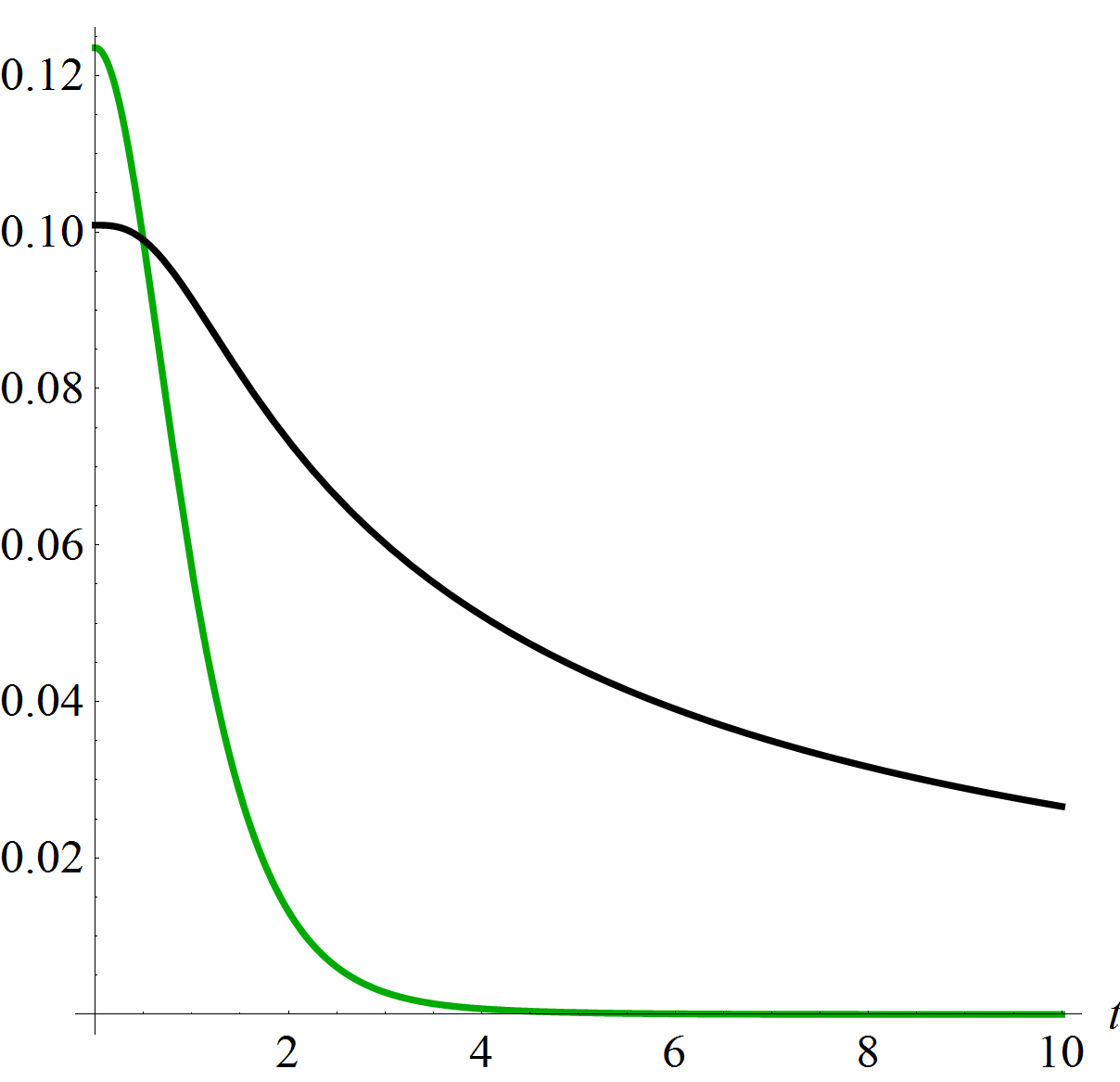}
\subcaption{Plot of $H(t)/\sqrt{M_0}$ and $H_c(t)/\sqrt{M_0}$ for the yellow trajectory.}
\label{fig:HPhi12}
\end{minipage}
\caption{Plot of $H(t)/\sqrt{M_0}$ (black) and $H_c(t)/\sqrt{M_0}$ (green) for the red and
  yellow trajectories with $\Phi=\Phi_1$ at $\alpha=\frac{M_0}{3}$.}
\label{fig:HPhi1}
\end{figure}

Figure \ref{fig:HPhi1} shows the time evolution
of the Hubble parameter for the red and yellow trajectories, both of
which lead to inflation for some (relatively short) cosmological time
interval starting at $t=0$.

\noindent The number of efolds ${\cal N}$ is given by integrating $H(t)$ 
over the first inflationary time interval:
\ben
\label{efolds}
{\cal N}=\int_{0}^{t_I}H(t)\dd t
\een
For all the four trajectories with initial conditions given in Table \ref{table:InCond}, 
calculating the number of efolds for those which start in inflationary regime (see 
Table \ref{table:InCondInf}) we find values between 0.05 and 3.55, which are smaller 
than the values of between 50-60 efolds expected by phenomenological measurements.
Nevertheless, we will show in the next subsection that we can also find
 trajectories with 50-60 efolds.

\subsection{Examples of cosmological trajectories with 50--60 efolds}

\paragraph{Trajectory for $\Phi_+$.}

\begin{figure}[H]
\centering
\begin{minipage}{.47\textwidth}
\centering \includegraphics[width=1.05\linewidth]{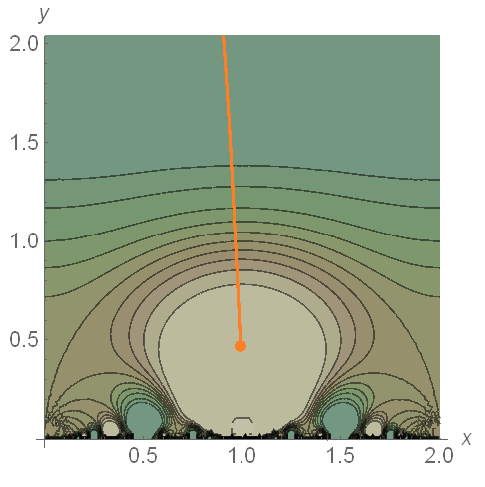}
\subcaption{Trajectory for $\tPhi_+/M_0$ on the Poincar\'e half-plane.}
\label{NTrajMphiP}
\end{minipage}\hfill
\begin{minipage}{.47\textwidth}
\centering \includegraphics[width=1.02\linewidth]{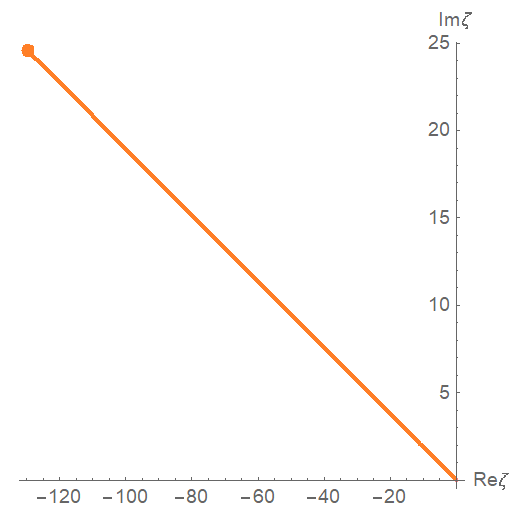}
\subcaption{Projection to $Y(2)$ of the trajectory shown at the left.}
\label{NProjTrajMphiP}
\end{minipage}
\caption{Numerical solution for the orange trajectory with  initial conditions 
$\tau_0=0.99+0.41\i$ and ${\tv}_0=0$ in the potential $\Phi=\Phi_+$ when $\alpha=\frac{M_0}{3}$. }
\label{MphiP}
\end{figure}

\begin{figure}[H]
\centering \includegraphics[width=63mm]{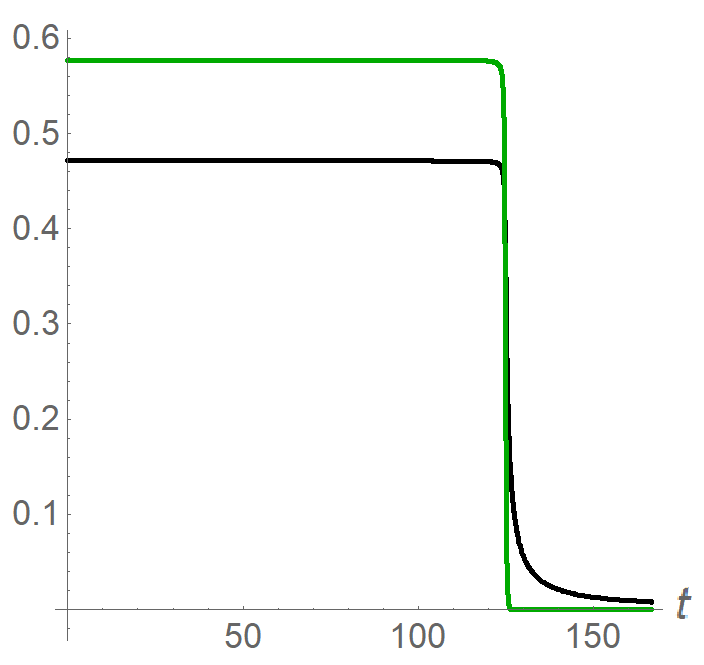}
\caption{Plot of $H(t)/\sqrt{M_0}$ (black) and $H_c(t)/\sqrt{M_0}$ (green) for the
orange trajectory in the potential $\Phi=\Phi_+$ when $\alpha=\frac{M_0}{3}$.}
\label{NHubMphiP}
\end{figure}

Choosing the initial conditions $\tau_0=0.99+0.41\i$\, and ${\tv}_0=0$, 
the trajectory obtained will have  ${\cal N}=58.8$ number of efolds. 
(See Figs. \ref{MphiP} and \ref{NHubMphiP}.)
With a small variation of $y_0$ we can vary the number of efolds between 50 and 60,
for example when  $\tau_0=0.99+0.415\i$ we get ${\cal N}=50.3$ efolds.

\paragraph{Trajectory for $\Phi_-$.}

\begin{figure}[H]
\centering
\begin{minipage}{.47\textwidth}
\centering \includegraphics[width=1.0\linewidth]{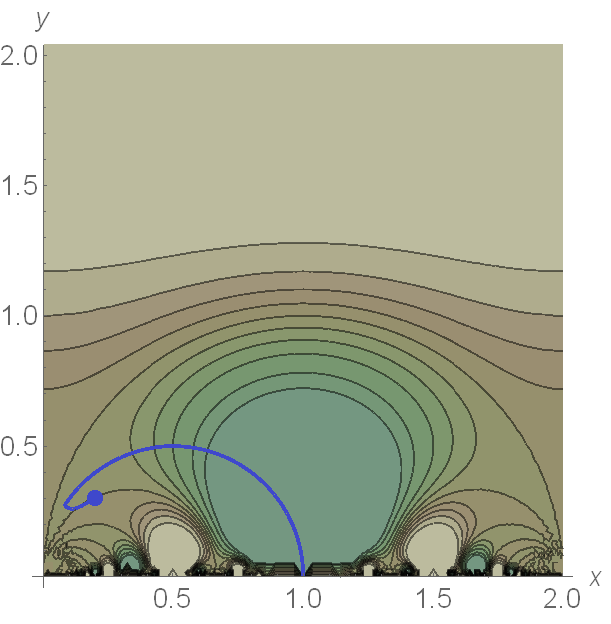}
%\vskip 0.5em
\subcaption{Trajectory for $\tPhi_-/M_0$ on the Poincar\'e half-plane.}
\label{NTrajMphiM}
\end{minipage}\hfill
\begin{minipage}{.47\textwidth}
\centering \includegraphics[width=1.03\linewidth]{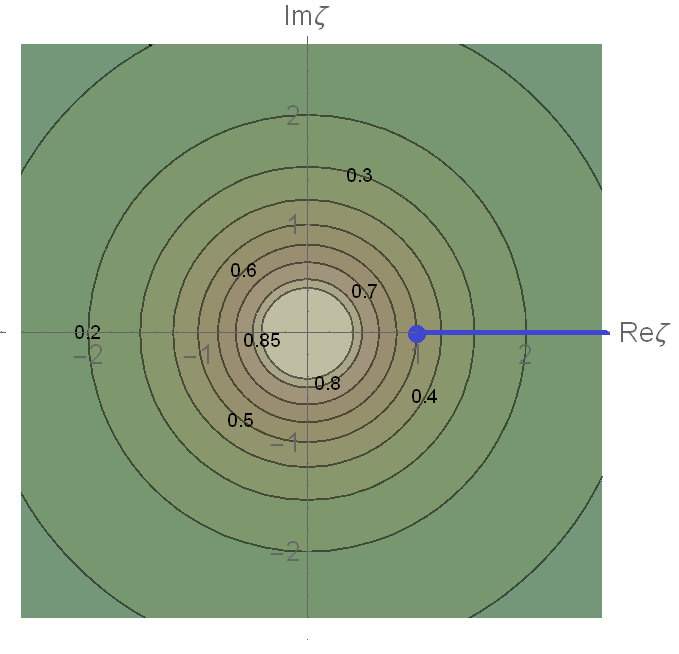}
\subcaption{Projection to $Y(2)$ of the trajectory shown at the left.}
\label{NProjTrajMphiM}
\end{minipage}
\caption{Numerical solution for the blue trajectory with  initial conditions 
$\tau_0=0.198+0.3\i$ and ${\tv}_0=0$ in the potential $\Phi=\Phi_-$ when $\alpha=\frac{M_0}{3}$.}
\label{MphiM}
\end{figure}

\begin{figure}[H]
\centering \includegraphics[width=62mm]{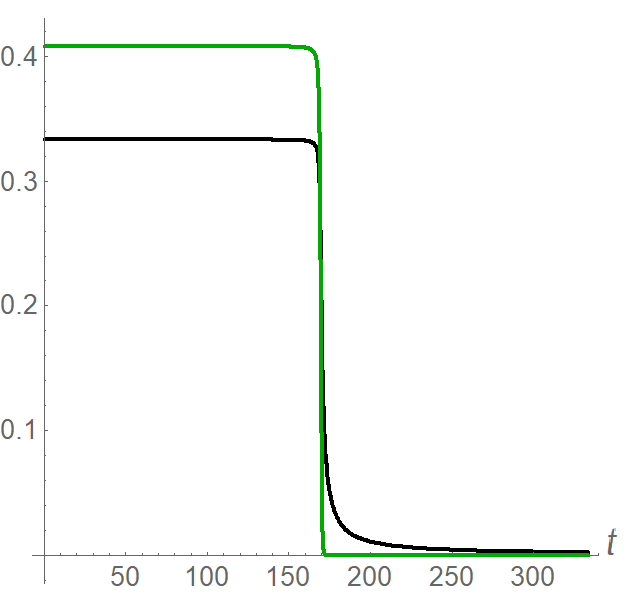}
\caption{Plot of $H(t)/\sqrt{M_0}$ (black) and $H_c(t)/\sqrt{M_0}$ (green) for the
blue trajectory in the potential $\Phi=\Phi_-$ when $\alpha=\frac{M_0}{3}$.}
\label{NHubMphiM}
\end{figure}

Choosing the initial conditions $\tau_0=0.198+0.3\i$\, and ${\tv}_0=0$, 
the trajectory obtained will have ${\cal N}=56$ efolds. (See Figs. \ref{MphiM} and \ref{NHubMphiM}.)

\section{The gradient flow approximation}
\label{sec:gradflow}

In this section, we explain how the gradient flow approximation of
\cite{genalpha} can be used to extract information about inflationary
trajectories and the number of efolds. In particular, we apply this
approximation in the vicinity of cusp ends, discussing a class of
scalar potentials for which certain special inflationary trajectories
produce any number of efolds within this approximation, including the
observationally favored number of $50-60$ efolds. Finally, we
construct explicit gradient flow trajectories of this type for those
well-behaved scalar potentials which have an asymptotic rotational
symmetry around cusp ends.

\subsection{The gradient flow approximation in general two-field models}

\noindent We first recall the {\em gradient flow approximation} of
\cite{genalpha} for a general two-field cosmological model (with flat
FLRW spatial section) having scalar manifold $(\Sigma,\cG)$. As
explained in loc. cit., this is the least restrictive of a ladder of
increasingly constraining approximations (of which the most
restrictive is the well-known SRST approximation). As shown in
\cite{genalpha}, the SRST approximation is of limited usefulness in
the study of generalized two-field $\alpha$-attractor models, since it
can fail near cusp ends. Since we are especially interested in the
vicinity of such ends, we cannot rely on SRST methods, so we will use
the gradient flow approximation instead.

As explained in \cite[Subsection 1.5]{genalpha}, the gradient flow
approximation applies to those (parts of) cosmological trajectories
along which the Hubble friction term of the cosmological equations of
motion dominates the acceleration term. In this approximation, one
replaces cosmological trajectories $t\rightarrow \varphi(t)$ by
appropriately reparameterized gradient flow curves of the scalar
potential $\Phi$, where the gradient of $\Phi$ is computed with
respect to the scalar manifold metric $\cG=3\alpha G$.

A gradient flow curve $\gamma$ of $\Phi$ (viewed as a curve oriented
{\em opposite} to the gradient vector field) admits three natural
parameters, namely:
\begin{itemize} \itemsep 0.0em
\item The {\em proper length parameter} $s$, which depends on the
origin chosen for $\gamma$ and on the metric $\cG$.
\item The {\em potential parameter}, obtained by restricting $\Phi$ to
$\gamma$. This parameter is strictly decreasing along the gradient flow
curve and depends on the origin chosen for $\gamma$ and on the scalar
potential $\Phi$.
\item The {\em gradient flow parameter} $q$, defined as the parameter
with respect to which the gradient flow equation\footnote{The first
equation of the system \eqref{gradflow}.} holds. This parameter
depends on the origin chosen for $\gamma$ and on both $\cG$ and $\Phi$.
\end{itemize}

\noindent Setting $\Phi(q)\eqdef \Phi(\gamma(q))$ and
$(\dd\Phi)(q)\eqdef (\dd\Phi)(\gamma(q))$, we have $||(\grad_\cG
\Phi)(\gamma(q))||_{\cG}=||(\dd \Phi)(q)||_\cG$ (since the
musical isomorphism of $(\Sigma,\cG)$ is an isometry).  The gradient
flow equation and the defining property $||\frac{\dd\gamma}{\dd
s}||=1$ of the proper length parameter imply:
\beqan
\label{gfparams}
&& s'(q)\eqdef \frac{\dd s(q)}{\dd q}~=+||(\dd \Phi)(q)||_\cG\Longrightarrow \dd q=+\frac{\dd s}{||\dd\Phi||_\cG}\nn\\
&&\Phi'(q)\eqdef \frac{\dd \Phi(q)}{\dd q}=-||(\dd \Phi)(q)||^2_{\cG}\Longrightarrow \dd q=-\frac{\dd \Phi}{||\dd \Phi||^2_{\cG}}~~.
\eeqan
In the gradient flow approximation, the cosmological trajectory is
approximated as $\varphi(t)\simeq \gamma(q(t))$, where $\dd t=3H \dd
q$. As a result, the cosmological equations of the model reduce to
(cf. \cite[eqs. (1.18)--(1.20)]{genalpha}):
\beqan
\label{gradflow}
&& \frac{\dd \gamma(q)}{\dd q}= -(\grad_{\cG}\Phi)(\gamma(q))~~\nn\\
&& \dd t=3H(q) \dd q\\
&& H(q)=\frac{1}{M\sqrt{6}}\Bigg(\Phi(q)+
\sqrt{\Phi(q)^2+ \frac{2}{3} M^2 ||(\dd \Phi)(q)||_\cG^2}~~\Bigg)^{1/2}~~,\nn
\eeqan
where as usual we assumed $H>0$. We shall also assume throughout that
$\Phi$ is positive everywhere. Formula \eqref{efolds} for the number
of efolds becomes:
\ben
\label{cNgrad}
\cN=3\int_\gamma H^2\dd q=\frac{1}{M\sqrt{6}}\int_\gamma \frac{1 +\sqrt{1+ f(s)^2}}{f(s)}\dd s~~,
\een
where we used the first of relations \eqref{gfparams} and we set:
\ben
\label{fdef}
f(s)\eqdef \sqrt{\frac{2}{3}} M\frac{||(\dd \Phi)(s)||_\cG}{\Phi(s)}=F(\gamma(s))~~,
\een
where $F:\Sigma\rightarrow \R_{\geq 0}$ is the smooth function defined through:
\ben
\label{Fdef}
F(p)\eqdef \sqrt{\frac{2}{3}} M\frac{||(\dd \Phi)(p)||_\cG}{\Phi(p)}~~,~~\forall p\in \Sigma~~.
\een
Hence in the gradient flow approximation 
the number $\cN$ of efolds realized on a
gradient flow curve $\gamma$ is controlled by the single function $F$
(which depends on the scalar potential $\Phi$ and on the metric
$\cG$). Notice that $F$ is everywhere non-negative and that it
vanishes precisely at the critical points of $\Phi$ (recall that we
assume $\Phi$ to be strictly positive everywhere). Inflation occurs
along the gradient flow curve $\gamma$ when $H(q)\leq
\frac{1}{M}\sqrt{\frac{\Phi(q)}{2}}$ (see eqs. \eqref{infcond0} and
\eqref{critHubble}). Using the last relation in \eqref{gradflow}, this
condition reads (cf.  \cite[Subsection 1.5]{genalpha}):
\ben
\label{infcondgrad}
f(s)\leq \sqrt{3}~~
\een
and provides a criterion for identifying the inflationary portions of
a gradient flow curve. In the gradient flow approximation, the
inflationary region $\cR\subset T\Sigma$ of the tangent bundle (see
Subsection \ref{subsec:infregion}) is `squeezed' to a closed subset
$\cR_o$ of $T\Sigma$ which projects onto the following subset of
$\Sigma$:
\be
\mathcal{Q}_o\eqdef \{p\in \Sigma| F(p)\leq \sqrt{3}\}~~,
\ee
 since in this approximation the velocity becomes a function
of the position:
\be
\dot{\gamma}(q)=-\frac{1}{3H(q)}(\grad_\cG \Phi)(\gamma(q))~~,
\ee
with $H(q)=H(\gamma(q))$ given by \eqref{gradflow}. 

\paragraph{Remark.}
We have $F=\sqrt{\frac{2}{3}} M ||\dd V||$, where $V=\log \Phi$. An
easy computation gives:
\ben
\label{dF}
\dd F= \sqrt{\frac{2}{3}} M \iota_T \Hess_\cG(V)~~,
\een
where $\Hess_\cG(V)\eqdef \nabla \dd V$ is the Hessian tensor\footnote{Here
$\nabla$ is the covariant derivative of differential forms induced by
the Levi-Civita connection of $\cG$.} of $V$ with respect to
$\cG$ and $\iota_T$ denotes contraction with the normalized gradient vector field:
\ben
\label{T}
T\eqdef \frac{\grad_\cG \Phi}{||\grad_\cG \Phi||}=\frac{\grad_\cG V}{||\grad_\cG V||}
\een
of $\Phi$. Notice that $T$ is well-defined except at the critical
points of $\Phi$.

\subsection{The number of efolds in the gradient flow approximation when $\Phi$ is Morse}
\label{subsec:cNgrad}

\noindent Suppose that the (everywhere positive) scalar potential
$\Phi$ is a Morse function on $\Sigma$ and let $\gamma$ be an
inextensible gradient flow curve which connects two critical points of
$\Phi$ lying on $\Sigma$. Thus $\gamma$ is an open curve which starts
at a critical point $p_+\in \Sigma$ and ends at a critical point
$p_-\in \Sigma$, while passing through no other critical point.  The
gradient flow parameter $q$ along $\gamma$ runs from $-\infty$ to
$+\infty$ and we have $\lim_{q\rightarrow \mp
\infty}\gamma(q)=p_\pm$. On the other hand, the length
$\ell(\gamma)\eqdef \int_\gamma \dd s$ of $\gamma$ is necessarily
finite and we can choose the proper length parameter $s$ along
$\gamma$ such that $\lim_{s\rightarrow 0}\gamma(s)=p_+$ and
$\lim_{s\rightarrow \ell(\gamma)}\gamma(s)=p_-$. In this situation, we
have $\lim_{s\rightarrow 0} f(s)=\lim_{s\rightarrow \ell(\gamma)}f(s)
=0^+$ since the function $F$ of equation \eqref{Fdef} vanishes at the
critical points of $\Phi$.

Let $\tau(s)\eqdef \frac{\dd \gamma(s)}{\dd s}=-T(\gamma(s))$ be the
normalized tangent vector to $\gamma$, where $T$ is the normalized
gradient vector field defined in equation \eqref{T}. Let $\tau_+\eqdef
\tau(0^+)$ and $\tau_-\eqdef \tau(0^-)$. The function
$f(s)=F(\gamma(s))$ has the following Taylor expansions at $s=0$ and
$s=\ell(\gamma)$:
\beqan
\label{fexp}
f(s)&=&_{\tiny{\frac{s}{\ell(\gamma)}\ll 1}} (\dd_{p_+} F)(\tau_+)s+\mathrm{o}(s)=\sqrt{\frac{2}{3}} M \Hess_\cG(V)_{p_+}(\tau_+,\tau_+)+\mathrm{o}(s) \\
f(s)&=&_{\tiny{1-\frac{s}{\ell(\gamma)}\ll 1}} (\dd_{p_-} F)(\tau_-)(\ell(\gamma)-s)+\mathrm{o}(\ell(\gamma)-s)=
\sqrt{\frac{2}{3}} M \Hess_\cG(V)_{p_-}(\tau_-,\tau_-)+\mathrm{o}(\ell(\gamma)-s)~~,\nn
\eeqan
where we used \eqref{dF} and the fact that $F(p_+)=F(p_-)=0$.
Since $f$ is continuous, the $f$-preimage of the
interval $(0,\sqrt{3})$ is an open connected set and hence:
\be
\{s\in (0,\ell(\gamma))|f(s)<\sqrt{3}\}=\cup_{j=1}^\kappa I_j~~,
\ee
where $I_j=(s'_j,s''_j)$ are open disjoint intervals such that
$s''_j<s'_{j+1}$. Here $\kappa$ is a strictly positive integer or
$\kappa=+\infty$ and we necessarily have $s'_1=0$ and
$s''_\kappa=\ell(\gamma)$\footnote{Notice that it can happen that
$f(s)<\sqrt{3}$ for all $s\in (0,\ell(\gamma))$. In that case, the
entire inextensible gradient curve is inflationary and we have
$\kappa=1$ and $I_1=(0,\ell(\gamma))$.}. Let $\gamma_j$ denote the
portion of $\gamma$ corresponding to $s\in I_j$ and let (cf. equation
\eqref{cNgrad}):
\ben
\label{cNj}
\cN_j\eqdef 3\int_{\gamma_j} H^2\dd q=\frac{1}{M\sqrt{6}}\int_{\gamma_j} \frac{1 +\sqrt{1+ f(s)^2}}{f(s)}\dd s=
\frac{1}{M\sqrt{6}}\int_{\gamma_j} \left[\frac{1}{f(s)} +\sqrt{1+\frac{1}{f(s)^2}}\right]\dd s~~.
\een
Relation \eqref{infcondgrad} (which holds with strict inequality on
each open curve $\gamma_j$) implies $\cN_j> \frac{1}{\sqrt{2}}
\frac{\ell(\gamma_j)}{M}$, where $\ell(\gamma_j)\eqdef \int_\gamma\dd
s$ is the proper length of $\gamma_j$. This gives the following upper
bound on $\ell(\gamma_j)$, which can be viewed as a constraint on the
allowed gradient flow curves (and hence on the allowed scalar
potentials and allowed scalar manifold metrics) in terms of the
observationally relevant quantity $\cN_j$:
\be
\ell(\gamma_j) < \sqrt{2} M \cN_j~~.
\ee
The first and last inflationary intervals $(0,s''_1)$ and
$(s'_N,\ell(\gamma))$ are especially interesting, since the
inextensible gradient flow curve $\gamma$ starts and ends at critical
points of $\Phi$. The following argument shows that
$\cN_1=\cN_\kappa=+\infty$, which implies that restricting $\gamma$ to
appropriate sub-intervals of either of the intervals $(0,s''_1)$ and
$(s'_\kappa,\ell(\gamma))$ allows one to produce any desired number of
efolds. To see this, notice that the expansions \eqref{fexp} imply
$\int_{I_1} \frac{\dd s}{f(s)}=\int_{I_\kappa}\frac{\dd
s}{f(s)}=+\infty$.  Since $\cN_1 \geq \frac{1}{M\sqrt{6}} \int_{I_1}
\frac{\dd s}{f(s)}$ and $\cN_\kappa \geq \frac{1}{M\sqrt{6}}
\int_{I_\kappa} \frac{\dd s}{f(s)}$, this gives:
\be
\cN_1=\cN_\kappa=+\infty~~.
\ee
Hence in the gradient flow approximation one can obtain any desired
number $\cN_o$ of efolds by considering extensible gradient flow
curves of one of the following types:
\begin{itemize}
\itemsep 0.0em
\item Extensible gradient flow curves of the form $[s_+,s'_1]\ni
s\rightarrow \gamma(s)$, where $s_+\in (0,s'_1)$ is chosen such that
$\frac{1}{M\sqrt{6}}\int_{\gamma_j} \frac{1 +\sqrt{1+
f(s)^2}}{f(s)}\dd s=\cN_o$~.
\item Extensible gradient flow curves of the form $[s''_\kappa,
s_-]\ni s\rightarrow \gamma(s)$, where $s_-\in (s''_1,\ell(\gamma))$
is chosen such that $\frac{1}{M\sqrt{6}}\int_{\gamma_j} \frac{1
+\sqrt{1+ f(s)^2}}{f(s)}\dd s=\cN_o$~.
\end{itemize} In particular, one can choose $s_+$ and $s_-$ for such
trajectories such that $\cN_o$ lies in the observationally favored
range of $50-60$ efolds.

\subsection{Inflation in the canonical neighborhood of a cusp end}
  
\noindent Let $(\Sigma,G)$ be any geometrically-finite hyperbolic
surface which has at least one cusp end and let $\dot{\cD}$ be a
canonical punctured neighborhood of such an end in the
Ker\'ekj\'art\'o-Stoilow compactification $\hSigma$. Such a
neighborhood is diffeomorphic with a punctured disk, where the cusp
end corresponds to the puncture. The restriction of the target space
metric $\cG=3\alpha G$ to $\dot{\cD}$ takes the form (see
eq. \eqref{ccuspmetricsg} or \cite[eq. (D.5)]{genalpha}):
\be
\dd s^2_{\cG}|_{\dot{\cD}}=3\alpha\left[\dd r^2+\frac{e^{-2r}}{(2\pi)^2}\dd \theta^2\right]~~,
\ee
where $r\in (0,+\infty)$ and $\theta$ is an angular variable of period
$2\pi$. The cusp end corresponds to $r\rightarrow +\infty$ and we have:
\ben
\label{dpncusp}
||\dd \Phi||_\cG^2|_{\dot{\cD}}=\frac{1}{3\alpha}\left[(\partial_r \Phi)^2+(2\pi)^2 e^{2r} (\partial_\theta\Phi)^2\right]~~.
\een
On this neighborhood of the cusp end, consider an extensible gradient flow
trajectory which starts at $r=r_0$ and $\theta=\theta_0$. The gradient
flow equation (the first equation in \eqref{gradflow}) is equivalent
with the following system of first order ODEs:
\beqan
\label{gradflowcusp}
\frac{\dd r}{\dd q} &=& -\frac{1}{3\alpha}(\partial_r \Phi)(r(q),\theta(q))\nn\\
\frac{\dd \theta}{\dd q} &=& -\frac{(2\pi)^2e^{2r}}{3\alpha}(\partial_\theta \Phi)(r(q),\theta(q))~~.
\eeqan
On the other hand, equation \eqref{Fdef} gives:
\ben
\label{Fcusp}
F=\sqrt{\frac{2}{9\alpha}} M \frac{\sqrt{(\partial_r \Phi)^2+(2\pi)^2 e^{2r} (\partial_\theta\Phi)^2}}{\Phi}~~.
\een

\paragraph{Solution by quadratures close to a cusp end.}

Suppose that $\Phi$ has the following asymptotic expansion for large
$r$ (cf. \cite[Subsection 2.3]{genalpha}):
\ben
\label{Phias}
\Phi(r,\theta)=_{r\gg 1} a-b(\theta)e^{-r} +\mathrm{O}(e^{-2r})=_{x\ll 1} a-b(\theta) x+
\mathrm{O}(x^2)~~
\een
where\footnote{The
constant $a$ should not be confused with the conformal factor $a(t)$
of the FLRW universe.} $a>0$, $b(\theta)>0$ for all $\theta$ and $x\eqdef
e^{-r}$. When the function $b(\theta)$ is non-constant, such a
potential leads to spiral inflation nearby the cusp end
(cf. \cite[Subsection 2.6]{genalpha}). It is easy to see that all
globally well-behaved scalar potentials considered in Section
\ref{sec:traj} have such asymptotic expansions around those
cusp ends of $Y(2)$ at which the corresponding extended potential has
a maximum. To first order in the expansion \eqref{Phias}, equation
\eqref{dpncusp} gives:
\be
||\dd \Phi||_\cG^2|_{\dot{\cD}}=\frac{1}{3\alpha}\left[b(\theta)^2 e^{-2r}+(2\pi)^2 b'(\theta)^2\right]~~,
\ee
where $b'(\theta)\eqdef \frac{\dd b(\theta)}{\dd \theta}$.  For $r\gg
1$, the gradient flow system \eqref{gradflowcusp} reduces to:
\beqan
\label{gradflowas}
\frac{\dd r}{\dd q} &=& -\frac{1}{3\alpha} b(\theta) e^{-r}\nn\\
\frac{\dd \theta}{\dd q} &=& +\frac{(2\pi)^2}{3\alpha}b'(\theta) e^r~~,
\eeqan
while relation \eqref{Fcusp} becomes:
\ben
\label{Fas}
F=\sqrt{\frac{2}{9\alpha}} M \frac{\sqrt{b(\theta)^2 e^{-2r}+(2\pi)^2 b'(\theta)^2}}{\Phi}~~.
\een

\

\noindent The gradient flow system \eqref{gradflowas} can be written as:
\ben
\label{qelim}
\dd q=\frac{3\alpha}{(2\pi)^2}e^{-r}\frac{\dd\theta}{b'(\theta)}=-3\alpha \frac{\dd (e^r)}{b(\theta)}~~.
\een
The second equality allows us to express $r=r(\theta)$ or
$\theta=\theta(r)$ using quadratures:
\ben
\label{rtheta}
e^{r}=e^{r_0}\sqrt{1-\frac{2e^{-2r_0}}{(2\pi)^2} Q(\theta)}~~,~~Q(\theta)=\frac{(2\pi)^2}{2} \left(e^{2r_0}-e^{2r}\right)~~,
\een
where:
\ben
\label{Qdef}
Q(\theta)\eqdef \int_{\theta_0}^\theta \frac{b(\theta')}{b'(\theta')}\dd\theta'
\een
is the primitive of $\frac{b(\theta)}{b'(\theta)}$ which vanishes at
$\theta=\theta_0$ and we let $\theta$ vary in $\R$, extending $b$ to a
$2\pi$-periodic function of $\theta$. Notice that $Q(\theta)$ is
$2\pi$-periodic.

\paragraph{Remark.}{The expressions above make sense provided that the
partially-defined smooth function $\frac{b(\theta)}{b'(\theta)}$ has
{\em isolated} singularities at the zeroes of $b'\eqdef \frac{\dd
  b}{\dd\theta}$ and has a partially-defined primitive on the interval
$[\theta_0,\theta]$. This excludes the case when $b$ is constant,
which must be treated separately (see below). Since $b(\theta)$ is
periodic of period $2\pi$, its derivative $b'(\theta)$ is also
$2\pi$-periodic and necessarily has at least two zeros within any
compact interval of length $2\pi$, because $b$ attains its absolute
maximum and absolute minimum within any such interval. If $\theta_c$
is a critical point of $b$ and $b''(\theta_c)\neq 0$, then the
primitive $Q(\theta)$ has a logarithmic singularity at
$\theta=\theta_c$ because $b(\theta_c)>0$ (recall that we assume 
$b$ to be strictly positive everywhere). Since $r$ decreases starting
from $r_0$ during the gradient flow, the second equation in
\eqref{rtheta} requires $Q(\theta)\geq 0$. Since we restricted to the
punctured neighborhood $\dot{\cD}$ of the cusp end, we have
$r>0$ and hence the first equation in \eqref{rtheta} requires
$Q(\theta)<\frac{(2\pi)^2}{2} e^{2r_0}$. This second condition shows
that $\theta(q)$ cannot reach a critical point of $b$ (i.e. a
singularity of $Q$) along a connected piece of gradient flow
trajectory restricted to lie within $\dot{\cD}$. In particular,
$Q'(\theta)$ has fixed sign (and hence $Q(\theta(q))$ is strictly
monotonous as a function of $q$) along such a piece of gradient flow
trajectory. This implies that the function $Q$ is non-singular and
invertible for $\theta$ varying within the interval allowed on such a
trajectory piece.}

\

\noindent Using \eqref{qelim} and \eqref{Fas}, we can write \eqref{cNgrad}
in two equivalent forms:

\begin{enumerate}[1.]
  \itemsep 0.0em
  \item As the following integral over $\theta$:
\ben
\label{cN1}
\cN =\frac{3\alpha}{2(2\pi)^2 M^2}\int_\gamma \frac{ x\left(a-b(\theta)x +\sqrt{[a-b(\theta)x]^2+
\frac{2M^2}{9\alpha} \left[b(\theta)^2 x^2+(2\pi)^2
  b'(\theta)^2\right]}\right)}{b'(\theta)}\dd \theta~~,
\een
where $x=e^{-r}$ is expressed as follows in terms of $\theta$ using
the first equation in \eqref{rtheta}:
\ben
\label{x}
x=x_0\left[1-\frac{x_0^2}{2\pi^2} Q(\theta)\right]^{-1/2}~~.
\een
The condition for inflation \eqref{infcondgrad} becomes:
\be
M\sqrt{b(\theta)^2 x^2 +(2\pi)^2 b'(\theta)^2} \leq \sqrt{\frac{27\alpha}{2}}  \left[a-b(\theta) x\right]~~,
\ee
where $x$ is given by \eqref{x}.

\item As the following integral over $y\eqdef e^r$:
\ben
\label{cN2}
\cN= -\frac{3\alpha}{2M^2}\int_\gamma \frac{ ay-b(\theta) +\sqrt{[ay-b(\theta)]^2+
\frac{2M^2}{9\alpha} \left[ b(\theta)^2+(2\pi)^2 b'(\theta)^2 y^2\right]}}{b(\theta) y}\dd y~~,
\een
where $\theta$ is expressed as follows in terms of $y$ using the
second equation in \eqref{rtheta}:
\ben
\label{theta}
\theta=Q^{-1}\left(2\pi^2(y_0^2-y^2)\right)~~.
\een
The condition for inflation \eqref{infcondgrad} becomes:
\be
M\sqrt{b(\theta)^2 +(2\pi)^2 b'(\theta)^2y^2} \leq \sqrt{\frac{27\alpha}{2}}  \left[ay-b(\theta) \right]~~,
\ee
where $\theta$ is given by \eqref{theta}.
\end{enumerate}

\noindent Relations \eqref{cN1} and \eqref{cN2} allow one to compute
$\cN$ by quadratures for any portion of gradient flow trajectory along
which $r\gg 1$, once the primitive $Q(\theta)$ has been determined
from equation \eqref{Qdef}.

\paragraph{Remark.}
The system \eqref{gradflowas} can also be written as:
\beqan
\label{gfsys}
\frac{\dd}{\dd q} (e^r) &=&-\frac{1}{3\alpha} b(\theta)\nn\\
 \frac{1}{b'(\theta)} \frac{\dd\theta}{\dd q} &=& \frac{(2\pi)^2}{3\alpha} e^r~~.
\eeqan
Differentiating the second relation with respect to $q$ and using the
first gives the following non-linear second order equation for the
function $\theta(q)$:
\ben
\label{thetaeq}
\frac{\dd}{\dd q}\left[\frac{1}{b'(\theta)} \frac{\dd\theta}{\dd q}\right]+\frac{(2\pi)^2}{9\alpha^2} b(\theta)=0~~.
\een
The initial conditions $r|_{q=0}=r_0$ and $\theta|_{q=0}=\theta_0$ amount to:
\be
\theta|_{q=0}=\theta_0~~,~~\frac{\dd \theta}{\dd q}\Big\vert_{q=0}= \frac{(2\pi)^2}{3\alpha} b'(\theta_0) e^{r_0}~~.
\ee
If $\theta(q)$ is the unique solution of \eqref{thetaeq} which satisfies these two conditions,
then the first equation of \eqref{gfsys} determines $r(q)$ as:
\be
r(q)=\log\left[e^{r_0}-\frac{1}{3\alpha} \int^q_0 b(\theta(q'))\dd q'\right]~~.
\ee

\paragraph{Example: Potentials with asymptotic rotational symmetry at a cusp end.}

Let us consider the degenerate case when $b$ is independent of
$\theta$, which means that the scalar potential $\Phi$ has an {\em
asymptotic} $\SO(2)$ symmetry at the chosen cusp end. In this situation,
we say that $\Phi$ is {\em asymptotically symmetric} at that end. Notice
that such a potential need not have any strict symmetry even when
restricted to an arbitrarily small neighborhood of the cusp end. Instead, such
a potential $\Phi$ has only an {\em approximate} $\SO(2)$ symmetry
near that end, up to $\theta$-dependent corrections which are of order at
least two in the quantity $e^{-r}$. It is obvious that there exists a
continuous infinity of smooth, globally well-behaved and compactly
Morse scalar potentials on the modular curve $Y(2)$ which are
asymptotically symmetric at each of the three cusp ends of $Y(2)$. As
we demonstrate below, such scalar potentials allow one to obtain
explicit inflationary trajectories which lead to the observationally
favored range for the number of efolds. One should notice that the
situation considered in this paragraph is rather special. In
particular, the scalar potentials of Section \ref{sec:traj} are {\em
not} asymptotically symmetric at the cusp ends of $Y(2)$. By contrast,
the special class of scalar potentials discussed in this paragraph
allows one to obtain good values for $\cN$ by considering a
`universal' class of inflationary trajectories.

When $\Phi$ is asymptotically symmetric at a cusp end, the gradient
flow system \eqref{gradflowas} reduces to:
\beqa
\frac{\dd r}{\dd q} &=& -\frac{b}{3\alpha} e^{-r}\nn\\
\frac{\dd \theta}{\dd q} &=& 0~~,
\eeqa
with the solution:
\ben
\label{rotinvsol}
y=y_0-\frac{b}{3\alpha} q~~,~~\theta=\theta_0~~,
\een
where $y=e^r$. Hence gradient flow trajectories proceed radially away
from the cusp end.  The first relation in \eqref{rotinvsol} gives $\dd
q=-\frac{3\alpha}{b}\dd y$, whereby \eqref{cNgrad} becomes:
\be
\cN=\frac{1}{3c}\int_\gamma \frac{ay -b +\sqrt{(ay-b)^2+
c b^2}}{by} \dd y~~,
\ee
where $c\eqdef \frac{2M^2}{9\alpha}$. This agrees with
\eqref{cN2}. The condition for inflation takes the form: \be M b \leq
\sqrt{\frac{27\alpha}{2}} (ay-b) \Longleftrightarrow y \geq y_I\eqdef
\left(1+\sqrt{\frac{2}{27 \alpha}}
M\right)\frac{b}{a}=\left(1+\sqrt{\frac{c}{3}}\right)\frac{b}{a}~~,
\ee which means that inflation stops at $r=r_I\eqdef\log y_I$.  The
integral above can be performed by an Euler substitution, with the
result $\cN=\bN(y_0)-\bN(y_I)$, where:
\beqa
3c \bN(y) &=& \frac{a y+\sqrt{(a y-b)^2+b^2 c}}{b} + \left(\sqrt{c+1}-1\right) \log\left[ay\right] +
\log \left[b-ay+\sqrt{(a y-b)^2+b^2 c}\right]-\nn\\
&-& \sqrt{c+1} \log \left[b(c+1) -a y+\sqrt{c+1} \sqrt{(a y-b)^2+b^2 c}\right]~~.
\eeqa
Of course, the asymptotic expansion \eqref{Phias} only holds for
sufficiently large $y=e^r$, hence consistency requires $y_I\gg 1$,
i.e.:
\ben
\label{consistency}
\frac{a}{b} \ll \sqrt{\frac{c}{3}} \Longleftrightarrow \frac{a}{b}\ll  M\sqrt{\frac{2}{27\alpha}}~~.
\een
This consistency condition constraints the asymptotic coefficients of
the scalar potential $\Phi$ at the cusp end in terms of $M$ and
$\alpha$. We have:
\be
\bN(y_I)=\frac{1}{3c}\left[1+\sqrt{3c}+\sqrt{c+1} \log \left(\frac{1-\sqrt{3c} +2 \sqrt{c+1}}{c}\right)-(1+\sqrt{c+1})\log \left(1+\sqrt{\frac{3}{c}}\right)\right]
\ee
and:
\be
\lim_{y_0\rightarrow +\infty}\bN(y_0)=+\infty~~,
\ee
as could be expected\footnote{For a globally well-behaved and
compactly Morse scalar potential $\Phi$ defined on any
geometrically-finite hyperbolic surface $\Sigma$ which has only cusp
ends, one can show that the argument of Subsection \ref{subsec:cNgrad}
can also be applied to any inextensible gradient flow trajectory which
connects two critical points of the extended scalar potential
$\hPhi$.} from the discussion of Subsection \ref{subsec:cNgrad}. Hence
we can obtain any desired value of $\cN$ by choosing an appropriate
initial value $r_0>r_I$ for $r$.  It particular, we can always choose
$r_0$ such that $\cN$ lies in the observationally favored range of $50-60$
efolds. This is similar to what happens for ordinary one-field
$\alpha$-attractor models.

\section{Conclusions and further directions}
\label{sec:conclusions}

We studied classical cosmological trajectories in generalized
two-field $\alpha$-attractor models defined by the non-compact modular
curve $Y(2)=\C\P^1\setminus \{0,1,\infty\}=\C\setminus \{0,1\}$,
endowed with its unique complete hyperbolic metric. For a few natural
choices of globally well-behaved scalar potentials, we computed field
trajectories using the methods proposed in \cite{genalpha}, finding
that the models display intricate behavior. When the scalar potential
is invariant under the action of the anharmonic group $\fB\simeq S_3$,
we showed that such models provide a geometric interpretation of the
$\PSL(2,\Z)$-invariant cosmological models of \cite{Sch1,Sch2}, in
which the Poincar\'e half-plane is replaced by the surface $Y(2)$
while the infinite discrete group $\PSL(2,\Z)$ is replaced by the
finite group $\fB$. The relation between our model and that considered
in \cite{Sch1,Sch2} is induced by the elliptic modular lambda
function, which provides an infinite to one field redefinition
eliminating most of the countably infinite unphysical ambiguity
present in the Poincar\'e half-plane description. As explained in the
main text, this relation between the two types of models is somewhat
subtle, due to the presence of the residual symmetry provided by the
anharmonic action of the symmetric group $S_3$ on $Y(2)$, which
permutes the three cusp ends of the hyperbolic triply-punctured
sphere.

The present work can be extended in a few directions. For example, it
would be interesting to compute cosmological perturbations and
observables in generalized two-field $\alpha$-attractor models defined by
$Y(2)$, for various choices of the scalar potential; this should relate to the 
results obtained in \cite{Sch2} for the lifted model. Since the SRST
approximation in curved field space \cite{PT1,PT2} cannot be applied to most
trajectories, such computations can be done numerically using the
methods developed in \cite{Dias1, Dias2, Mulryne}. The models
discussed in \cite{elem} and in this paper could serve as a testing
ground for such methods and for probing the corresponding range of
phenomenological predictions.

Given any finite index subgroup $\Gamma$ of $\PSL(2,\Z)$, the quotient
$Y(\Gamma)=\H/\Gamma$ is called a (non-compact) {\em modular
curve}. When $\Gamma$ is torsion-free (i.e. contains no elliptic
elements), the non-compact hyperbolic surface $Y(\Gamma)$ is
geometrically-finite, smooth and has only cusp ends.  In particular,
$Y(\Gamma)$ has finite area and $\Gamma$ is a Fuchsian group of the
first kind.  The classical (but non-generic) case arises when $\Gamma$
is a congruence subgroup of $\PSL(2,\Z)$. All torsion-free congruence
subgroups of $\PSL(2,\Z)$ whose modular curve has genus zero were
classified in reference \cite{Sebbar}, where it was shown that there
exist $33$ possibilities up to conjugation inside $\PSL(2,\Z)$. The
simplest of these is the group $\Gamma(2)$, whose non-compact modular
curve $Y(2)\eqdef Y(\Gamma(2))$ is the triply-punctured sphere
considered in the present paper. For any of the other $32$ subgroups,
$Y(\Gamma)$ is a sphere with more than $3$ punctures. It would be
interesting to study generalized two-field $\alpha$-attractor models
defined by these modular curves.

When the scalar potential is preserved by the action of the anharmonic
group on $Y(2)$, the generalized two-field $\alpha$-attractor model
has a finite $S_3$ symmetry. This symmetry cannot be eliminated
directly within the framework of \cite{genalpha}, since (as explained
in Subsection \ref{subsec:singmetric}) taking the Riemannian quotient
through the anharmonic group action would lead to a singular
hyperbolic metric on the complex plane $\C=Y(2)/\fB$, which cannot be
used as a scalar manifold metric in the usual construction of
non-linear sigma models. However, it is likely that the definition of
non-linear sigma models admits a physically-appropriate generalization
to the case when the scalar manifold $\Sigma$ is replaced by a
Riemannian orbifold\footnote{In the context of models with $\cN=1$
supersymmetry, proposals in this direction were made in
\cite{HS}. Here we are interested in the non-supersymmetric case.}. It
would be interesting to investigate this problem in connection with
the larger subject of discrete symmetries of theories coupled to
gravity \cite{BS}, especially in view of applications to string theory
compactifications \cite{D1,D2}.

Finally, it would be interesting to look for embeddings of such models
in string theory, perhaps along the lines suggested in \cite{MSU}. We
hope to report on this and related problems in future work.

\acknowledgments{\noindent The work of E.M.B. and C.I.L. was supported by grant
  IBS-R003-S1.}

\appendix 

\section{The anharmonic action on $Y(2)$}
\label{app:anharmonic}

\paragraph{The group $\PSL(2,\Z_2)$.}

\noindent 
The group $\PSL(2,\Z_2)$ has six elements, which we denote by: 
\beqan
&& a_1=\left[\begin{array}{cc} {\hat 1} & {\hat 0} \\ {\hat 0} & {\hat 1} \end{array}\right]~~,
~~a_2=\left[\begin{array}{cc} {\hat 0} & {\hat 1} \\ {\hat 1} & {\hat 1} \end{array}\right]~~,
~~a_3=\left[\begin{array}{cc} {\hat 1} & {\hat 1} \\ {\hat 1} & {\hat 0} \end{array}\right]~~,\nn\\
&& a_4=\left[\begin{array}{cc} {\hat 0} & {\hat 1} \\ {\hat 1} & {\hat 0} \end{array}\right]~~,
~~a_5=\left[\begin{array}{cc} {\hat 1} & {\hat 1} \\ {\hat 0} & {\hat 1} \end{array}\right]~~,
~~a_6=\left[\begin{array}{cc} {\hat 1} & {\hat 0} \\ {\hat 1} & {\hat 1} \end{array}\right]~~.
\eeqan 
It is isomorphic with the permutation group $S_3$ on three elements,
where $a_1$, $a_2$ and $a_3$ (the last two of which have order $3$)
correspond to the alternating subgroup $A_3$ (which for $S_3$ consists of 
the cyclic permutations) while $a_4$, $a_5$ and $a_6$ (which have order $2$)
correspond to the transpositions. 

\paragraph{The anharmonic action of $\PSL(2,\Z_2)$ on $Y(2)$.}

Given $A\in \PSL(2,\Z)$, we have $\lambda(A\tau)=\lambda(\tau)$ for all
$\tau\in \H$ iff $A$ belongs to the subgroup $\Gamma(2)$. Hence the
orbits of $\Gamma(2)$ on $\H$ consist of the $\lambda$-preimages of
the points of $Y(2)$. The short exact sequence: 
\be
1\rightarrow \Gamma(2)\hookrightarrow \PSL(2,\Z)\stackrel{\mu_2}{\longrightarrow} \PSL(2,\Z_2)
\ee
gives the isomorphism of groups: 
\ben
\label{is}
\PSL(2,\Z)/\Gamma(2)\simeq \PSL(2,\Z_2)~~.
\een
The group $N(\Gamma(2))/\Gamma(2)=\PSL(2,\Z)/\Gamma(2)$ acts on the
quotient $Y(2)=\H/\Gamma(2)$ as in \eqref{adef}. Composing this with
the isomorphism \eqref{is} gives the {\em anharmonic action} of
$\PSL(2,\Z_2)$ on $Y(2)$:
\be
a\zeta=\lambda(A\tau)~~,~~\forall a=\mu_2(A)\in \PSL(2,\Z_2)~~,~~\forall \zeta=\lambda(\tau)\in Y(2)~~,
\ee
where $A\in \PSL(2,\Z)$ and $\tau\in \H$. A collection of coset
representatives for $a_j$ in the quotient $\PSL(2,\Z)/\Gamma(2)$ is
provided by the following elements of $\PSL(2,\Z)$:

\beqan
\label{Alifts}
&& A_1=\left[\begin{array}{cc} 1 & 0 \\ ~0~ & 1 \end{array}\right]~~,
~~A_2=\left[\begin{array}{cc} 0 & ~~1 \\ -1 & ~~1 \end{array}\right]~~,
~~A_3=\left[\begin{array}{cc} 1 & -1 \\ 1 & 0 \end{array}\right]~~,\nn\\
&&A_4=\left[\begin{array}{cc} 0 & 1 \\ -1 & 0 \end{array}\right]~~,
~~A_5=\left[\begin{array}{cc} ~1~ & 1 \\ 0 & ~1~ \end{array}\right]~~,
~~A_6=\left[\begin{array}{cc} ~1~ & 0 \\ 1 & 1 \end{array}\right]~~,
\eeqan 
which act on $\H$ as follows: 
\beqa
&& A_1\tau=\tau~~~~,~~A_2\tau=\frac{1}{1-\tau}~~,~~A_3\tau=\frac{\tau-1}{\tau}~~,\nn\\
&& A_4\tau=-\frac{1}{\tau}~~,~~A_5\tau=\tau+1~~,~~A_6\tau=\frac{\tau}{\tau+1}~~.
\eeqa
Notice that $A_2,A_3$ and $A_4$ are elliptic, while $A_5$ and $A_6$
are parabolic. Using these lifts of $a_j$, we find that the
anharmonic action of $\PSL(2,\Z_2)$ on $Y(2)$ is given by:
\beqan
\label{anharmonic}
&& a_1\zeta=\zeta~~~~~,~~~~a_2\zeta=\frac{1}{1-\zeta}~~,~~a_3\zeta=\frac{\zeta-1}{\zeta}~~,\nn\\
&& a_4\zeta=1-\zeta~~,~~a_5\zeta=\frac{\zeta}{\zeta-1}~~,~~a_6\zeta=~~\frac{1}{\zeta}~~,
\eeqan
thus coinciding with the restriction to $Y(2)$ of the action on
$\C\P^1$ of the following elements of the M\"obius group $\PSL(2,\C)$:
\beqan
\label{B}
&& B_1=\left[\begin{array}{cc} 1 & 0 \\ ~0~ & 1 \end{array}\right]~~,
~~B_2=\left[\begin{array}{cc} 0 & 1 \\ -1 & 1 \end{array}\right]~~,
~~B_3=\left[\begin{array}{cc} 1 & -1 \\ 1 & 0 \end{array}\right]~~,\nn\\
&&B_4=\left[\begin{array}{cc} -\i & \i \\ 0 & \i \end{array}\right]~~,
~~B_5=\left[\begin{array}{cc} ~\i & ~~0 \\ ~\i & -\i \end{array}\right]~~,
~~B_6=\left[\begin{array}{cc} ~0~ & \i \\ \i & 0 \end{array}\right]~~.
\eeqan

\paragraph{The anharmonic group.}
 
The six elements \eqref{B} form a classical subgroup $\fB$ of
$\PSL(2,\C)$ known as the {\em anharmonic group}. The correspondence
$a_j\rightarrow B_j$ gives an isomorphism of groups between
$\PSL(2,\Z_2)$ and $\fB$ showing, in particular, that $\fB$ is
isomorphic with the permutation group $S_3$. The anharmonic group
coincides with the group of biholomorphisms of the complex curve
$Y(2)=\C\P^1\setminus \{0,1,\infty\}$, consisting of those elements of the
M\"obius group $\PSL(2,\C)$ which stabilize the set of punctures
$\{0,1,\infty\}\subset \C\P^1$. The elements of $\fB$ permute the set
of punctures as shown in Table \ref{table:A} (which in particular
gives an explicit isomorphism between $\fB$ and the permutation group
$S_3$).

\

\begin{table}[H]
\centering
\begin{tabular}{|c|c|c|c|}
\toprule
$B_j$ & $0$ & $1$ & $\infty$\\
\midrule\midrule
$B_1$ & $0$ & $1$ & $\infty$ \\
\hline
$B_2$  & $1$  & $\infty$ & $0$ \\
\hline
$B_3$ & $\infty$ & $0$ & $1$ \\
\hline
$B_4$ & $1$ & $0$ & $\infty$  \\
\hline
$B_5$ & $0$  & $\infty$ & $1$\\
\hline
$B_6$ & $\infty$  & $1$ & $0$ \\
\bottomrule
\end{tabular}
\caption{Action of the anharmonic group on the punctures
  $\{0,1,\infty\}$.  Each of the punctures is fixed by one of the
  subgroups $\{1,B_j\}$ (with $j=4,5,6$), which corresponds to the $\Z_2$
  subgroup of $S_3$ generated by the corresponding transposition.}
\label{table:A}
\end{table}

\

\noindent 
The elements $B_4, B_5$ and $B_6$ respectively fix the points $1/2$,
$2$ and $-1$ on $Y(2)$, while each of $B_2$ and $B_3$ fixes both of
the points $e^{\frac{\i\pi}{3}}$ and $e^{- \frac{\i\pi}{3}}$. On the
entire Riemann sphere, the elements $B_4$, $B_5$ and $B_6$ also
respectively fix the punctures $\infty$, $0$ and $1$ (see Table
\ref{table:Afix}).

\

\begin{table}[H]
\centering
\begin{tabular}{|c|c|c|c|c|c|}
\toprule Element & $B_2$ & $B_3$ & $B_4$ & $B_5$ &
$B_6$\\ \midrule\midrule $Y(2)$ & $e^{\pm\frac{\i\pi}{3}} $ & $e^{\pm
  \frac{\i\pi}{3}}$ & $1/2$ & $2$ & $-1$\\ \hline $\C\P^1\setminus
Y(2)$ & none & none & $\infty$ & $0$ & $1$\\ \bottomrule
\end{tabular}
\caption{Fixed points for the action of $B_j$ ($j=2,\ldots, 5$) on
  $\C\P^1$. For each element, the fixed points lying on $Y(2)$ and
  those that belong to the set $\{0,1,\infty\}$ of punctures are shown
  on separate lines.}
\label{table:Afix}
\end{table}

\

\noindent The points $\{e^{+\frac{\i\pi}{3}},e^{-\frac{\i\pi}{3}}\}$ form an
orbit of $\fB$, as do the points $\{-1,1/2,2\}$. The action of
$\fB$ on these orbits is given in Tables \ref{table:Afix2} and
\ref{table:Afix3}.

\begin{table}[H]
\centering
\begin{tabular}{|c||c|c|c|c|c|c|}
\toprule  & $B_1$ & $B_2$ & $B_3$ & $B_4$ & $B_5$ & $B_6$\\ 
\midrule\midrule 
$e^{\frac{\i\pi}{3}}$ & $e^{\frac{\i\pi}{3}}$ & $e^{\frac{\i\pi}{3}}$ & $e^{\frac{\i\pi}{3}}$ & $e^{-\frac{\i\pi}{3}}$ & $e^{-\frac{\i\pi}{3}} $ & $e^{-\frac{\i\pi}{3}}$\\ 
\hline 
$e^{\frac{-\i\pi}{3}}$ & $e^{-\frac{\i\pi}{3}}$ & $e^{-\frac{\i\pi}{3}}$ & $e^{-\frac{\i\pi}{3}}$ &$e^{\frac{\i\pi}{3}}$ & $e^{\frac{\i\pi}{3}}$ & $e^{\frac{\i\pi}{3}} $ \\ 
\bottomrule
\end{tabular}
\caption{Action of $\fB$ on the orbit
  $\{e^{+\frac{\i\pi}{3}},e^{-\frac{\i\pi}{3}}\}$. Each point of this
  orbit is fixed by the subgroup $\{B_1,B_2,B_3\}$ of $\fB$,
  which corresponds to the alternating subgroup $A_3\simeq \Z_3$ of $S_3$.}
\label{table:Afix2}
\end{table}

\begin{table}[H]
\centering
\begin{tabular}{|c||c|c|c|c|c|c|}
\toprule  & $B_1$ & $B_2$ & $B_3$ & $B_4$ & $B_5$ & $B_6$\\ 
\midrule\midrule 
$-1$ & $-1$ & $1/2$ & $2$ & $2$ & $1/2$ & $-1$\\ 
\hline 
$1/2$ & $1/2$ & $2$ & $-1$ &$1/2$ & $-1$ & $2$ \\ 
\hline
$2$ & $2$ & $-1$ & $1/2$ &$-1$ & $2$ & $1/2$ \\ 
\bottomrule
\end{tabular}
\caption{Action of $\fB$ on the orbit $\{-1,1/2,2\}$. Each point of
  this orbit is fixed by one of the subgroups $\{1,B_j\}$ (with $j=4,5,6$),
  which corresponds to the $\Z_2$ subgroup of $S_3$ generated by the
  corresponding transposition.}
\label{table:Afix3}
\end{table}

\noindent The anharmonic group coincides with the group of
orientation-preserving isometries of the hyperbolic surface
$(Y(2),G)$:
\be
\Iso^+(Y(2),G)=\fB\simeq \PSL(2,\Z_2)\simeq S_3~~.
\ee
Since $\fB\simeq \PSL(2,\Z_2)\simeq \PSL(2,\Z)/\Gamma(2)$, the
topological quotient $Y(2)/\fB$ can be identified with the
topological quotient $\H/\PSL(2,\Z)$. 

\section{Fundamental polygons and local cusp coordinates for $Y(2)$}
\label{app:fpolygons}
\subsection{Fundamental polygons}

\paragraph{On the Poincar\'e half-plane.}

\begin{figure}[H]
\centering
\begin{minipage}{.47\textwidth}
\centering
\includegraphics[width=0.8\linewidth]{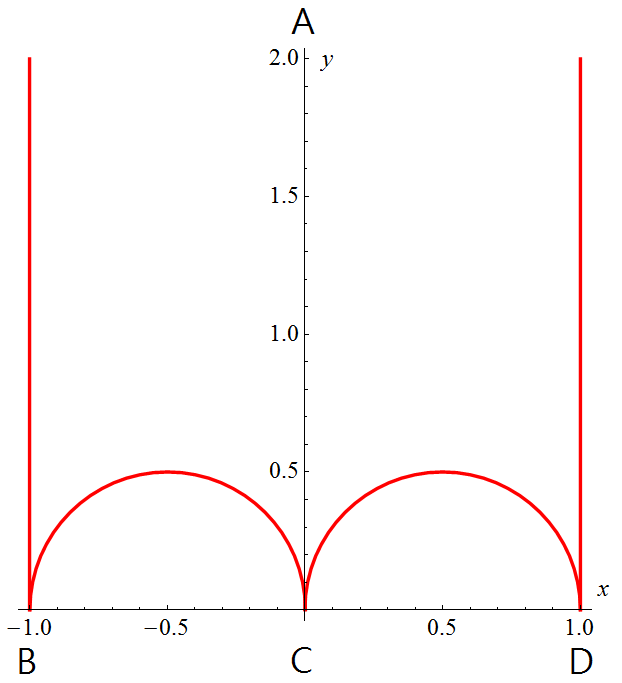}
\subcaption{A fundamental polygon $\fD_\H$ on the Poincar\'e half-plane.}
\label{fig:FunDomH}
\end{minipage}\hfill
\begin{minipage}{.47\textwidth}
\centering
\vspace{-3mm}
\includegraphics[width=0.9\linewidth]{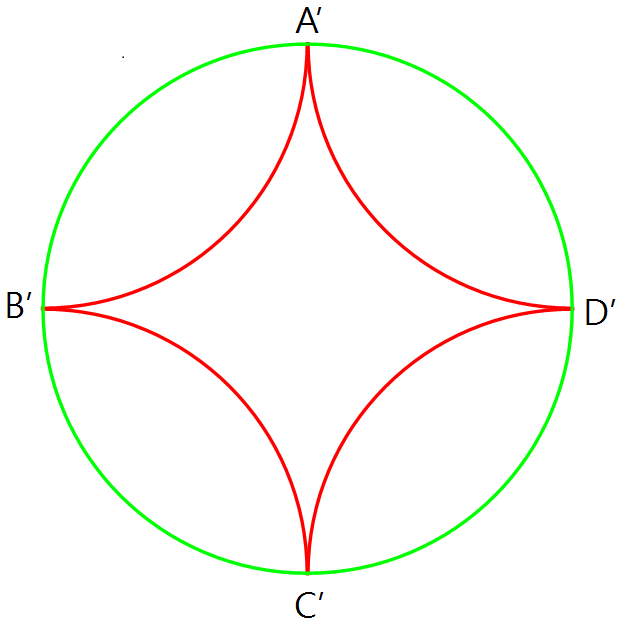}
\subcaption{A fundamental polygon $\fD_\mD$ on the hyperbolic disk.}
\label{fig:FunDomD}
\end{minipage}
\caption{Fundamental polygons for the uniformization of $Y(2)$ to
  the Poincar\'e half-plane and to the hyperbolic disk.}
\label{fig:FunDom}
\end{figure}

\noindent A fundamental polygon for the action of $\Gamma(2)$ on $\H$
is given \cite[Sec. 7]{Ahlfors} by the hyperbolic quadrilateral (see
Figure \ref{fig:FunDomH}):
\be
\fD_{\mathbb{H}} =\{\tau \in \H| -1 < \Re \tau < 0, |\tau+\frac{1}{2}|> \frac{1}{2} \}\cup \{\tau \in \H| 0 \leq \Re \tau < 1, |\tau-\frac{1}{2}|> \frac{1}{2}\}~~,
\ee
whose ideal vertices are located at the following points on the
conformal boundary of the Poincar\'e half-plane:
\be
A: \tau=\infty~,~B:\tau=-1~,~C:\tau=0~,~D:\tau=1~~.
\ee
The translation $\tau\rightarrow \tau+2$ gives the Poincar\'{e}
pairing between the sides (AB) and (AD), while the transformation
$\tau\rightarrow \frac{\tau}{2\tau +1}$ gives the Poincar\'{e} pairing
between the sides $(BC)$ and $(CD)$, which are the two bounding circle
arcs. We have:
\be
\lim_{\tau\rightarrow \pm 1}\lambda(\tau)=\infty~~,~~\lim_{\tau\rightarrow 0}\lambda(\tau)=1~~,~~\lim_{\tau\rightarrow \infty}\lambda(\tau)=0~~,
\ee
where the limits are taken from within $\fD_\H$. 
The function $\lambda$ maps the relative frontier of
$\fD_{\mathbb{H}}$ onto the region $(-\infty,0)\cup (1,+\infty)$ of
the real axis, taking each of the vertical sides into the interval
$(-\infty,0)$ and each of the two half-circles into the interval
$(1,+\infty)$. The vertical half-line which cuts $\fD$ through the
middle (and which is part of the $y$ axis) is mapped to the interval
$(0,1)$.

\paragraph{On the hyperbolic disk.}

The fundamental polygon $\fD_{\mD}$ on the hyperbolic disk with
complex coordinate $u$ which corresponds to $\fD_\H$ through the
Cayley transform is the ideal hyperbolic quadrilateral with vertices
located at the points (see Figure \ref{fig:FunDomD}):
\be
A': u=\i ~,~B':u=-1~,~C':u=-\i~,~D':u=1~~.
\ee
The diagonals of this quadrilateral are orthogonal to each
other. Decomposing $\fD_\mD$ into ideal triangles gives the hyperbolic
area of $Y(2)$:
\be
\area(Y(2))=2\pi~~.
\ee

\subsection{Local holomorphic cusp coordinates on $Y(2)$}

\noindent Local holomorphic coordinates $u_0,u_1,u_\infty$ can be constructed
around each of the punctures $\zeta=0,1,\infty$ as explained in
\cite[Sec. 5.4]{genalpha}:
\begin{enumerate}[1.]
\itemsep 0.0em
\item The point $\tau=\infty$ (which projects to $\zeta=0$) is
  invariant under the action of the cyclic subgroup of $\gamma$
  generated by the parabolic element $P_\infty$ of \eqref{gens}, which
  acts as the translation $\tau\rightarrow \tau+2$. The element
  $T_\infty=\left[\begin{array}{cc}\frac{1}{\sqrt{2}}& 0\\0
      &\sqrt{2}\end{array}\right]\in \PSL(2,\R)$ acts as
  $T_\infty\tau=\frac{\tau}{2}$ and satisfies $(T_\infty P_\infty
  T_\infty^{-1})(\tau)=\tau+1$. Hence the holomorphic cusp coordinate
  around $\zeta=0$ is given by \cite{genalpha}:
\be
u_0=e^{2\pi \i T_\infty \tau}=e^{\i \pi \tau}=e^{\i \pi\mu(\zeta)}~~.
\ee
In the expression above, it is understood that one chooses a branch of
the multivalued inverse \eqref{mu} for which $\tau=\mu(\zeta)$ belongs
to the cusp domain:
\be
\cC_\infty=\{\tau\in \H \, | \, \Im \tau > 2\}~~.
\ee
This insures that $u_0$ covers the punctured disk $0<|u_0|<e^{-2\pi}$
(see \cite{genalpha,elem}).
\item The point $\tau=0$ (which projects to $\zeta=1$) is invariant
  under the action of the cyclic subgroup of $\Gamma$ generated by the
  parabolic element $P_0$ of \eqref{gens}. The element
  $T_0=\left[\begin{array}{cc} 0 &\frac{1}{\sqrt{2}} \\ -\sqrt{2} &
      0\end{array}\right]\in \PSL(2,\R)$ acts as
  $T_0\tau=-\frac{1}{2\tau}$ and satisfies $(T_0 P_\infty
  T_0^{-1})(\tau)=\tau+1$. Hence the holomorphic cusp coordinate
  around $\zeta=1$ is given by:
\be
u_1=e^{2\pi \i T_0 \tau}=e^{-\frac{\i \pi}{\tau}}=e^{-\frac{\i \pi}{\mu(\zeta)}}~~.
\ee
In the expression above, one must choose a branch of $\mu$ for which $\tau=\mu(\zeta)$ 
belongs to the cusp domain: 
\be
\cC_0=\{\tau\in \H \, | \, \Im \frac{1}{\tau}<-2\}=\{\tau\in \H \, | \, \Im \tau> 2|\tau|^2\}~~.
\ee
This insures that $u_1$ covers the punctured disk $0<|u_1|<e^{-2\pi}$
.
\item The point $\tau=1$ (which projects to $\zeta=\infty$) is
  invariant under the action of the cyclic subgroup of $\Gamma$
  generated by the parabolic element:
\be
P_1\eqdef P_\infty P_0=\left[\begin{array}{cc} -3 & 2\\ -2&
    1\end{array}\right]~~,
\ee
which acts as $P_1\tau=\frac{2-3\tau}{1-2\tau}$.  The element
$T_1=\left[\begin{array}{cc} 0 &-\frac{1}{\sqrt{2}}\\ \sqrt{2} &
    -\sqrt{2} \end{array}\right]\in \PSL(2,\R)$ acts as
$T_1\tau=\frac{1}{2(1-\tau)}$ and satisfies $(T_1 P_1
T_1^{-1})(\tau)=\tau-1$.  Hence the holomorphic cusp coordinate around
$\zeta=\infty$ is given by:
\be
u_\infty=e^{2\pi \i T_1 \tau}=e^{\frac{\i \pi}{1-\tau}}=e^{\frac{\i \pi}{1-\mu(\zeta)}}~~.
\ee
In the expression above, one must choose a branch of
$\mu$ for which $\tau=\mu(\zeta)$ belongs to the cusp domain:
\be
\cC_1=\{\tau\in \H \, | \, \Im \left(\frac{1}{1-\tau}\right) >2\}=\{\tau\in \H \, | \, \Im \tau > 2 |1-\tau|^2\}~~.
\ee
This insures that $u_\infty$ covers the punctured disk
$0<|u_\infty|<e^{-2\pi}$.
\end{enumerate}

\

\noindent Let $c\in \{0,1,\infty\}$ be any of the punctures, $u_c$ be
the local holomorphic cusp coordinate introduced above and 
$\dot{\cD}_c$
be the punctured neighborhood of $c$ in $Y(2)$ defined by $0<|u_c|<e^{-2\pi}$. 
The restriction of the hyperbolic metric \eqref{metric} to
$\dot{\cD}_c$ is given by:
\ben
\label{cuspmetric}
\dd s^2|_{\dot{\cD}_c}=\Lambda(u_c,\bar{u}_c)^2|\dd u_c|^2~~,~\mathrm{with}~~\Lambda(u_c,\bar{u}_c)=\frac{1}{|u_c|\log(1/|u_c|)}~~.
\een
This is the standard form of the hyperbolic cusp metric
\cite{Borthwick,elem}. It can be brought to the
form:
\ben
\label{ccuspmetricsg}
\dd s^2|_{\dot{\cD}_c}=\dd r^2+\frac{e^{-2r}}{(2\pi)^2}\dd\theta^2~~
\een
upon passing to semi-geodesic coordinates $(r_c,\theta_c)$ around the
puncture $c$ through the transformation \cite{elem}:
\be
u_c=e^{2\pi e^{r_c}+\i \theta_c}~~.
\ee
In these coordinates, the puncture $c$ is located at $r_c\rightarrow
+\infty$ and $\theta$ runs from $0$ to $2\pi$.

\end{document}